\documentclass{article}
\usepackage{changes} 
\usepackage{graphicx,subcaption}
\usepackage{amsmath,natbib,verbatim,booktabs, longtable,multirow,lscape}
\usepackage{enumerate, color, url}
\usepackage{dsfont,mathtools}
\usepackage{txfonts}
\usepackage{algorithm}
\usepackage{bm}
\usepackage{paralist}
\usepackage{setspace}
\usepackage{listings}
\usepackage{amsfonts}
\usepackage[utf8]{inputenc}
\usepackage[english]{babel}
\usepackage{algpseudocode}
\usepackage{fixmath}
\usepackage{pifont}
\usepackage{longtable}
\usepackage{lscape}
\usepackage{ulem}
\usepackage{footnote}
\usepackage{hyperref}
\usepackage[toc,page]{appendix}
\usepackage[flushleft]{threeparttable}

\newtheorem{innercustomlem}{Lemma}
\newenvironment{customlem}[1]
  {\renewcommand\theinnercustomlem{#1}\innercustomlem}
  {\endinnercustomlem}

\newcommand{\indep}{\rotatebox[origin=c]{90}{$\models$}} 
\algnewcommand{\Initialize}[1]{%
  \State \textbf{Initialize:}
  \Statex \hspace*{\algorithmicindent}\parbox[t]{.8\linewidth}{\raggedright #1}
}

\usepackage{lineno}
\date{}
\usepackage{authblk}

\begin{document}

\def\spacingset#1{\renewcommand{\baselinestretch}%
{#1}\small\normalsize} \spacingset{1}

  \title{\bf Bayesian Edge Regression in Undirected Graphical Models to Characterize Interpatient Heterogeneity in Cancer}
  \author[1,2]{Zeya Wang }
    \author[3]{Veera Baladandayuthapan}
  \author[4]{Ahmed O. Kaseb}
  \author[5]{Hesham M. Amin}
  \author[6]{Manal M. Hassan}
  \author[2]{Wenyi Wang}
  \author[7]{Jeffrey S. Morris \thanks{Correspondence: Jeffrey.Morris@pennmedicine.upenn.edu}}

  \affil[1]{Department of Statistics, Rice University}
  \affil[2]{Department of Bioinformatics and Computational Biology, University of Texas MD Anderson Cancer Center}
  \affil[3]{Department of Biostatistics, University of Michigan}
  \affil[4]{Department of Gastrointestinal Medical Oncology, The University of Texas MD Anderson Cancer Center}
  \affil[5]{Department of Hematopathology, The University of Texas MD Anderson Cancer Center}
  \affil[6]{Department of Epidemiology, The University of Texas MD Anderson Cancer Center}
  \affil[7]{Department of Biostatistics, Epidemiology and Informatics, University of Pennsylvania}

  \maketitle

\bigskip
\begin{abstract}
\noindent  Graphical models are commonly used to discover associations within gene or protein networks for complex diseases such as cancer.  Most existing methods estimate a single graph for a population, while in many cases, researchers are interested in characterizing the heterogeneity of individual networks across subjects with respect to subject-level covariates. Examples include assessments of how the network varies with patient-specific prognostic scores or comparisons of tumor and normal graphs while accounting for tumor purity as a continuous predictor.  In this paper, we propose a novel edge regression model for undirected graphs, which estimates conditional dependencies as a function of subject-level covariates. Bayesian shrinkage algorithms are used to induce sparsity in the underlying graphical models. We assess our model performance through simulation studies focused on comparing tumor and normal graphs while adjusting for tumor purity and a case study assessing how blood protein networks in hepatocellular carcinoma patients vary with severity of disease, measured by \textit{HepatoScore}, a novel biomarker signature measuring disease severity.  
\end{abstract}

\noindent%

\spacingset{1.45} 
\section{Introduction} \label{s::intro}  
The proliferation of new technologies that can simultaneously measure genetic, transcriptomic, and proteomic markers have revolutionized biomedical research and contributed to the advent of \textit{precision therapy}, whereby medical treatment strategies are tailored to individual patients on the basis of molecular characteristics of their disease.  While certain individual genes have key biological roles in healthy and/or diseased cells, molecular processes relevant to the functional behavior of multi-cellular organisms or complex diseases are not determined by individual genetic factors, but rather complex interactions of various molecules at various molecular resolution levels.  \textit{Network Biology} is a nascent and burgeoning subfield of systems biology that involves the discovery and characterization of molecular interactions underlying complex diseases, including cancer.  Graphical models, which characterize the conditional dependency structure among random variables, are widely used in genomic studies to build networks representing interactions among different biological units, including genes and proteins.

There has been a great deal of work on graphical models over the past decade.  One model class shown to be useful for discovering biological networks is the \textit{undirected graphical model}, for which nodes index random variables and edges connecting nodes represent the global conditional dependency structure among the variables \citep{ref13}. A popular tool in studying undirected graphs is the \text{Gaussian graphical model}, for which conditional independence, the absence of an edge, corresponds to a zero entry in the \textit{precision} (or \textit{concentration}) matrix of multivariate Gaussian distribution \citep{ref5}, which also attracts growing interest in the recent development of distributed statistical learning \citep{lee2017communication}.

However, most of the graphical model work in existing literature involves estimation of a single network for a population, while inter-patient heterogeneity in many complex diseases, including cancer, suggests that these networks may vary across patients.  Characterization of this heterogeneity has the potential to reveal insights into the differences in molecular processes across patients that can lead to the discovery of novel precision therapy strategies.  One way of characterizing inter-patient network heterogeneity is to assess how the networks vary across patient-level covariates.  Two specific examples that have motivated this work include tumor purity and prognostic indices explaining inter-patient heterogeneity in cancer.

\textbf{Accounting for tumor purity in biological networks.} Tumor samples are inherently heterogeneous, with different types of cells present in a clinically derived sample, potentially confounding, to a large extent, the downstream analysis of gene expression or protein profiling of solid tumors \citep{ref38,ref31}. In practice, tumor samples invariably contain some contaminating normal tissue, and the proportion of a sample that is pure tumor, called \textit{tumor purity}, varies from sample to sample.  For this reason, many of the standard tumor versus normal comparisons are biased, typically attenuated, because they do not adjust for tumor purity and assume that the tumor samples are pure tumors. This principle also holds true in more advanced analyses including gene or protein networks, as any comparison of normal and tumor networks would be similarly biased by this factor. Deconvolution models such as \textit{DeMixT} \citep{ref9} can be fit to molecular data for tumor and normal samples in order to obtain an estimate of the tumor purity, $\pi_i$ for each sample $i=1,\ldots,N$.  Including this measurement as a continuous covariate in a graph regression model enables estimation of pure normal and pure tumor networks in a way that adjusts for this heterogeneous contamination.

\textbf{Differential biological networks by severity of disease.} The characterization of inter-patient heterogeneity within a given cancer can contribute to new precision therapy strategies.  For example, important biological mechanisms can be revealed by assessments of how various gene-gene or protein-protein networks strengthen or weaken with advancing disease.  In hepatocellular carcinoma (HCC), \cite{HepatoScore} has developed a novel prognostic signature computed from a patient's plasma protein profile called the \textit{HepatoScore}. The \textit{HepatoScore} for a given patient is a score $\pi_i \in [0,1]$ that quantifies the degree of aberration in the patient's blood protein profile relative to healthy subjects, with $\pi \approx 0$ indicating a protein profile essentially no different from a healthy subject, $\pi \approx 1$ indicating that patient's profile is maximally aberrant, and $\pi$ in between (e.g., $\pi \approx 0.5$) indicating a moderate level of aberration relative to healthy subjects.  Although determined without consideration of any patient-level clinical factors (i.e., unsupervised), \textit{HepatoScore} has demonstrated remarkable prognostic separability (low/medium/high \textit{HepatoScore} with median survival of 38.2/18.3/7.1 months) in a set of 766 HCC patients. This biological score contains more prognostic information than standard factors such as metastasis or nodal involvement, and provides a significant refinement of existing staging systems, e.g. with metastatic HCC patients with low HepatoScore having substantially better prognosis than non-metastatic HCC patients with high \textit{HepatoScores}.  \textit{HepatoScore} can be shown to be driven by a number of key proteins, including some in key pathways relevant to HCC such as growth hormone (GH), angiogenesis, and immune response.  By modeling how protein networks vary across the continuous covariate \textit{HepatoScore}, we can assess which protein-protein connections characterize advanced disease and provide molecular insights into this inter-patient heterogeneity.

\textbf{Literature on heterogeneous graphical models.}  There are a number of papers in the existing literature on heterogeneous graphs, but none that precisely solves the problem underlying our motivating examples. There are a number of papers on “group graphs,” in which graphical models are jointly estimated for discretized groups of subjects in a way that estimates group-specific graphs, while also borrowing strength between groups on common edges. Two-sample inference can be used to test differential edges between groups. \cite{xia2015testing} developed a multiple testing procedure to detect gene-by-gene interactions with binary traits, while \cite{narayan2015two} proposed a novel resampling, random penalization, and random effects method for testing to identify the functional brain connections between two groups from neuroimages. Many other works focus on more than two groups \citep{adref7,ref3,cai2016joint,liu2017structural,saegusa2016joint}. \cite{liu2017structural} extended the two-sample test to capture the structural similarities and differences among multiple Gaussian graphical models. \cite{adref7} jointly estimated multiple graphical models by incorporating a hierarchical penalty for common factors and group-specific factors. \cite{ref3} developed a more general model by employing fused lasso or group lasso penalties to encourage shared edges across the estimated precision matrices. \cite{saegusa2016joint} applied a Laplacian shrinkage penalty to encourage similarity among estimates from related subpopulations, and further proposed a Laplacian penalty based on hierarchical clustering for unknown population structures. Most recently, Bayesian approaches have become popular in modeling group graphs to construct the differential biological networks \citep{ref2,tan2017bayesian,mitra2016bayesian,lin2017joint}. \cite{ref2} utilized a Markov random field prior for encouraging the common structure between different groups. \cite{tan2017bayesian} investigated metabolic associations with the effect of cadmium through inducing multiplicative priors on the graphical space. \cite{lin2017joint} proposed a Bayesian neighborhood selection method that jointly estimates multiple Gaussian graphical models for data with both spatial and temporal structure through naturally incorporating this structure. Motivated by the progress made for the joint estimation of multiple graphs, there is a growing development to have heterogeneous graphical models with a more relaxed assumption for observations \citep{yang2014mixed,hao2017simultaneous}. \cite{hao2017simultaneous} proposed a method to learn a cluster structure of data while estimating multiple graphical models that does not need to specify the membership of observations. \cite{yang2014mixed} developed a class of mixed graphical models, in which each node-conditional distribution with a graphical model belongs to a possibly different univariate exponential family, therefore allowing random variables to be from heterogeneous domain sets for complex data. While groups can be viewed as categorical covariates, these methods do not model graphical variation across the continuous covariates of primary interest as in our motivating examples and the present paper.

Other methods model heterogeneity across covariates with graphs or covariance matrices, but are not suitable for our setting. \cite{ref16} proposed a covariance regression model that regresses a covariance matrix on a set of explanatory variables using a factor analysis. \cite{zou2017covariance} studied different estimators to parameterize the covariance matrix as a function of predictors. \cite{cai2012covariate} proposed a covariate-adjusted Gaussian graphical model that regress a p-dimensional vector of responses on a q-dimensional vector of covariates, but the precision matrix does not depend on predictors, so this method only evaluates how the nodes change with covariates, but not node-to-node dependencies. \cite{ref17} and \cite{ref18} developed dynamic undirected graph models varying with time. \cite{ref19} modeled multivariate binary data using an Ising model to study the change of dependency with covariates. While some of these designs model covariance heterogeneity, these methods either do not provide node-specific inference, deal with covariance rather than precision matrices, or cannot be applied to general regression settings with multiple covariates for Gaussian graphical models.   

 \cite{ref15} proposed \textit{Graph-optimized classification and regression trees} to partition the covariate space and estimate the graph within each partitioned subspace.  While quite flexible, as reported by \cite{ref19}, this model lacks interpretation of the graphical model and covariates, and it has the undesirable property that graphs constructed for covariate values close to each are not necessarily similar. A machine learning method proposed by \cite{ref10} applied a penalized kernel smoothing approach and allowed the precision matrix to change with covariates. One weakness of this method is that it ignores the intrinsic symmetry of the precision matrix, which may result in contradictory, unclear results in neighborhood selection and subsequent interpretation. Similarly using a kernel regression-based approach, \cite{lee2018nonparametric} proposed another covariate dependent graphical model that utilizes a nonparametric mixture of Gaussian graphical models with a single scalar covariate controlling mixture probability and distribution. The finite mixture model effectively clusters the subjects into discrete subgroups based on a partitioning of the covariate space, and then estimates separate graphs for each partition point. This method shares similar limitations as found with kernel based methods: it lacks a clear interpretation of the change of graph structure with respect to the covariates. Plus, more fundamentally, it can only handle a single covariate, not multiple covariates as in a general graph regression modeling framework as we develop in this paper. \cite{ni2018bayesian} constructed Bayesian graphical regression models for directed acyclic graphs (\textit{DAG}), which enable directed graphs to vary with general covariates, but their approach does not work in the undirected graph setting, which poses additional challenging difficulties and is our primary interest here. To our knowledge, none of the existing literature has considered building a regression model for edges in undirected graphs allowing general linear model-based effects and multiple covariates, whose development is the primary goal of this manuscript.

\textbf{Outline:} In this paper, we present a Bayesian method to perform edge regression for undirected graphical models. We define edge-specific \textit{conditional precision functions} that allow the edge strengths of an undirected graphical model to vary with extraneous covariates. We estimate these elements of the precision matrix using a joint regression model while constraining the elements corresponding to a given node to be the same, and thus we guarantee symmetry in the corresponding edges of the precision matrix.  We induce sparsity on both the edges and covariates through Bayesian global-local priors that introduce nonlinear shrinkage, after which posterior edge selection occurs to generate predicted graphs for given sets of covariates while accounting for multiple testing across edges and covariates using Bayesian false discovery rate (FDR) considerations.  We demonstrate the performance of this method in a simulation study in the context of estimating gene networks that are specific for the tumor and stromal components by accounting for proportions of the two components in the observed mixed data, and by application to an HCC case study in which we assess heterogeneity of protein networks across the prognostic index \textit{HepatoScore}. 

The rest of the paper is structured as follows. In Section \ref{method}, we provide a formal description of edge regression with several theoretical properties for undirected graphical models. Then we present our models with sampling scheme and posterior inference technique. We present our simulation studies in Section \ref{simd} and our HCC case study in Section \ref{app}. Section \ref{discussion} contains a discussion and conclusions.
\section{Methods} \label{method}

\subsection{Edge regression}  \label{er}
A graphical model for a random $p$-vector $\mathbold{Y}$  is defined by a tuple $\mathcal{G_Y} = \{G, \mathcal{P}(\mathbold{Y})\}$, where $G$ is a graph and $\mathcal{P}(\mathbold{Y})$ denotes its associated distribution. $G = (V, E)$ represents a conditional independence structure among random variables by specifying a set of nodes $V = {1, 2, 3, \cdots, p}$ and a set of edges $E \in V \times V$. In this work, our intended focus for application is on moderate-sized graphs $G$ with nodes $V$ in the dozens to greater than 100 or so and thus edges $E$ from hundreds to thousands, which is useful in practice in studying genetic pathways, many of which are on that order of magnitude. Each node in graph $G$ corresponds to a random variable in $\mathbold{Y}$. In an undirected graph, we have undirected edges $E$, where $(i, j) \in E$ if and only if $(j, i) \in E$. For example, a Gaussian graphical model is defined by assuming $\mathcal{P}(Y)$ is a Gaussian distribution with mean $\mathbold{\mu}\in \mathbb{R}^p$ and covariance matrix $\mathbold{\Sigma} \in \mathbb{R}^{p\times p}$. $\mathbold{Y_n} \sim \mathcal{N}(\mathbold{\mu}, \mathbold{\Omega}^{-1}), n = 1, \cdots, N\noindent$, where $\mathbold{Y_n}$ is the observed data and $\mathbold{\Omega} = \mathbold{\Sigma}^{-1}  \in \mathbb{R}^{p\times p}$ is the inverse covariance matrix (a.k.a., precision matrix or concentration matrix). In a Gaussian graphical model, $\mathbold{\Omega}$ is a $p \times p$ symmetric positive definite matrix with elements $(\omega^{ij})$. If $\omega^{ij} = 0$, then the random variables $i$ and $j$ are conditionally independent given all the other variables of $\mathbold{Y}$, which indicates that there is no edge in $G$ between nodes $i$ and $j$. Therefore the conditional independence structure of graph $G$ can be inferred from models for the precision matrix $\mathbold{\Omega}$, which is well-known as the \textit{covariance selection} model.

In our proposed \textit{edge regression} model, given another $q$-dimension random vector $\mathbold{X} = (x_1, \cdots, x_q)^T$, we consider $\mathcal{G_Y}(\mathbold{X}) = \{G(\mathbold{X}), \mathcal{P}(\mathbold{Y|X})\}$,  and the precision matrix for each observation $\mathbold{Y_n}$ given $\mathbold{X} = \mathbold{x_n}$ is a function of $\mathbold{X}$, allowing the conditional independence structure to vary from observation to observation over different realizations of $\mathbold{X}$. In the following discussion, we use the term extraneous covariates to define $\mathbold{X}$. We denote the precision matrix dependent on $\mathbold{X}$ through $\Omega(\mathbold{x})$ with elements $\omega^{ij}(\mathbold{X})$.  Here we focus on linear assumptions in $X$, leaving extensions to nonparametric representations to future work.  \cite{ni2018bayesian} shows a functional pairwise Markov property for directed acyclic graphs, which implies  the pairwise Markov property still holds if given the covariates when modeling the graph $\mathcal{G_Y}$ with $\mathcal{P}(\mathbold{Y})$ as a function of the external covariates $\mathbold{X}$. Similarly, we have the following lemma for functional covariance selection that is used to represent edge regression for the \textit{covariance selection} problem.
\vskip -24pt
\begin{customlem}{1}\label{one}
(FUNCTIONAL COVARIANCE SELECTION RULE) Assume $\mathbold{Y}$ has a multivariate Gaussian distribution given extraneous covariates $\mathbold{X}$ with a precision matrix $\Omega(\mathbold{X})$. \\ $Y^i \indep Y^j|\mathbold{Y^{-(i,j)}},\mathbold{X} \Leftrightarrow \omega^{ij}(\mathbold{X}) = 0.$  This follows from the covariance selection rule when a set of extraneous covariates $\mathbold{X}$ is given. Edge regression includes the following special cases:\end{customlem}
\vskip -6pt
 (1) If $\mathbold{X} = \mathbold{\emptyset}$, then we have an ordinary undirected graphical model; \\
(2) If $\mathbold{X}$ is a set of discrete covariates (e.g., binary/categorical), then the edge regression model reduces to the problem of estimating multiple graphical models.\\

\subsection{Regression model for undirected graphs} \label{reg}

In this section, we introduce a sparse regression model to perform edge regression for undirected graphical models. From now on we assume the $\mathbold{\mu} = 0$ for simplicity. Denote the partial correlation between random variable $Y^i$ and $Y^j$ by $\rho^{ij} (1 \leq i \neq j \leq p)$, where $\rho^{ij} = -\frac{\omega^{ij}}{\sqrt{\omega^{ii}\omega^{jj}}}$.  Hence, from the covariance selection rule, the edge $(i,j) \in E$ is equivalent to the partial correlation $\rho_{ij} \neq 0$. A well-known lemma implies that when $y^i$ ($1 \leq i \leq p$) is expressed in a linear regression form of $\sum_{j \neq i} \gamma^{ij}y^j + \epsilon_i$, $\gamma^{ij} = -(\omega^{ij}/\omega^{ii})$ and $\rho^{ij}$ can be represented as $sign(\gamma^{ij})\sqrt{(\gamma^{ij} \gamma^{ji})} $ \citep{ref39}.  We can extend this lemma to a case of edge regression by including the extraneous covariates $\mathbold{X}$ into the regression method, which is stated formally in the following lemma.
\vskip -36pt

\begin{customlem}{2}\label{two}
For $1 \leq i \leq p$, considering predicting $y^i$ from other variables $y^{-i}$ given extraneous covariates $\mathbold{X=x}$ with a varying-coefficient model, we have $y^i = \sum_{j \neq i} \gamma^{ij}(\mathbold{x})y^j + \epsilon_i$, such that $\epsilon_i$ is uncorrelated with $y^{-i}$ given $\mathbold{X=x}$ if and only if the optimal prediction rule gives $\gamma^{ij}(\mathbold{x}) = -\frac{\omega^{ij}(\mathbold{x})}{\omega^{ii}(\mathbold{x})} = \rho^{ij}(\mathbold{x})\sqrt{\frac{\omega^{jj}(\mathbold{x})}{\omega^{ii}(\mathbold{x})}}$, where $\omega^{ij}(\mathbold{x})$ and $\omega^{ii}(\mathbold{x})$ respectively correspond to the off-diagonal and diagonal element of $\Omega(\mathbold{x})$.  Hence, $\rho^{ij}(\mathbold{x}) = sign(\gamma^{ij}(\mathbold{x}))\times$ $\sqrt{\gamma^{ij}(\mathbold{x})\gamma^{ji}(\mathbold{x})}$. Additionally, $var(\epsilon_i) = 1/\omega^{ii}(\mathbold{x})$. $\gamma^{ij}(.)$ is a \textit{conditional precision function} (\textit{CPF}) that defines $\rho^{ij}$ through $\mathbold{X}$.
\end{customlem}

\noindent Lemma 2 is also self-evident when the partial correlation is calculated given $\mathbold{X}$. From Lemma 2, $\mathbold{X}$ changes the partial correlation $\rho^{ij}$ as well as the regression coefficients of $y^i$ over $y^{j}$ through the function $\gamma^{ij}(.)$. We call $\gamma^{ij}(.)$ \textit{CPF} considering it defines the relationship between the partial correlation and extraneous covariates through $\omega^{ij}(.)$ and $\omega^{ii}(.)$ from the precision matrix (i.e., inverse covariance matrix). In this sense, $\gamma^{ij}(.)$ can be estimated to characterize the conditional dependency structure for a subject-level graph given $\mathbold{X}$. Under this setting, the covariance selection problem for a subject-level graph is transformed into a feature selection problem for regression with varying coefficients, i.e., the sparsity structure of an undirected graph can be learned through a sparse regression. Under the assumption of sparse edges, in a Bayesian regression framework we can use variable selection or nonlinear shrinkage on the $\gamma^{ij}(\mathbold{x})$ and perform thresholding on the posterior probabilities (posterior thresholding) to establish the nonzero entries of the graph and their magnitudes \citep{ref27}. $\omega^{ii}$ is determined by the multiple correlation of variable $Y_i$ with the remaining variables, $R_{i, -i}$, and the node variance, $\sigma_i^2$. Similarly to the assumption of homoscedastic node-level variances made in a \textit{DAG} setting for graphical regression \citep{ni2018bayesian}, we assume $R_{i, -i}$ and $\sigma_i^2$ do not change with $X$ and model $\omega_{ii}$ as constant for simplicity and parsimony. Since our primary interest is in the edge structure with pairwise correlation, in this way we allow \textit{CPF} to vary across $\mathbold{X}$ just through $\omega^{ij}$, which facilitates modeling and expedites computation. Thus, the learning how covariates affect edge selection is equivalent to learning how the sparsity structure of off-diagonal elements of the precision matrix varies with $\mathbold{X}$.

\subsection{Parameterization of the conditional precision function} \label{CPF}

In Lemma 2, we defined the \textit{conditional precision function}.  
Suppose we have a set of extraneous covariates $\mathbold{X}$, which can be continuous or discrete. According to our assumptions, $\omega^{ii}(\mathbold{\mathbold{x}}) = \omega^{ii}$. With $\gamma^{ij}(\mathbold{x}) = -\frac{\omega^{ij}(\mathbold{x})}{\omega^{ii}}$, $\gamma^{ij}(.)$ is functionally determined by $\omega^{ij}(.)$, which is constrained to be equal to $\omega^{ji}(.)$ in the precision matrix. Using a linear function to model the relationship between the partial correlations and extraneous covariates, we parameterize the dependence of $\omega^{ij}(.)$  on $\mathbold{X}$:
 \begin{equation}
\omega^{ij}(\mathbold{X}) = \sum_{s=1}^{q}{\beta_s^{ij}X_s} 
\label{eq:1}
\end{equation}
where $\beta^{ij}_s$ is the effect of discrete or categorical variable $X_s$ on the edge $(i,j)$. 
Note that the functional relationship between $\omega^{ij}$ and $\mathbold{X}$ is the same as that between $\rho^{ij}$ and $\mathbold{X}$.  In the regression for each edge $(i,j)$, the conditional precision function can be considered a type of \textit{varying coefficient model} \citep{ref37}.

\noindent \textbf{Joint regression models:}. By regressing $Y^{i}$ over $\mathbold{Y^{-i}}$ given $\mathbold{X}$, we can write our model as:
 \begin{equation}
\begin{aligned}
Y^i & = \sum_{j \neq i} \gamma^{ij}(\mathbold{X})Y^j + \epsilon^i, i = 1, \cdots, p\\
\gamma^{ij}(\mathbold{X}) & = -\frac{\omega^{ij}(\mathbold{X})}{\omega^{ii}}; \epsilon^i \sim N(0, \frac{1}{\omega^{ii}})
\end{aligned}\label{eq:2}
\end{equation}

As we previously mentioned, only $\omega^{ij}$ is assumed to vary across $\mathbold{X}$ for focusing on learning the dynamic sparsity structure of off-diagonal elements and reducing the computational complexity, so $\gamma^{ij}$ and $\epsilon^i$  share the same scaling parameter $\omega^{ii}$. Since $\omega^{ij}(.)$ is the off-diagonal element of precision matrix corresponding to vertex $i$ and vertex $j$, we have $\omega^{ji}(.) = \omega^{ij}(.)$. Hence we will constrain these two functions to be identical in the sampling scheme. Consequently, we have $\beta^{ij}_s = \beta^{ji}_s$ for every $i \neq j$.
We rewrite the full conditional probability of $Y^i$ as:

\begin{equation}
Y^i|\mathbold{Y^{-i}}, \{\mathbold{\beta_s^{i,-i}}\}_{s=1}^q, \omega^{i,i}, \{X_s\}_{s=1}^q  \sim N\left(-\frac{\sum_{j\neq i}^p\sum_{s=1}^{q} \beta^{ij}_sX_sy^{j} }{\omega^{ii}}, \frac{1}{\omega^{ii}}\right)
\label{eq:3}.
\end{equation} 

\subsection{Bayesian adaptive shrinkage} \label{bayshrk}
It has been widely observed that genomic and proteomic graphs tend to be sparse, and as previously discussed, the sparsity of a graph corresponds to sparsity in the estimated precision matrix given extraneous covariates.  We will induce sparsity in the subject-specific precision matrix using a Bayesian approach involving shrinkage priors on the coefficients corresponding to the off-diagonal precision matrix elements.

The spike-slab prior \citep{adref6}, consisting of a mixture of a spike at 0 and a continuous slab, is a popular choice as a Bayesian sparsity prior.  It provides true zero estimates for some variables in the model, yielding automatic edge selection in our graphical setting, plus it has some desirable theoretical properties \citep{adref2, adref3, adref4, adref5}.  However, in high dimensional settings involving a large number of variables or, as in our setting, a large number of potential graph edges, this prior can have computational problems in searching the enormously large underlying state space. Another alternative is to use global-local priors \citep{adref1} that involve scale mixtures of normals.  These priors are absolutely continuous, making them computationally easy to work with even in high dimensional settings, and with a shape that induces a type of nonlinear shrinkage in which small magnitude coefficients shrink strongly towards zero, while large magnitude coefficients are left largely unaffected.  As described below in Section \ref{pinf}, this nonlinear shrinkage effectively induces a type of sparsity in the graph edges, and posterior selection rules can be used to induce true zeros in the estimated graph structure for specific covariate levels \citep{adref1}.
 
\begin{figure}[H] 
\centering
\includegraphics[width=0.5\textwidth]{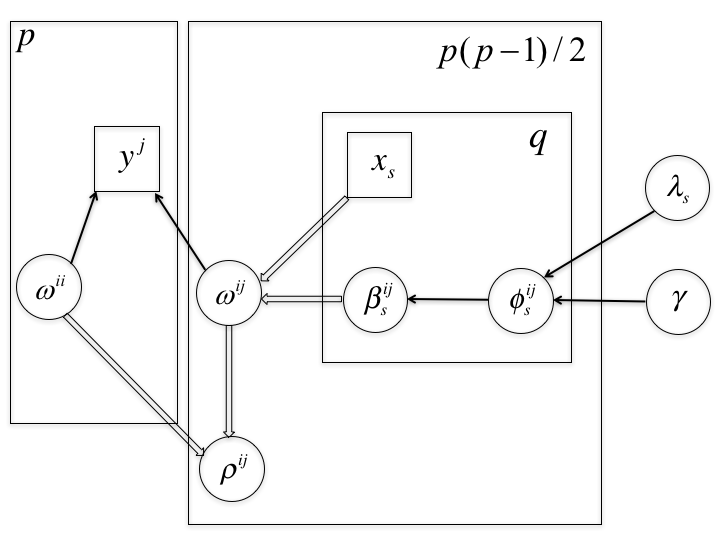}\\

\caption{\footnotesize A graphical representation of edge regression with normal-gamma prior. Single arrows are probabilistic edges; double arrows are deterministic edges; squares are observed data; circles are random variables. The total number of instances of each variable that is enclosed in the same plate is given by the constant in the corner of that plate. $\rho^{ij}$ is the partial correlation for edge $(i,j)$.}
\label{fig1}
\end{figure}

There are many potential global-local prior choices, including the Bayesian Lasso \citep{bsref1}, Horseshoe \citep{bsref2}, Dirichlet Laplace \citep{bsref3}, Normal-Exponential-Gamma \citep{bsref4}, and Normal-Gamma priors \citep{bsref5}.  Here, we will use the Normal-Gamma prior, which has been shown to have outstanding sparsity properties \citep{ref11}. This distribution is indexed by two parameters that together provide useful flexibility in capturing varying degrees of sparsity and heavy-tails in the distribution of coefficients. Furthermore, there are efficient Gibbs sampling schemes available for the Normal-Gamma prior \citep{ref11}. Specifically, assuming $\omega^{ij}(\mathbold{X}) =  \sum_{s=1}^{q} \beta^{ij}_sX_s $, we will assume the following \textbf{Normal-Gamma} prior for the coefficients $\beta^{ij}_s$: 
\begin{equation}
\begin{aligned}
\beta^{ij}_s & \sim N(0, \psi^{ij}_s) \hspace{.5in};
\psi^{ij}_s & \sim Gamma(\lambda_s, 1/(2\gamma^2)). \\
\end{aligned}
\label{eq:4}%
\end{equation}
The \textit{CPF} $\gamma^{ij}(x)= -\frac{\omega^{ij}(x)}{\omega^{ii}}= -\frac{ \sum_{s=1}^{q} \beta^{ij}_sX_s}{\omega^{ii}} = \sum_{s=1}^{q}- \frac{\beta^{ij}_s}{\omega^{ii}}X_s$ is still a linear function. For each $\beta^{ij}_s$ in edge regression,  The latent scale parameter $\psi_s^{ij}$ serves as an adaptive shrinkage parameter across both edges and covariates.  We allow the shape parameter $\lambda_s$ to vary across covariates, but borrow strength across edges, and set the scale parameter $\gamma$ to be common across covariates and edges. This hierarchical structure is constructed to have flexibility, yet borrow strength across edges within covariates, and then across covariates. For $\omega^{ii}$, which controls the variance parameter in the neighborhood selection model, we choose a vague prior such that $\omega^{ii} \propto 1$, as done by \cite{ref11}, for our following discussion. If $\omega^{ii}$ is given with a conjugate prior $Gamma(a^*,b^*)$, the full conditional distribution for  $\omega^{ii}$ keeps the same form, so our sampling scheme can still be implemented by a Gibbs step. A graphical representation of this hierarchical formulation is shown in Figure \ref{fig1}.

\noindent \textbf{Sampling scheme:}  
We adapt the scheme of \cite{ref11} to sample $\lambda_s$ and $\gamma$ simultaneously by specifying exponential and inverse-gamma hyper priors. Enabled by this hierarchical specification of the Normal-Gamma prior, we implement a block Metropolis-within-Gibbs sampling scheme to update each parameter sequentially.  The Gibbs steps involve a multivariate Gaussian for ${\beta}^{ij}$, generalized inverse Gaussians for $\omega^{ii}$ and $\psi^{ij}_s$, and a Metropolis-Hastings step is used to update the shape parameter $\lambda_s$ and scale parameter $\gamma^{-2}$.  After sampling the parameters in the edge regression model, we subsequently obtain posterior samples for the subject-specific precision matrices for all subjects in the dataset, and we could also produce posterior precision matrices for any other hypothetical subjects with specific levels of the covariates $\mathbf{x}$.  The steps of the sampler are summarized in Algorithm 1 (Supplementary Section B) and the corresponding computational details are given in the Supplementary Section B and C. Recall that the CPF $\gamma^{ij}(x)$ is used to define the edge strength $\rho^{ij}(x)$ through $X$. When the edge strength of the subject-level graph is varying across $x$, the adaptive shrinkage priors imposed on each $\beta_s^{ij}$ induce different degrees of shrinkage on $\gamma^{ij}(x)$ across $x$ (Equation \ref{eq:1}). Note that the Normal-Gamma prior induces a ridge-type prior $N(0,\psi_s^{ij})$ on each $\beta_s^{ij}$, so with a linear CPF, the prior induced to $\gamma^{ij}(X=x)$ will still be a ridge prior, of which the variance item is controlled by $x$. The intercept term is also given sparsity priors, inducing sparsity across the edges overall. This implies that, a priori, we expect most edges do not vary strongly with a given $x$, but only a subset of edges. With the shrinkage induced onto the edges across subjects, a Bayesian FDR control procedure will be proposed to select edges for each subject-level graph, which finally induces sparsity at each subject-level graph.

\subsection{Posterior inference and thresholding} \label{pinf}

We perform edge selection to estimate covariate-specific graphs by thresholding posterior probabilities of edge inclusion (PPI) for each edge based on the MCMC samples.  For a given set of covariate levels $x$, we have $L$ posterior samples of the precision matrix elements $\rho^{ij,l}_x, l=1\ldots,L$ after burn-in and thinning. Recall that our Normal-Gamma prior will not result in $\rho^{ij,l}_x \equiv 0$, but does nonlinearly shrink the $\rho$ towards zero, such that $\rho^{ij} \approx 0$ for a large number of $(i,j)$, and $\rho^{ij}$ is very large in magnitude for a relatively small number of $(i,j)$.  If we choose a minimum magnitude of interest $\kappa$ below which we consider the conditional dependence negligible, such that we consider $(i,j) \in E_{x,\kappa}$ if $|\rho^{i,j}_x|> \kappa$ \citep{ref28}, we can estimate the marginal PPI, $P^{i,j}_{x,\kappa} = Pr\{(i,j) \in E_{x,\kappa}|\mathbf{Y},x\}$, with $\sum_{l=1}^{L}I(|\rho^{ij,l}_x| > \kappa)/L$.   The quantity $q_{x,\kappa}^{i,j}=1 - P^{(i,j)}_{x,\kappa}$ can be considered an estimate of the Bayesian local FDR for selecting edge $(i,j)$ for the graph for covariate levels $x$ thus defined.  For a given global Bayesian FDR level $\alpha$, we will flag any edges for which $P^{i,j}_{x,\kappa}>\phi_{x,\alpha}$ as present in our inferred graph, chosen as follows: First, sort $\{q_{x,\kappa}^{i,j}\}$ in ascending order to obtain $\{q_{x,\kappa}^{(t)}, t = 1, \cdots, p(p-1)/2\}$; Second, for a given $\alpha$, find the largest $t^*$ such that $(t^*)^{-1}\sum_{t=1}^{t^*}q_{x,\kappa}^{(t)} < \alpha$; Thrid, set $\phi_{x, \alpha} = q_{x,\kappa}^{(t^*)}$, and select edges with $q_{x,\kappa}^{i,j} < \phi_{x,\alpha}$. This choice implies that we expect $\leq 100 \alpha\%$ of the edges in the estimated edge set $\hat{E}_{x,\kappa}$  will result in false positives, as defined above. 

\textbf{Selection of $\alpha$ and $\kappa$:} In order to apply this selection rule, choices must be made for $\alpha$ and $\kappa$.  $\alpha$ should be chosen to correspond to the desired expected FDR, and $\kappa$ the minimal value of partial correlation below which we consider the association practically negligible in the context of the given application.  In practice, it is a good idea to assess the sensitivity of results to a choice of $\kappa$ and $\alpha$, with edges that persist even with smaller $\alpha$ and larger $\kappa$ are prioritized more highly for any subsequent follow up.  In any simulation studies, the average area under the curve (bAUC) of the receiver operating curves (ROC) \citep{ref26} can be used to assessed model performance over the entire range of possible choices for $\alpha$ and $\kappa$.

\section{Simulations} \label{simd}

In this section, we present simulation studies to investigate the performance of our Bayesian edge regression, designed to mimic the setting of tumor heterogeneity discussed in the introduction.  As previously stated, most researchers interested in contrasting tumor and normal networks would not account for tumor purity, but instead would estimate normal and tumor graphs from the respective samples either using independent or group graphical models.  Thus, we will compare our Bayesian edge regression method with three commonly used approaches for estimating multiple graphical models, the fused graphical lasso; the group graphical lasso \citep{ref3}; as well as the Laplacian shrinkage for inverse covariance matrices from heterogeneous populations (\textit{LASICH}) \citep{saegusa2016joint}, and an approach for solving covariate-dependent graphical model, the nonparametric finite mixture of Gaussian graphical model (\textit{NFMGGM}) \citep{lee2018nonparametric}. Additionally, we include a comparison with a method of Bayesian inference of multiple Gaussian graphical models (\textit{BIMGGM}) \citep{ref2}, which is a Bayesian approach to inference on group graphs. We further run the proposed Bayesian edge regression model with binary covariates mimicking the group definition used when applying the other group models, which is denoted as \textit{Bayesian edge regression (group case)} in the following discussion. For each simulation, we run $20,000$ MCMC iterations, in which the first $10,000$ iterations are discarded as a ``burn-in" period, and thin out the chain using every $10$-th sample.

\textbf{Data generation for simulation.} 
In order to test the ability of our Bayesian edge regression method to account for tumor purity in network estimation, we simulate data in a way to mimic the real-life setting of normal contamination in tumor samples. In our simulation study, we use a similar setting to construct precision matrices from \cite{ref2} and include $20$ nodes to represent $20$ genes, which produces a proper degree of sparsity with around 20\% of possible edges included for the generated precision matrix. From here on, we will use the more general term ``normal'' to represent the stroma component. According to \cite{ahn2013demix} and \cite{ref9}, observed expressions $\mathbold{Y_n}$ from the clinically derived tumor sample $n$ are well-modeled as a linear mixture of the expressions from the normal and the tumor components before log2-transformation of gene expression data. It follows that
\vskip -12pt
 \begin{equation}
2^{Y_n} = (1 - \pi_n)2^{N_n} + \pi_n2^{T_n},
\label{eq15}%
\end{equation}
\vskip -12pt
\noindent where the expressions from the normal component $\mathbold{N_n} \sim \mathbold{\mathcal{N}(\mu_N, \Omega^{-1}_N)}$ and  those from the tumor component $\mathbold{T_n} \sim \mathbold{\mathcal{N}(\mu_T, \Omega^{-1}_T)}$. $\pi_n$ is the proportion of the tumor component before log2-transformation, i.e., the measured tumor purity for sample $n$, which we generated from the range $[0,1]$. Following Equation \ref{eq15}, we generated observed gene expression levels $\mathbold{Y_n}$ for each sample $n$ from the the simulated expressions $\mathbold{N_n}$ and $\mathbold{T_n}$.
For simplicity, we set $\mathbold{\mu_N = 0}$ and $\mathbold{\mu_T = 0}$ in our simulation.  We also generate $\mathbold{N'_n} \sim \mathbold{\mathcal{N}(0, \Omega^{-1}_N)}$ as a reference group for normal component. In our following simulation, we provide two simulations with different set-up of precision matrix.

\textbf{Simulation 1.  Low overlap in tumor and normal graphs.}   $\mathbold{\Omega_T}$, where off-diagonal elements $\omega_T^{i,i+2} = \omega_T^{i+2,i}$ uniformly sampled from $[-0.5, -0.3] \cup [0.3, 0.5]$ for $i = 1, \cdots, 18$. $\mathbold{\Omega_N}$, where off-diagonal elements $\omega_N^{i,i+1} = \omega_N^{i+1,i}$ uniformly sampled from $[-0.5, -0.3] \cup [0.3, 0.5]$ for $i = 1, \cdots, 19$. For both $\mathbold{\Omega_T}$ and $\mathbold{\Omega_N}$, all the diagonal elements are $1$ and all the other elements are left with zero.  $\mathbold{\Omega_T}$ and $\mathbold{\Omega_N}$ are truly sparse with just $18$ and $19$ edges. They do not have any overlapping edges by construction, and the edge strengths are relatively weak. We simulate reference normal samples of size $N_{N'} = 50$ and mixed tumor samples of size $N_Y = 150$ with $\{\pi_n\}_{n=1}^{150}$ generated from an arithmetic sequence from $0.01$ to $0.99$. $\pi_n$ will be considered as the extraneous covariate to our edge regression model and also taken as a fixed value in our following experiments. We randomly generated 100 datasets for this simulation.

\textbf{Simulation 2.  Higher overlap in tumor and normal graphs.}  $\mathbold{\Omega_T} $, where diagonal elements $\omega_T^{i,i} = 1$ for $i = 1, \cdots, 20$ and off-diagonal elements $\omega_T^{i,i+1} = \omega_T^{i+1,i} =0.5$ for $i = 1, \cdots, 19$, $\omega_T^{i,i+2} = \omega_T^{i+2,i} =0.4$ for $i = 1, \cdots, 18$. All the other elements are left with zero. $\mathbold{\Omega_N}$, where we remove $30$ edges randomly from $\mathbold{\Omega_T}$ by substituting these $30$ nonzero elements with zero and randomly add $30$ edges to $\mathbold{\Omega_T}$ by substituting these $30$ zero elements with values uniformly sampled from $[-0.6, -0.4] \cup [0.4, 0.6]$. To ensure $\mathbold{\Omega_N}$ is positive definite, following \cite{ref3}, we divide each off-diagonal element by $1.5$ times the sum of the absolute value of all the off-diagonal elements in its row. Then we average the transformed matrix with its transpose to guarantee it is symmetric. Although this procedure is able to guarantee the generated matrix will be positive definite, it can bring weak signals to $\mathbold{\Omega_N}$, which makes the estimation even more difficult. We allow $\mathbold{\Omega_T}$ and $\mathbold{\Omega_N}$ to have seven overlapping edges, and $\mathbold{\Omega_N}$ has relatively weak edge strengths. We simulate reference normal samples of size $N_{N'} = 100$ and mixed tumor samples of size $N_Y = 200$ and generate $\pi_n$ as in Simulation 1. We randomly generated 100 datasets for this simulation. The graph structures for Simulation 1 and 2 are shown in Supplementary Figure 1.%

We compare the results of our edge regression with application of the fused and group graphical lassos\footnote{available in the R package \textit{JGL}}, \textit{LASICH} \footnote{available in the R package \textit{LASICH}}, \textit{NFMGGM} \footnote{Implementation requested from the authors}, and \textit{BIMGGM} \footnote{\url{https://odin.mdacc.tmc.edu/~cbpeterson/software.html}}, respectively, to the tumor and normal measurements, $\mathbold{Y}$ and $\mathbold{N'}$, respectively. The application of these group graph methods corresponds to what might be the usual practice of estimating tumor graphs from tumor samples without adjusting for tumor purity and normal contamination, and estimation of the normal graph from normal controls, so has practical scientific relevance. For running our method and \textit{NFMGGM}, all the genes are normalized to have a mean of zero and a standard deviation of one with all the samples. For running all the group graphical models, the data are normalized to have a mean of zero and a standard deviation of one, respectively, within the tumor and normal group. Details of the implementations for the competing method are in the Supplementary Section D.

In our Bayesian edge regression model, following Equation (\ref{eq:1}) we parameterize the dependence of $\omega^{ij}(.)$  on $\mathbold{X}$:
\vskip -24pt
 \begin{equation}
 \begin{aligned}
\omega^{ij}(\pi) = \beta^{ij}(1 - \pi) + \alpha^{ij}(\pi).
\end{aligned}
\label{eq:15}%
\end{equation}
\vskip -12pt 
\noindent Under this parameterization, $\alpha^{ij}$ represents the precision element for pure tumor samples, and $\beta^{ij}$ the precision element for a pure normal sample, with the sample specific edges given by a linear combination as determined by their tumor purity $\pi$.  This model allows us to reweight samples based on tumor purity to estimate the pure normal and pure tumor graphs. For an additional comparison of our approach using discrete predictors, we also ran our Bayesian edge regression model with binary covariates mimicking the group definition used when applying the other group models (i.e., \textit{Bayesian edge regression (group case)}). In this application, we use two binary covariates to encode the membership of tumor and normal groups and add an additional covariate to capture the interaction effects between tumor and normal groups (see model details in Supplementary Section D).

In our MCMC sampling scheme, we use $M_\beta$ and $M_\alpha$ as hyperparameters to control the scale of $\lambda_s\gamma^2$ when $s$ is corresponding to $\beta$ and $\alpha$. Following \cite{ref11}, we suggest setting $M_\beta$ and $M_\alpha$ by using $\hat \Omega_{MLE}$ for simulated samples with $\pi_T < 0.5$ and $\pi_T \ge 0.5$ (see Supplementary Section B and C).
We implement these methods across $100$ simulated datasets for both the first and second simulations. 
We compare the methods in terms of accuracy in estimating the graph structure using the area under the ROC curve (AUC) and true positive rate (TPR) and false positive rate (FPR).  We have two regularization parameters for both our graph edge regression ($\kappa$ and $\alpha$), \textit{NFMGGM} ($\lambda$ and $h$), \textit{LASICH}, and the two graphical lasso methods ($\lambda_1$ and $\lambda_2$). Thus, for all these methods we compute a bivariate AUC \citep{ref26} by varying both two parameters at the same time, and then computing the expected AUC (bAUC) by binning results on a grid of 1- \textit{specificity}, and computing the average sensitivity within those values.  \textit{BIMGGM} is reported with a univariate AUC given it only requires one tuning parameter. For better observing how each tuning parameter affects the model performance for these methods with two hyperparameters, we further report the best univariate AUC over one hyperparameter given the other is fixed. To compare TPR and FPR for a single choice of regularization parameters, we use $\kappa=0.1$ and $\alpha=0.1$ for the Bayesian edge regression methods that are respectively built with continuous covariates and discrete covariates, and we use $\lambda_1$ and $\lambda_2$ or $\lambda$ and $h$ for the compared methods following the previously mentioned guidelines (Supplementary Section D). We report the results for \textit{BIMGGM} with the same $\alpha=0.1$ for posterior thresholding with Bayesian local FDR. In addition to the TPR and FPR reported with the selected model, we also report a TPR corresponding to FPR$\approx 0.01$ for each method, providing a fair comparison of methods using common criteria. The ROC curves for these two simulations are given in Figure \ref{figroc1}, and tables showing the AUC, TPR, and FPR for normal, tumor, and overall are given in Supplementary Section D.

In \textit{Simulation 1}, we see that all methods yield relatively high bAUC for the normal graphs, but the Bayesian edge regression with continuous covariates has much better bAUC (0.916) for the tumor graph than the other group graph methods (0.706, 0.793, 0.808, and 0.812) (see Supplementary Table 2). This is related to the fact that, unlike the group graph methods, our edge regression can adjust for continuous variables, such as the tumor purity, and thus reduce biases in parameter estimations for the tumor graph. The proposed method outperforms the kernel regression-based method \textit{NFMGGM} (0.810) for the tumor graph as well, which suggests a parametric edge regression leads to better performance in this case. Also, note that the FPRs for the Bayesian edge regression method are all below 0.100, while the \textit{LASICH}, \textit{NFMGGM} and graphical lasso methods have high FPR for the choices of their regularization parameters. We can further find that the proposed method is reported with the highest overall TPR (0.918 versus 0.727, 0.746, 0.763, 0.589, 0.773, 0.750) among all the methods when the model is chosen such that $FPR \approx 0.1$ (see Supplementary Table 3).

\begin{figure}[H] 
\centering
\includegraphics[width=0.3\textwidth]{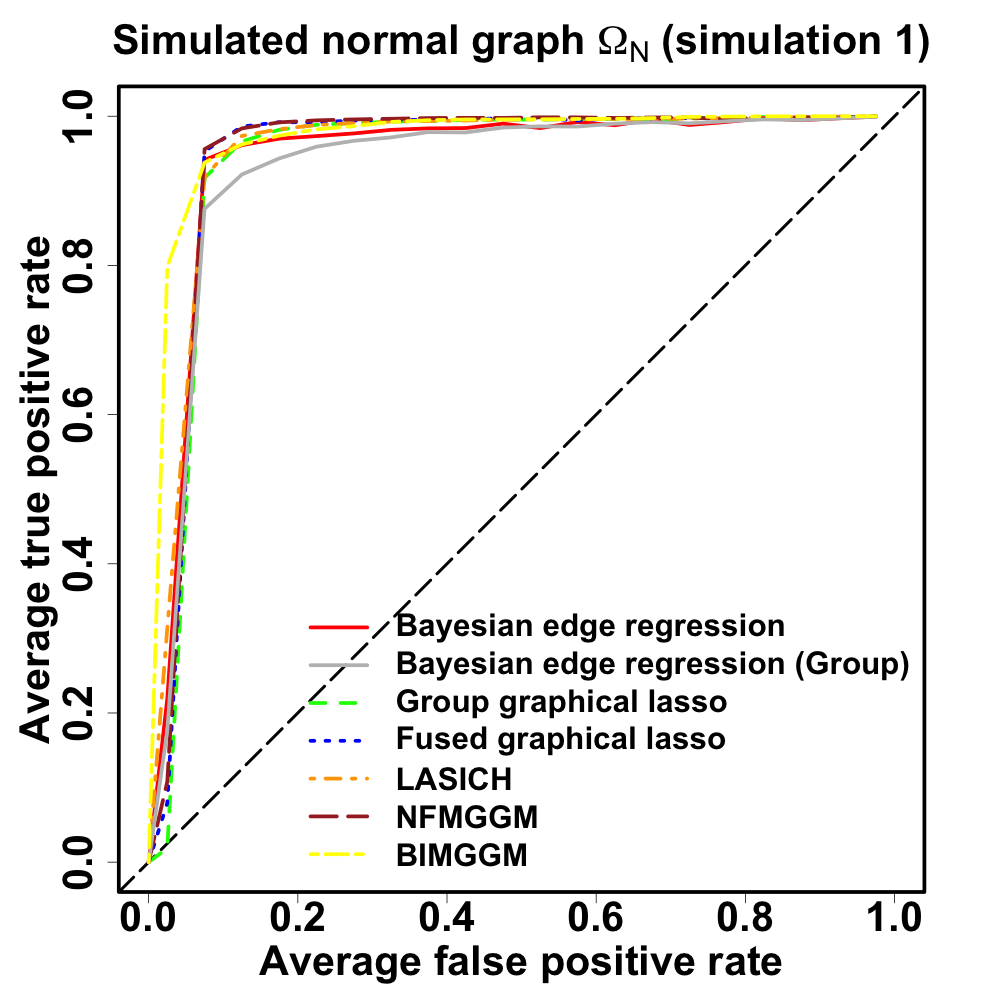}
\includegraphics[width=0.3\textwidth]{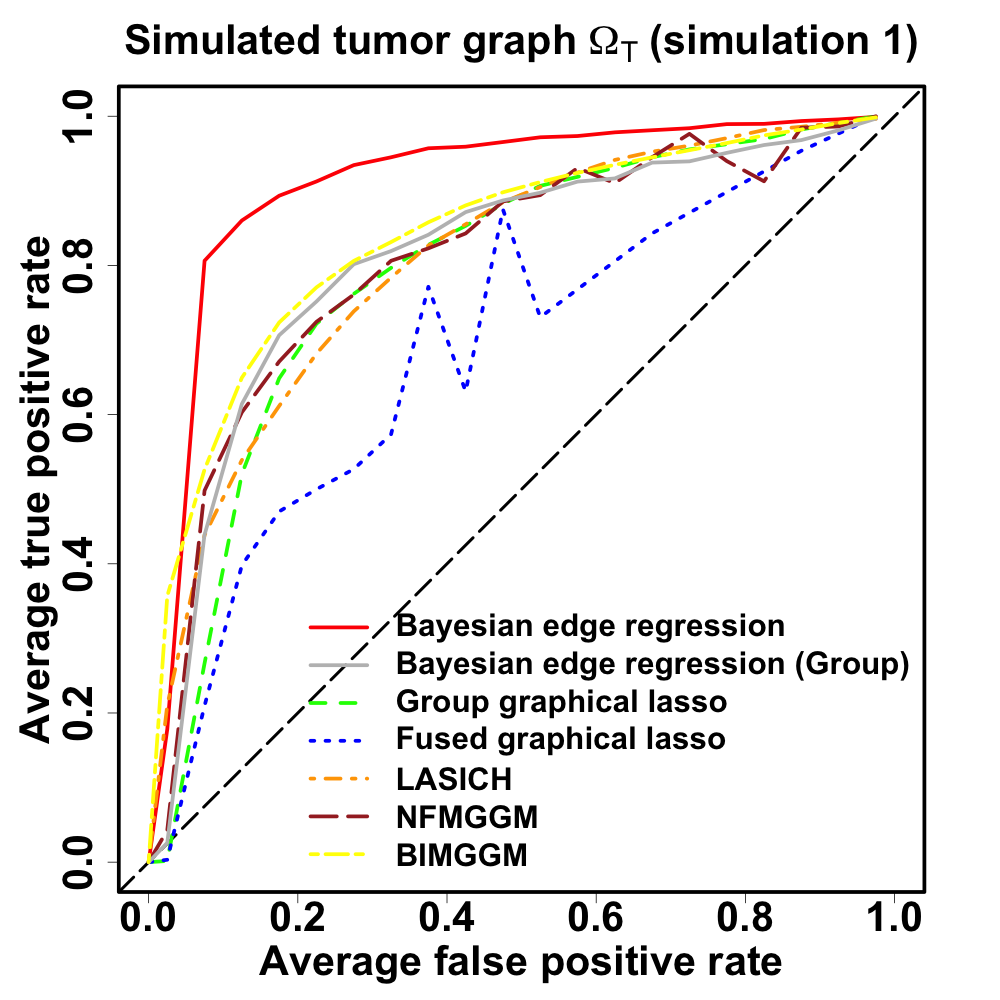}\\
\includegraphics[width=0.3\textwidth]{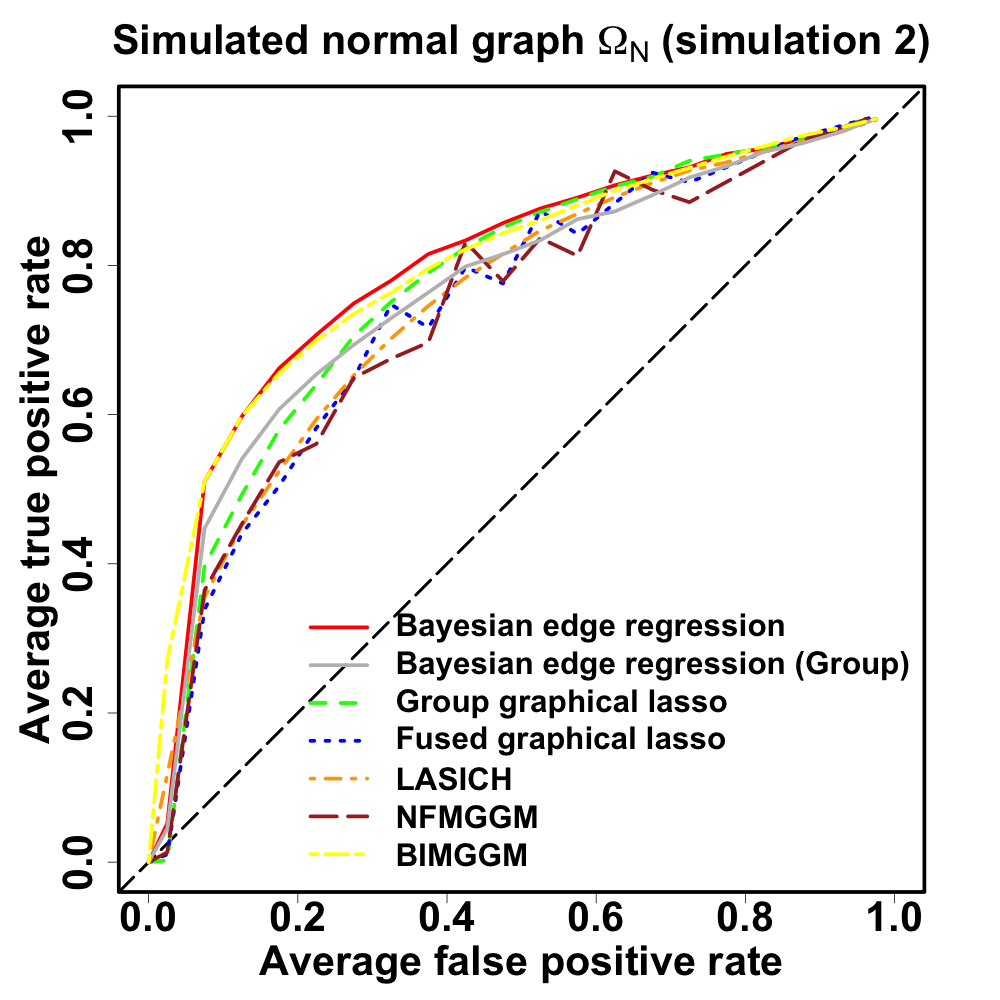}
\includegraphics[width=0.3\textwidth]{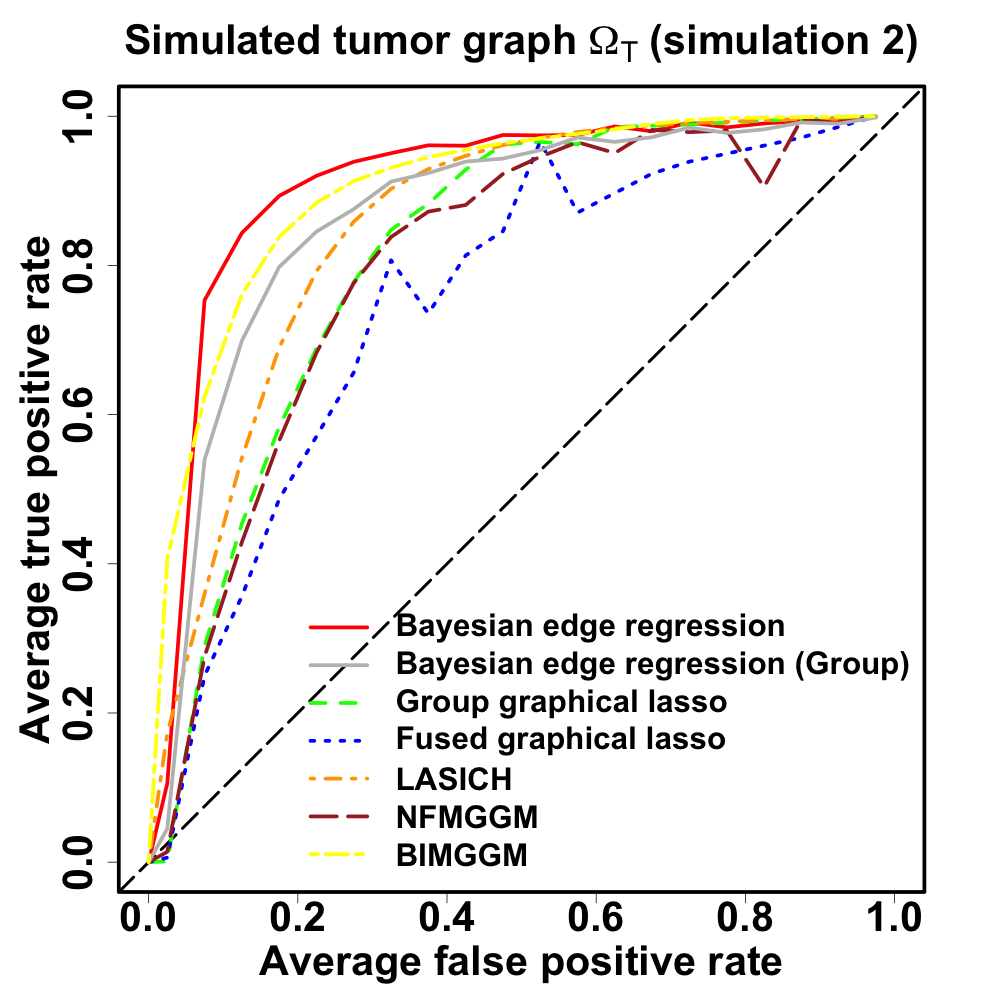}
\caption{\small Simulation results for Section \ref{simd}. ROC curves for the structure learning of simulated normal (w.r.t $\mathbold{\Omega_N}$) and tumor graphs (w.r.t $\mathbold{\Omega_T}$) in  \textit{Simulation 1} and  \textit{Simulation 2}.}
\label{figroc1}
\end{figure}
In \textit{Simulation 2}, a more challenging setting with weaker signal, we also observe that our Bayesian edge regression method with continuous covariates produces higher bAUC for both the normal (0.803 versus 0.748, 0.770, 0.751, 0.738, 0.773) and tumor (0.913 versus 0.758, 0.813, 0.852, 0.799, 0.870) graphs (see Supplementary Table 4).  Once again, the graph lasso methods with regularization parameters chosen by AIC resulted in much higher FPRs ($>0.500$), while the Bayesian edge regression was much lower (0.047 for normal and 0.269 for tumor; see the results in Supplementary Section D). Similarly, the proposed method leads to the highest overall TPR (0.720 versus 0.469, 0.458, 0.534, 0.383, 0.625, 0.596) when $FPR \approx 0.1$ (see Supplementary Table 5). From the two simulations, we find that implementing a group graphical model on our Bayesian edge regression framework (i.e., \textit{Bayesian edge regression (group)}) obtains a performance as similar as other group graph methods, but worse than the regression model with continuous covariates (overall AUC: 0.877 versus 0.932 in Simulation 1; 0.822 versus 0.858 in Simulation 2) (Supplementary Table 2 and 4), which again highlights the importance of adjusting for tumor purity as a continuous covariates when estimating tumor and normal networks.

\section{Proteomic Networks in Hepatocellular Carcinoma}\label{app}

\textbf{Markers in HCC:}  Hepatocellular carcinoma (HCC) is the fifth most common cancer worldwide and the third-leading cause of cancer-related death, and it has an increasing incidence in developing countries.  In the USA, it is the fastest growing cause of cancer-related mortality in men, and with the alarming increase of hepatitis C, it is expected to continue to grow in incidence in the coming years.  Over 80\% of patients present with advanced disease and underlying cirrhosis \citep{ref20,ref25}, which prevents curative treatment options.  There are only a few approved systemic therapies for HCC, e.g., sorafenib, and various targeted therapies are being assessed in combination, including therapies targeting angiogenesis pathways.  New, targeted therapies and precision therapy strategies clearly are needed.  Cytokines are blood proteins secreted by various types of cells in the immune system that have an effect on other cells. There is significant evidence that numerous cytokines mediate processes involved in the liver, including inflammation, necrosis, cholestasis, fibrosis, and regeneration, and are a key factor in many liver diseases, including HCC \citep{Tilg2001}. Other biological pathways instrumental for HCC include inflammation, metabolic pathways, immune response, growth factor, and angiogenesis \citep{ref22, ref23}.  A deeper characterization of the molecular basis of interpatient heterogeneity of HCC, including behavior within these pathways, has the potential to contribute to new, targeted precision therapy strategies for HCC.

A recently developed novel prognostic measure characterizing inter-patient heterogeneity in HCC from blood protein profiles is the \textit{HepatoScore} \citep{HepatoScore}. This biological prognostic score has been shown to dramatically refine HCC staging systems, e.g. accurately delineating a subset of metastatic patients with low \textit{HepatoScore} who have substantially better prognosis than non-metastatic patients with high \textit{HepatoScore}.  Although it is a biological score based only on blood protein levels including no clinical factors, \textit{HepatoScore} by itself outperforms all existing staging systems and prognostic factors \citep{HepatoScore}.  While a global score computed from the entire panel of circulating proteins, \textit{HepatoScore} is primarily driven by a subset of key circulating proteins within various molecular pathways relevant to HCC, most notably the immune response, GH, and angiogenesis pathways.  It is thought that these pathways play a major role in characterizing the patient's cancer and prognosis, and deeper characterization of the interrelationships across these proteins may yield important biological insights.

The data set analyzed in this paper involves measurements of proteins that are obtained using CytokineMAP (Myriad RBM, Austin, TX) on plasma samples from 766 HCC patients \citep{HepatoScore}.  The proteins considered here include 71 proteins from immune system, GH, and angiogenesis pathways plus alpha-fetoprotein (AFP), an important protein for HCC used for early detection and prognosis. After scaling all proteins to have a mean of zero and variance of one, we ran our graph edge regression model as outlined above.  Given \textit{HepatoScore} $\pi \in [0,1]$, our model for the conditional precision edge $(i,j)$ is given by $\omega^{ij}(\pi) = \beta^{ij}(1 - \pi) + \alpha^{ij}(\pi)$.  

Under this parameterization $\beta^{ij}$ represents the edge strength for low \textit{HepatoScore} ($\pi=0$), $\alpha^{ij}$ the edge strength for high \textit{HepatoScore} ($\pi=1$), and a linear combination assumed for a moderate \textit{HepatoScore}, e.g. with the edge strength for $\pi=0.5$ given by $0.5\beta^{ij}+0.5\alpha^{ij}$.  The graph edge strengths for any continuous \textit{HepatoScore} $\pi \in [0,1]$ can be computed from this model. We used the same prior setting as in Section \ref{simd}, and chose a tuning parameter $\sigma_\lambda$ that provides an acceptance rate of the Metropolis step at around $20\% \sim 30\%$, determined after the burn-in period. We ran the MCMC sampler for $10,000$ iterations after a burn-in of $10,000$, then thinning to keep every $10^{th}$ sample.  
\begin{figure}[H]
\begin{subfigure}[H]{0.475\textwidth}
\includegraphics[width=0.75\textwidth]{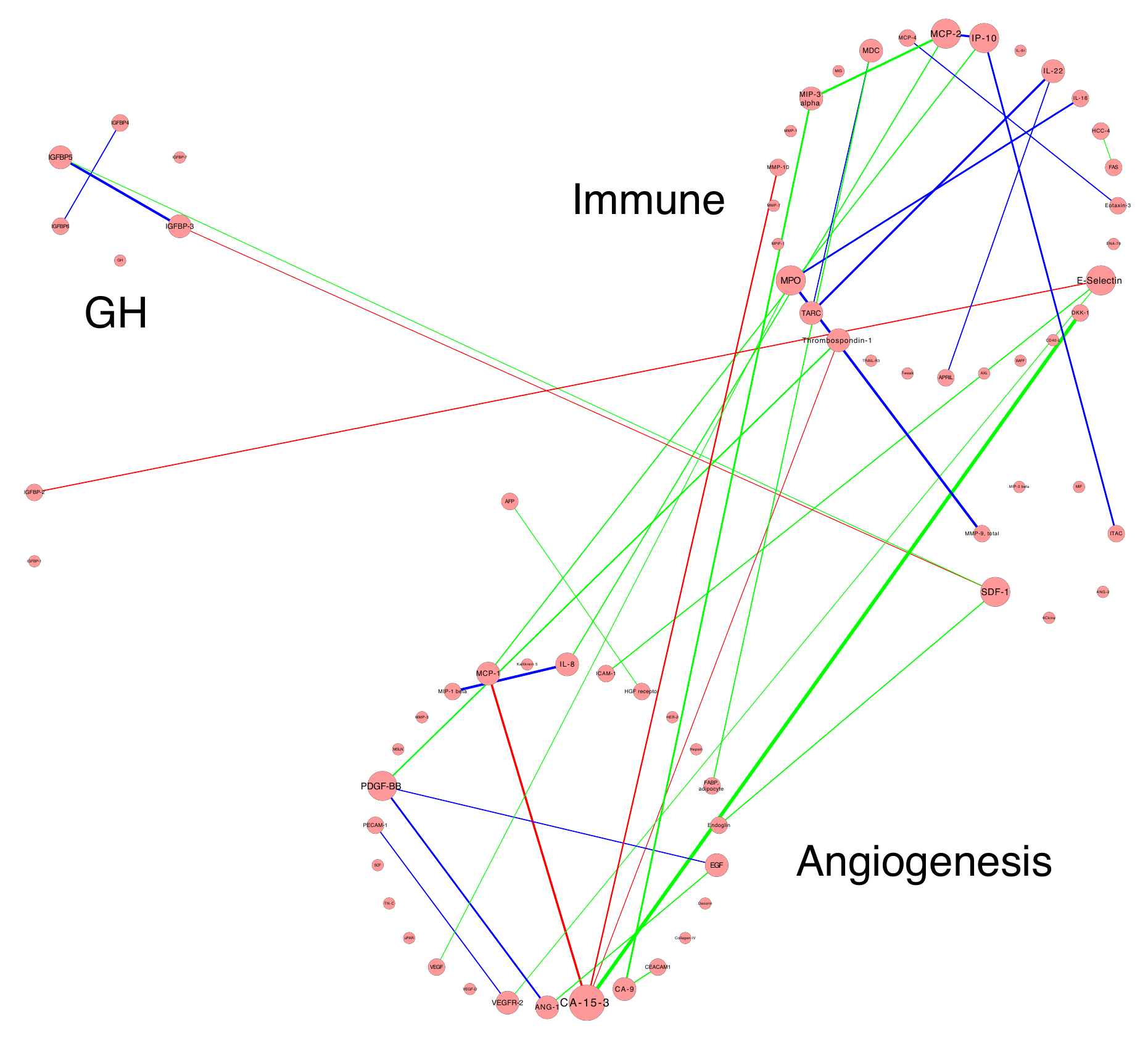}
\caption{ Low \textit{HepatoScore} ($\pi=0$)}
\end{subfigure}\hfill%
\quad
\begin{subfigure}[H]{0.475\textwidth}
\includegraphics[width=0.75\textwidth]{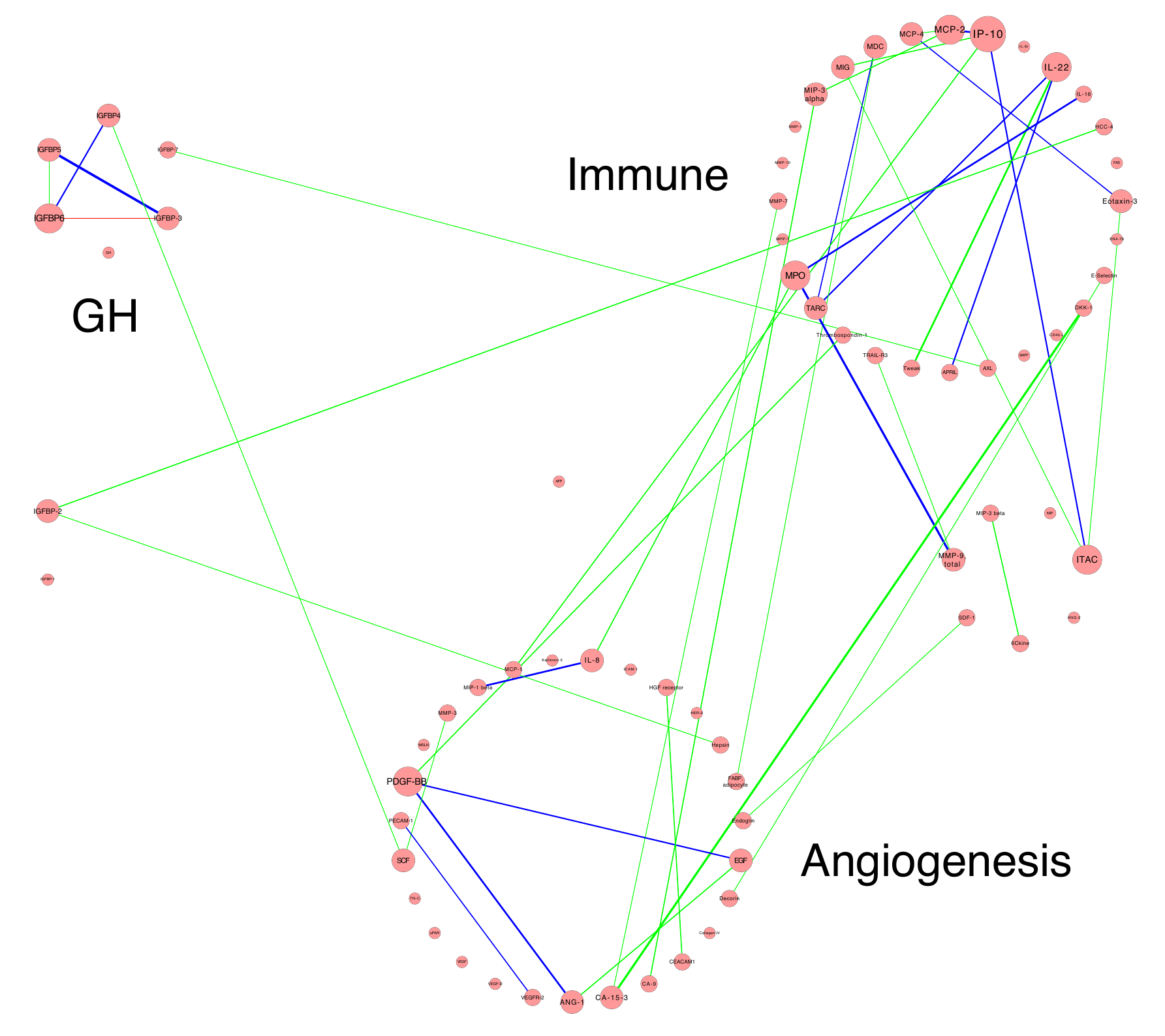}
\caption{ Medium \textit{HepatoScore} ($\pi=0.5$)}
\end{subfigure}%
\\
\begin{subfigure}[H]{0.475\textwidth}
\includegraphics[width=0.75\textwidth]{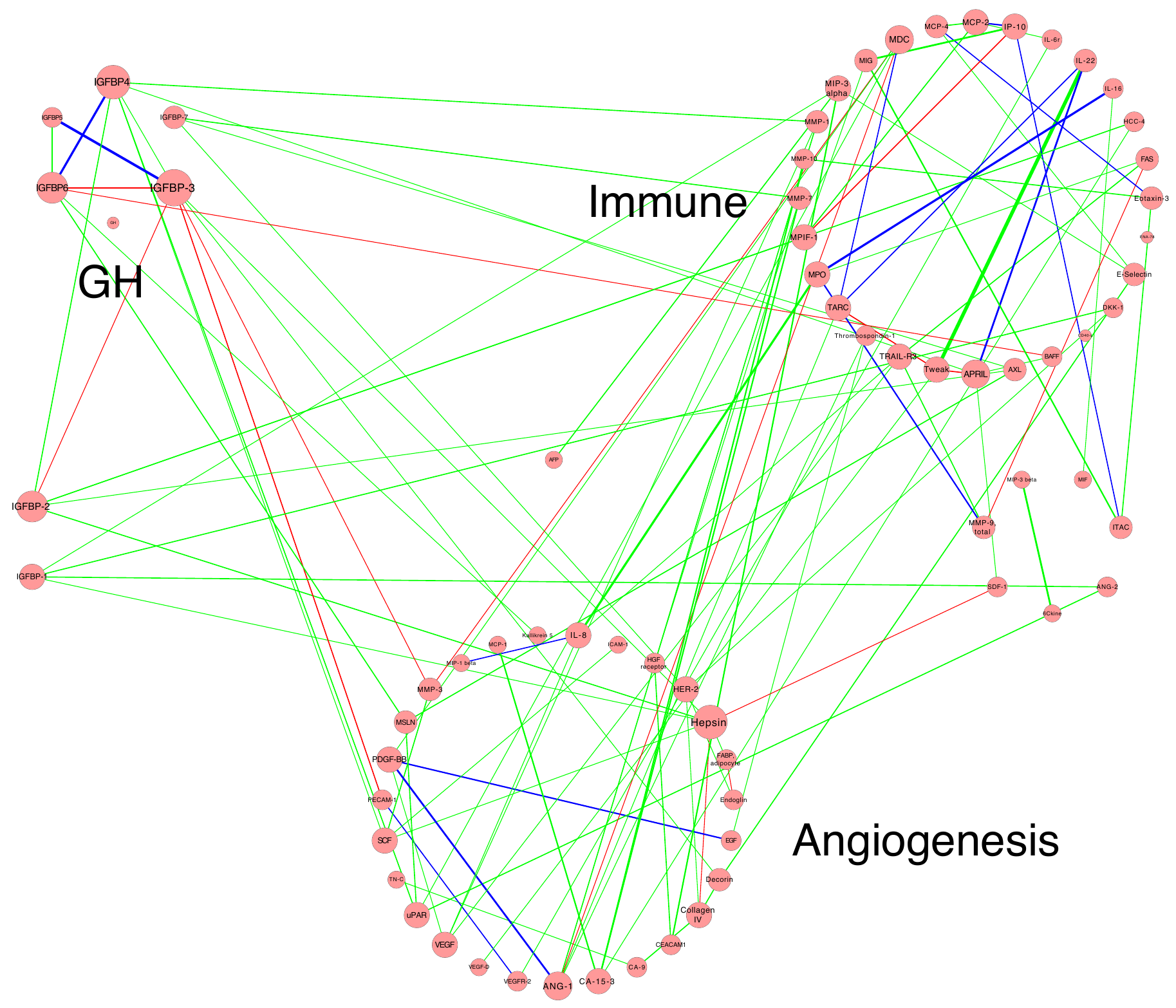}
\caption{ High \textit{HepatoScore} ($\pi=1$)}
\end{subfigure}\hfill%
\begin{subfigure}[H]{0.475\textwidth}
\includegraphics[width=0.75\textwidth]{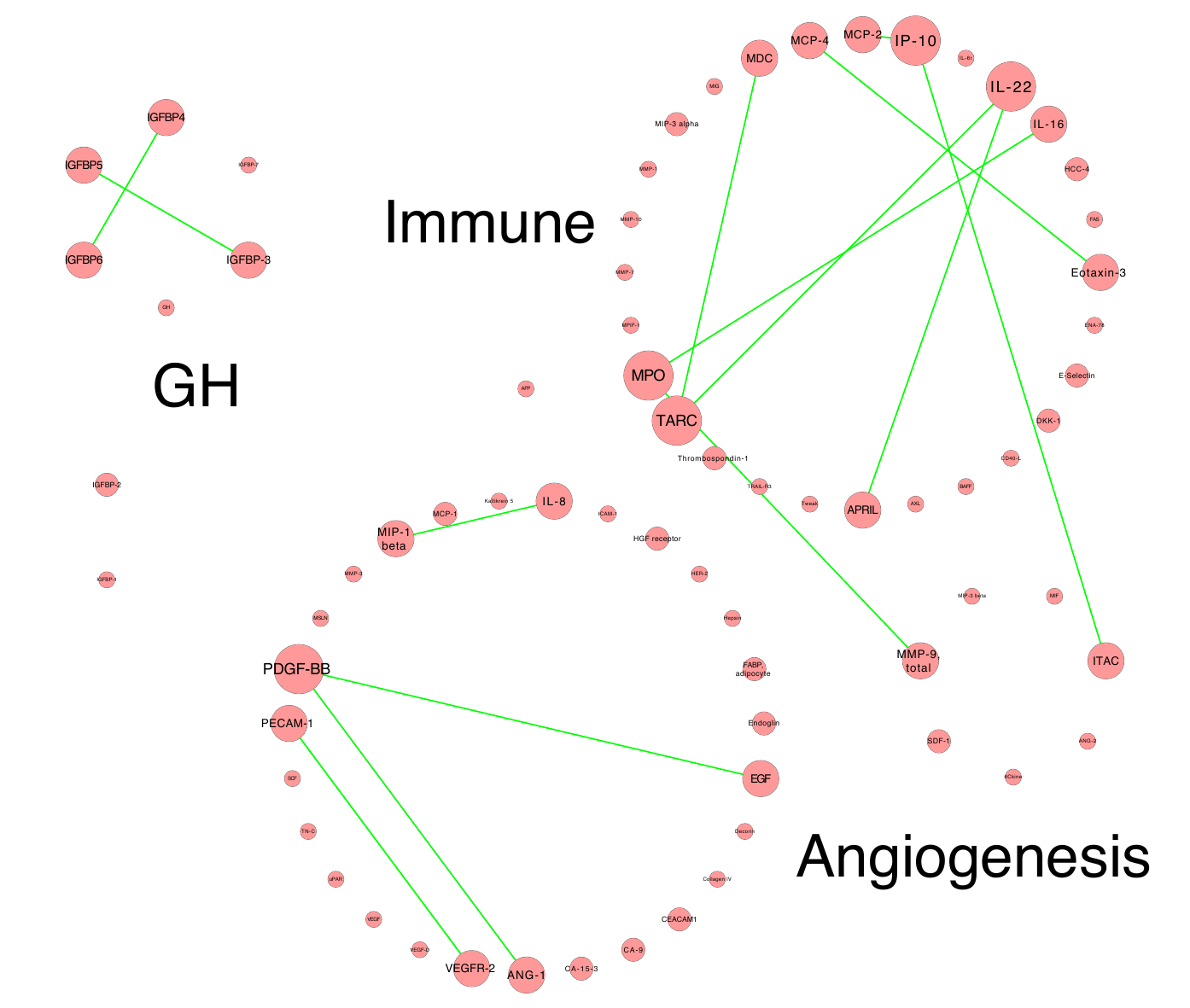}
\caption{Shared edges}
\end{subfigure}%
\caption{\small Estimated graphs ($\alpha=0.10, \kappa=0.15$) from the Bayesian edge regression for the GH, immune, and angiogenesis pathways with (a)  $\pi = 0$; (b) $\pi = 0.5$; (c) $\pi = 1$; (d) Common edges. Colors indicate positive (green), negative (red), and common (blue) edges. The thickness of edge is proportional to $\hat \rho^{ij}$ for the edge $(i, j)$, and the size of node is proportional to its degree.}
\label{fighccm}
\end{figure}

To assess convergence, we observed trace plots and ran a Geweke convergence diagnostic for all parameters.  The histogram of the Geweke p-values suggests that the chain converged satisfactorily (Supplementary Figure 3).  Within the sampler, we also obtained posterior samples for the predicted precision matrices corresponding to a low ($\pi=0$), medium ($\pi=0.5$), and high ($\pi=1$) \textit{HepatoScore} as described above, and applied our posterior edge selection approach based on $\alpha=0.1$ and $\kappa=0.15$, also considering $\kappa=0.1$ and $\kappa=0.2$ for sensitivity. We also ran \textit{NFMGGM} for comparison, with results in Supplementary Section E.

\begin{figure}[H] 
\centering
\includegraphics[width=1\textwidth]{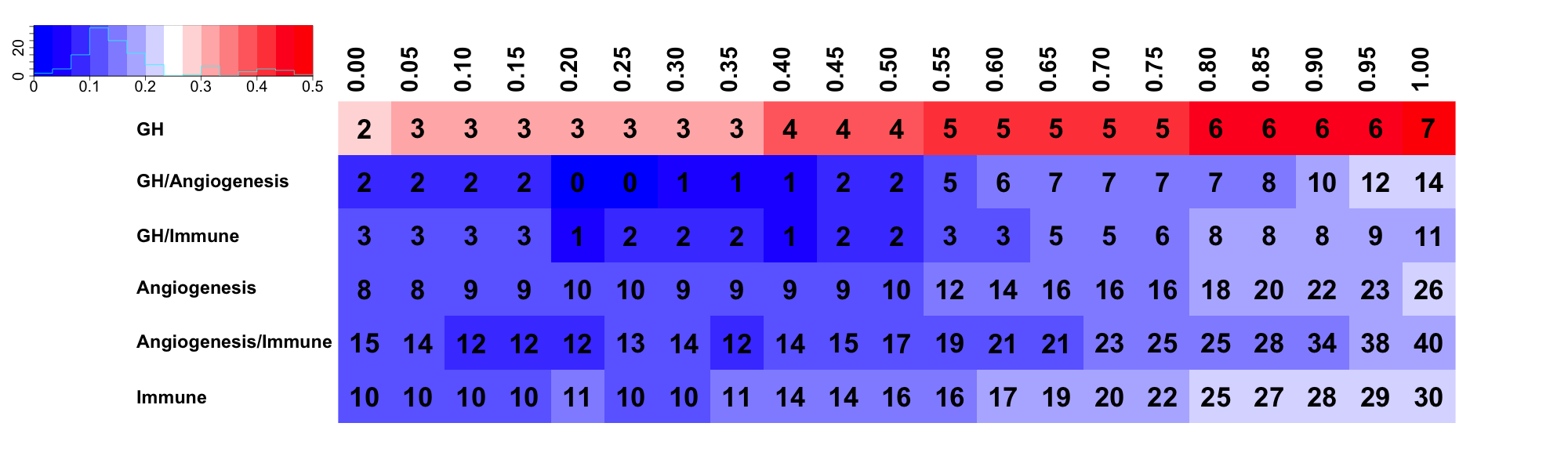}

\caption{\small The number of connected edges within each pathway and across different pathways varies with different \textit{HepatoScores} (from $\pi = 0$ to $1$). The color gradient for each table cell changes with the square root of the proportion of connected. $\kappa = 0.15$.}  \label{figheat}
\end{figure}

While our method produces predicted graphs for any $\pi \in [0,1]$, for interpretation we focus on three levels of $\pi \in \{0, 0.5, 1\}$. Figure \ref{fighccm} contains the estimated graphs for a low \textit{HepatoScore} ($\pi=0$), medium \textit{HepatoScore} ($\pi=0.5$), and high \textit{HepatoScore} ($\pi=1$), with edge direction indicated by color (green = positive, red = negative), edge strength by line width, and node size indicating the number of connecting edges.  Blue lines indicate edges shared across all levels of $\pi$, and their direction is given by panel (d). It is clear that the number of graph edges increases with \textit{HepatoScore} values, indicating that the protein network connectivity increases for more invasive forms of HCC. Figure 4 summarizes the number of connections within each pathway and between each pair of pathways, as a function of \textit{HepatoScore} $\pi$, and Supplementary Table 11 shows the number of edges within and between different pathways in the respective graphs for $\pi=0, 0.5$ and $1$.  We see that the number of intra-pathway connections within each of the three pathways strongly increases with \textit{HepatoScore}, especially for a high \textit{HepatoScore} ($\pi>0.8$), with more than twice the number of edges than a low \textit{HepatoScore}.  The number of inter-pathway edges increases with the \textit{HepatoScore} even more strongly, with a three to fourfold increase, much notably between the angiogenesis and immune pathways with 40 edges for $\pi=1$ and only 15 edges for $\pi=0$.  The increased connectivity could correspond to increased activity within these important pathways, and increased cross talk between them.  This could have important implications for the underlying molecular biology, and it needs to be followed up to validate and assess the biological implications of these associations. Supplementary Section E contains the results using $\kappa=0.1$ and $0.2$, which demonstrate the same substantive effects, although of course with a greater and fewer number of total edges, respectively, in the graphs. 

Hub proteins are proteins with many connections in the graph, and they may be involved in multiple regulatory activities. Different hub genes are identified for these three graphs (Supplementary Table 12). \textit{IGFBP-3} has been identified as a hub gene in the high \textit{HepatoScore} graph and with a moderate degree of connectivity in the low \textit{HepatoScore} graphs, where the connected nodes are different. \textit{IGFBP-3} has been considered as an effective predictor for HCC patients with chronic HCV infections, and it is a transcription factor encoding proteins to suppress HCC cell proliferation, so the reduction of \textit{IGFBP-3} is significantly associated with the development of HCC \citep{bref7, bref8}. There are many edges that vary over the \textit{HepatoScore}.  We highlight a few notable ones here and present the rest in Supplementary Section E.  Figure \ref{figedge1} contains a plot for three edges, which presents the edge strength as a function of \textit{HepatoScore} $\pi$ along with 95\% credible intervals and the corresponding $PPI_{\alpha=0.1,\kappa=0.15}$. 

\textit{6Ckine} is strongly associated with $\textit{MIP-3, $\beta$}$ for medium and high \textit{HepatoScores}, whereas this edge is not apparent in the graphs for a low \textit{HepatoScore}. The regulation between \textit{6Ckine} and \textit{MIP-3, $\beta$} has been previously reported to play a determinant role in accumulating antigen-loaded mature dendritic cells \citep{bref4}. \textit{AFP}/\textit{MIP-3, $\alpha$} is another pair that shows no correlation in the graph for low \textit{HepatoScore}, but a positive correlation for high \textit{HepatoScore}. \textit{AFP} ($\alpha$-fetoprotein) is a tumor marker for liver cancer. The levels of \textit{AFP} have been reported to relate with \textit{MIP-3, $\alpha$} levels in HCC, where the serum levels of \textit{MIP-3, $\alpha$} are increased \citep{bref5}. \textit{CA-15-3} is well known to detect breast cancer and distinguish from non-cancerous lesions, and its level has been shown to be increased for end-stage liver disease patients \citep{bref1,bref2}. \textit{MCP-1} is a protein secreted by the HCC microenvironment that can promote progression, angiogenesis, and metastasis in cancer through recruiting and modifying mesenchymal stromal cells (MSCs). \textit{CA-15-3}/\textit{MCP-1} shows negative correlation for $\pi = 0$ and positive correlation for $\pi = 1$, which corresponds to these previous empirical findings of elevated levels of \textit{CA-15-3} and \textit{MCP-1} in liver disease.

  \begin{figure}[H] 
  \centering
  \includegraphics[width=0.25\textwidth]{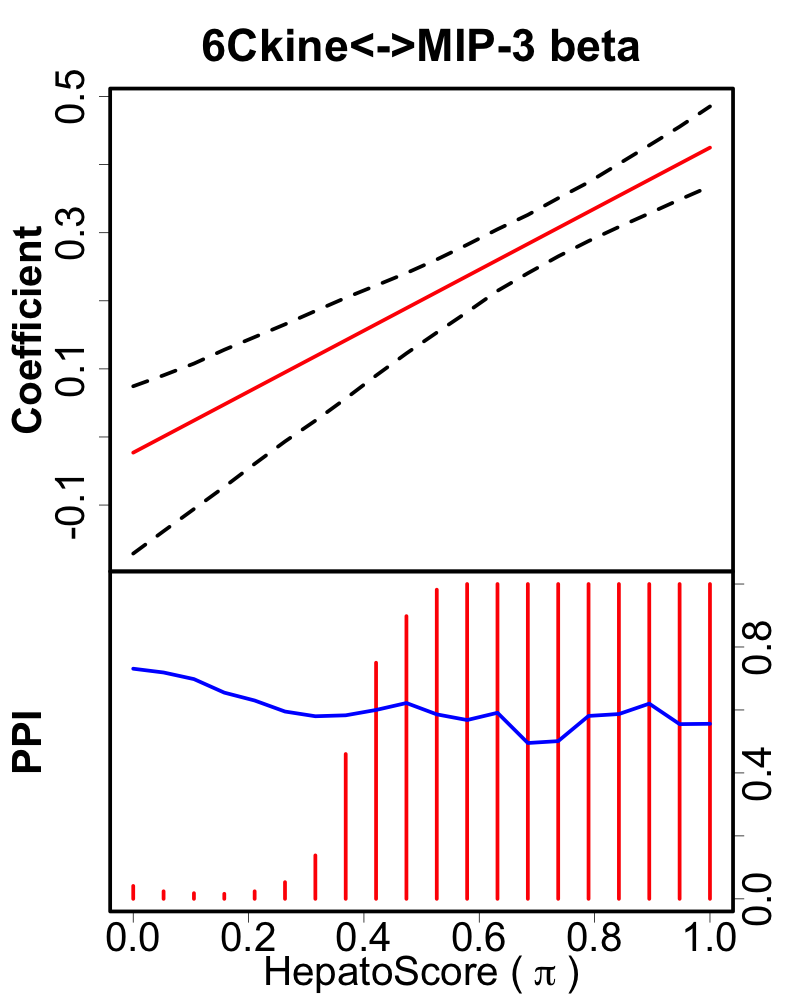}
  \includegraphics[width=0.25\textwidth]{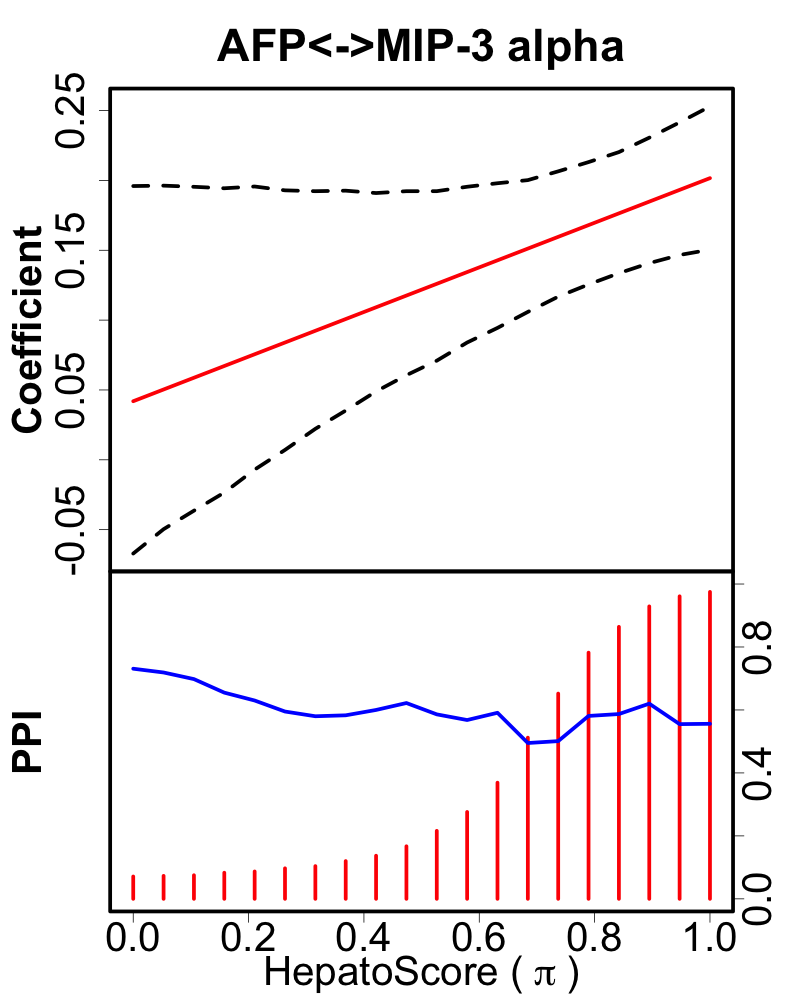}
  \includegraphics[width=0.25\textwidth]{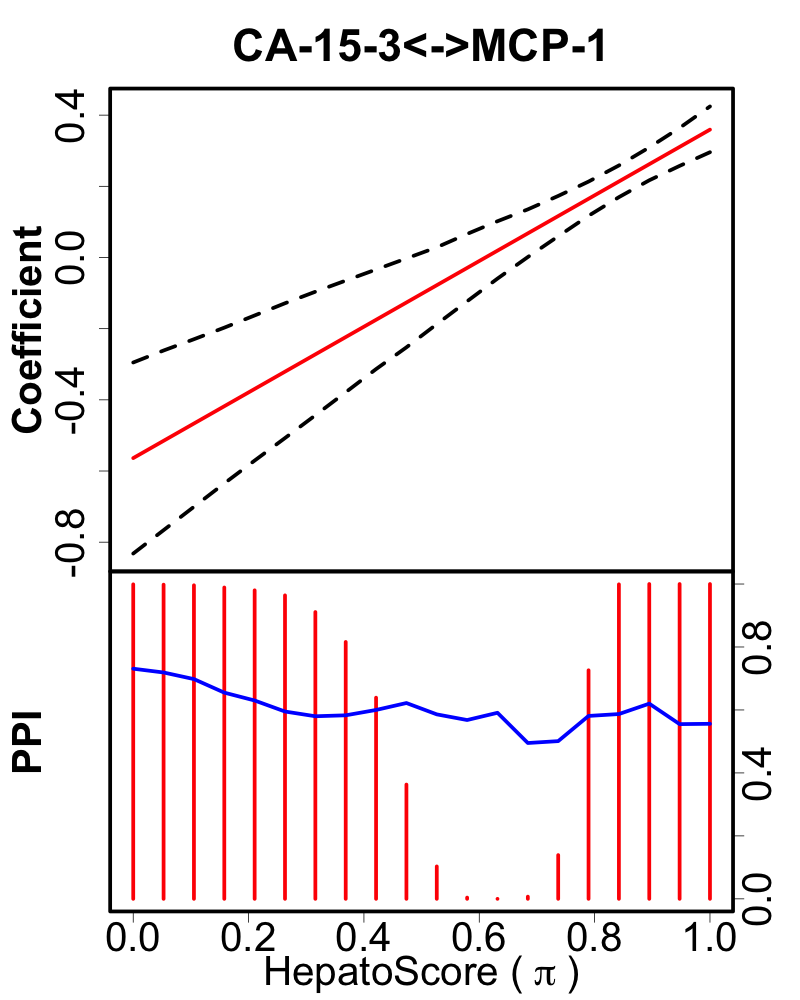}
\caption{\small Predicted Edges. Top panels show predicted edge strengths, with bottom panels plotting posterior probabilities of inclusion (PPI) with blue lines indicating FDR=0.10 thresholds. For a value of $\pi$, the head of the corresponding red line above the blue curve indicates a nonzero edge strength.}
  \label{figedge1}
  \end{figure}
 
 In Supplementary Figure 7, we show several additional connections we consider biologically meaningful, as we discuss in Supplementary Section E. We further summarize how edge connectedness varies with the \textit{HepatoScore} in Figure 4 (see Supplementary Figures 8 and 9 for $\kappa = 0.1$ and $0.2$). These results suggest that the protein networks in these important pathways differ in HCC patients with more and less advanced stages of the disease, with connectivity increasing with \textit{HepatoScore}, with a dramatically greater number of connections for the HCC patients with higher \textit{HepatoScore} and the most poor prognoses.  Biological studies investigating these differences have the potential to reveal insights into the molecular heterogeneity distinguishing these patients from those with a much better prognosis, and this knowledge can contribute towards our efforts to identify sorely needed new precision therapy strategies for HCC. These discoveries were made possible by the novel modeling framework we have introduced in this paper.


\section{Discussion} \label{discussion}

In this article, we introduce a \textit{Bayesian edge regression} model for construction of non-static undirected graphs with edge strengths varying with extraneous covariates. To deal with potential high dimensionality, we use global-local priors to effectively induce sparsity into the underlying graphs, and we use posterior probabilities to infer important graph edges for a given set of covariates. Based on node-wise regressions, that have been shown with good performance for graph reconstruction \citep{ref4,ref6,ha2020bayesian}, our method is primarily focused and recommended for applied settings in which the focus is on edge detection rather than estimation of the full precision or covariance matrix. This modeling framework allows researchers to study how clinical and biological factors lead to heterogeneous genomic or proteomic networks varying across patients. We demonstrate how this method could be used to incorporate tumor purity in estimating structural differences between tumor and normal graphs, and to assess how graphs vary across a continuous prognostic factor \textit{HepatoScore} to explain interpatient heterogeneity in HCC. Our construction is general and  can be used in any setting with multivariate data and covariates for which assessment of how conditional dependencies across the variables vary across  continuous or discrete covariates are of interest.  While motivated by the setting of continuous covariates, the method is based on a general regression framework in which any number of continuous or discrete covariates or interactions can be included.  We also provide freely available code for fitting our models.

Our sampling scheme employs a Gibbs sampler, which yields posterior samples for the model and predicted graphs for any set of covariates that can be used for posterior inference. Our hyperpriors for the normal-gamma shrinkage prior are set to allow the borrowing of information for regularization parameters of covariate coefficient across different edges. Our sampling procedure is also easy to implement and requires only minimal tuning of shrinkage hyperparameters. We show the parameterization of our \textit{conditional precision function} and its practicality through simulation study, and we demonstrate that our method is able to provide a reasonable sensitivity and specificity in edge selection. The parameterization is flexible and shown to be able to borrow strength in a group-specific setting by introducing interactions. 

Our fully Bayesian method is designed for application to moderate-sized graphs from dozens to over 100 nodes, the scale of our motivating examples, and our method scales well to these sizes.  It is not intended for enormous graphs with 1000s or 10,000s of nodes, a setting that would require enormous sample sizes for estimability anyway.  As is commonly done in genomic settings \citep{ref2, telesca2012modeling, chun2015gene}, we recommend researchers select a subset of genes of interest from pre-specified pathways of interest. For example, if one does not have an {\it a priori} list of 100 or so genes to look at, they could first download pathway genes from a public database such as KEGG, Reactome, BioCyc, or Pathway Commons, and then perform manual curation to come up a gene list in the order of dozens to 100s of genes or less for each model fit.

In this paper, we focus on the setting for which the graph edge strengths are linearly related to covariates.  In some settings, one may wish to relax this linearity assumption and use nonparametric regression approaches, such as generalized additive models in this setting, so that edge strengths can vary nonlinearly with the covariates using the well-known association between penalized splines and random effect models.  This involves significant changes to the methodological framework and computational schemes. 

Additionally, we note that like many previous works in graphical modeling and especially in graphical regression \citep{ni2018bayesian, ref10} that care most about graph structures, our current implementation focuses on regression for the off-diagonal elements of the precision matrices after standardizing all variables, which does not account for the change of diagonal elements. Our simulation studies are generated such that the diagonals vary with the covariates, and the outstanding performance of our method in the simulations demonstrate robustness of performance to this type of heteroscedasticity.  In principle, we could avoid standardization and regress the diagonal variances on covariates as well, although this would require substantially reworking the modeling framework, and is outside the scope of this paper.

We also acknowledge that, like numerous other node-wise regression methods in the literature \citep{ref4,ref6,ref39,ref10,ha2020bayesian}, our model does not explicitly constrain positive definiteness  for all possible covariate levels. One possible route is to construct a joint prior (and hence a generative model) on the entire precision matrix elements,  $\Omega(\mathbold{X}),~\forall \mathbold{X}$, such that it lies in the cone of positive definite matrices. Although theoretically sound, this would, in principle,  invoke a joint sampling scheme to generate precision matrix (and its elements) for each subject and  would add considerable computational expense. Specifically, in contrast to fitting one population level graphical model or a few in the case of multiple graphical models, this scenario involves a fitting schema that scales both in the number of nodes and subjects, thus significantly increasing the complexity of the problem and enforcement of sparsity, and without additional structural assumptions, would make it untenable for many practical settings, including ours. 
Instead, we focus on identifying edges whose strengths vary across covariates, and we have found in practice our method tends to yield positive definite predicted precision matrices in a vast majority of the cases, e.g., for 99.8\% of covariate levels for Simulation 1 (Supplementary Table 6).  Given the general utility of our model and outstanding performance in simulations, we believe that our method is a substantial addition to the literature even without this explicit constraint, so we leave its consideration for future work.

\begin{singlespace}
\bibliographystyle{authordate1}
\bibliography{ref}
\end{singlespace}
\newpage
\appendix 
\renewcommand{\thefigure}{S\arabic{figure}}
\renewcommand{\thetable}{S\arabic{table}}

\setcounter{figure}{0}
\setcounter{table}{0}

\begin{center}
{\Large \bf Supplementary Information for Bayesian Edge Regression in Undirected Graphical Models to Characterize Interpatient Heterogeneity in Cancer}
\end{center}

\section*{A: Summary of Notation}
The context of edge regression enables the use of an index set for exogenous covariates, which makes it complicated in notation. Herein, we summarize the notation of random vectors and their matrix form.\\
\vskip -16pt
\begin{table}[H]
\centering
\small
\caption{\small Summary of notation}
\begin{tabular}{ll}
  \hline
Symbol & Description  \\ 
  \hline
  $s$ & Index of exogenous covariate\\
  $i, j$ & Index of vertex\\
  $n$ & Index of sample\\
  $q$ & Number of exogenous covariates\\
  $p$ & Number of vertices\\
  $N$ & Sample size\\
 $\beta^{ij}_s$& edge regression coefficient of $s$-th covariate for edge $(i,j)$ \\ 
 $\mathbold{\beta^{ij}} = (\beta^{ij}_1, \beta^{ij}_2, \cdots, \beta^{ij}_q)^T$ &  $q$-dimensional random vector of $\beta^{ij}_s$\\
$\mathbold{\beta = (\beta^{1,2}, \beta^{1,3}, \cdots, \beta^{p-1,p})}$ & $q\times \frac{p(p-1)}{2}$-matrix of $\mathbold{\beta^{ij}}$\\
$X_{s,n}$ & the $s$-th covariate of sample $n$\\
$\mathbold{X_n} = (X_{1,n}, X_{2,n}, \cdots, X_{q,n})^T$& $q$-dimensional random vector of exogenous covariates for sample $n$\\
$\mathbold{X = (X_1^T, X_2^T, \cdots, X_N^T)^T}$ & $N \times q$-matrix of exogenous covariates\\
$Y_n^i$ & random variable of vertex $i$ in graph for sample n\\
$\mathbold{Y_n} = (Y_n^1, Y_n^2, \cdots, Y_n^p)^T$&$p$-dimensional random vector for sample $n$\\
$\mathbold{Y = (Y_1^T, Y_2^T, \cdots, Y_N^T)^T}$& $N \times p$-matrix of observed data\\
$\psi^{ij}_s$ &  scale parameter of normal prior for the $s$-th covariate of edge $(i,j)$\\
$\mathbold{\psi^{ij}} = diag(\psi^{ij}_1, \psi^{ij}_2, \cdots, \psi^{ij}_q)$ & $q \times q$-matrix of scale parameter of posterior probability for edge $(i,j)$\\
$\mathbold{{\widetilde{\mu}}^{ij}}$,  $\mathbold{{\widetilde{\Sigma}}^{ij}}$ & parameter of normal prior for $\mathbold{\beta^{ij}}$\\
 $S_{1,n}$ & element of calculation for $\mathbold{{\widetilde{\mu}}^{ij}}$,  $\mathbold{{\widetilde{\Sigma}}^{ij}}$\\
  $S_{1,n}$ & element of calculation for $\mathbold{{\widetilde{\mu}}^{ij}}$,  $\mathbold{{\widetilde{\Sigma}}^{ij}}$\\
$\mathbold{S_1} = diag(S_{1,1}, S_{1,2}, \cdots, S_{1,N})$ & $N\times N$-matrix of $S_{1,n}$\\
$\mathbold{S_2} = \{S_{2,1}, S_{2,2}, \cdots, S_{2,N}\}^T$ & $N$-dimensional vector of $S_{2,n}$ \\
  \end{tabular}
\end{table}

\section*{B: MCMC Details}
\begin{algorithm}
\caption{\small MCMC sampling scheme under normal-gamma prior for edge regression 
}
\label{Algorithm1} 
\begin{algorithmic}[1]
\Initialize{ $\{\beta^{ij}_s\}_{s\in S}^{ 1<i \neq j <p}$, $\{\omega^{ii}\}_{i=1}^p$, $\{\psi_s^{ij}\}_{s\in S}^{ 1<i \neq j <p}$, $\{\lambda_s\}_{s\in S}$, $\{\gamma_s\}_{s\in S}$}
\For{iteration $l = B+1, \cdots, L$, ($B$ is the burn-in period)}
\State a. update $\beta^{ij,l}_s$, $\omega^{ii,l}$, $\psi_s^{ij,l}$ by a Gibbs step
\State b. update $\lambda_s$ and $\gamma_s$ by a Metropolis-Hasings step 
\For{each sample $n = 1, \cdots, N$}
\State calculate $\omega^{ij,l}(x_{(n)})$ from $\beta^{ij,l}_s$ and $ x_{(n)}$
\State calculate $\rho^{ij,l}(x_{(n)})$ from $\omega^{ij,l}(x)$ and  $\omega^{ii,l}$ given  $X = x_{(n)}$
\EndFor
\EndFor
\State Output thinned posterior samples of $\rho^{ij}(x_{(n)})$
\end{algorithmic}
\end{algorithm}

\begin{itemize}
\item Update $\beta^{ij}$ for every pair $(i,j), i < j$.\\
For $\mathbold{\beta^{ij}} = \{\beta^{ij}_s\}^{S}$ of any given pair of vertex $(i, j)$, the full conditional distribution follows a multivariate Gaussian distribution with mean\\
\begin{equation}
\begin{aligned}
\mathbold{{\widetilde{\mu}}^{ij}}& = -(\mathbold{X}^T\mathbold{S_1}\mathbold{X}+(\mathbold{\psi^{ij}})^{-1})^{-1}\mathbold{X}^T\mathbold{S_2}\\
\end{aligned}
\end{equation}
and variance\\
\begin{equation}
\begin{aligned}
\mathbold{{\widetilde{\Sigma}}^{ij}}& = (\mathbold{X}^T\mathbold{S_1}\mathbold{X}+(\mathbold{\psi^{ij}})^{-1})^{-1}
\end{aligned}
\end{equation}
where $\mathbold{X = (X_1^T, X_2^T, \cdots, X_N^T)^T}$ is an $N\times q$ matrix with each row describing observed $X$ for each sample. $\mathbold{S_1} = diag(S_{1,1}, S_{1,2}, \cdots, S_{1,N})$ is an $N\times N$-vector and $\mathbold{S_2} = \{S_{2,1}, S_{2,2}, \cdots, S_{2,N}\}^T$ is an $N$-dimensional vector. $S_{1,n}$ and $S_{2,n}$ are given as:\\
\begin{equation}
\begin{aligned}
S_{1,n} & = \frac{(Y_n^j)^2}{\omega^{ii}} + \frac{(Y_n^i)^2}{\omega^{jj}}\\
S_{2,n} & = 2Y_n^iY_n^j +\mathbold{X_n}^T\mathbold{\beta^{i,-j}Y_n^{-(i,j)}}\frac{\mathbold{Y_n^{j}}}{\omega^{ii}} + \mathbold{X_n}^T\mathbold{\beta^{j,-i}Y_n^{-(j,i)}}\frac{\mathbold{Y_n^{i}}}{\omega^{jj}}
\end{aligned}
  \end{equation}
\item Update $\omega^{ii}, i = 1, 2, \cdots, p$.\\
The full conditional distribution of $\omega^{ii}$ is:
\begin{equation}
\begin{aligned}
GIG(\frac{n}{2}+1,  \sum_{n=1}^N(Y_n^i)^2 , diag(\mathbold{X}\mathbold{\beta^{i\cdot}}(\mathbold{Y^{-i}})^T)diag(\mathbold{X}\mathbold{\beta^{i\cdot}}(\mathbold{Y^{-i}})^T)^T)\end{aligned}
  \end{equation}
where $GIG(m,a,b)$ is the generalized inverse Gaussian (GIG) distribution. It has the density\\
\begin{equation}
\begin{aligned}
f(x) = \frac{(a/b)^{m/2}}{2 K_m(\sqrt{ab})} x^{(m-1)} e^{-(ax + b/x)/2}.
\end{aligned}
  \end{equation}
\item Update $\psi^{ij}_s$.\\
$\psi^{ij}_s$ for the edge $(i,j)$ and $s$-covariate can be effectively updated in a block, since their full conditional distributions are independent. The full conditional distribution also follows a GIG distribution with:\\
\begin{equation}
\begin{aligned}
GIG(\lambda_s - \frac{1}{2}, 1/\gamma_s^2, (\beta^{ij}_s)^2).
\end{aligned}
  \end{equation}
\item Update hyper-parameters of the normal-gamma prior.\\
We assigned prior  $\pi(\lambda_s) = exp(1)$ for the shape parameter $\lambda_s$, then the full conditional $\lambda_s$ is proportional to:\\
\begin{equation}
\begin{aligned}
 \propto \pi(\lambda_s)\frac{1}{(2\gamma^2)^{\frac{p(p-1)}{2}\lambda_s}(\Gamma(\lambda_s))^\frac{p(p-1)}{2}}(\prod_{i \neq j} \psi^{ij}_s)^{\lambda_s}.
\end{aligned}
\end{equation}
For the scale parameter $\gamma$, we specify a prior $\sum_s{\lambda_s}\gamma^{2} \sim Ga(2, \sum{M_s}))$. $M_s$ is a hyper-parameter to approximately control the scale of $\lambda_s\gamma^2$ for the $s$-th covariate. The calculation of $M_s$ is discussed in our supplementary materials and each specific problem. We have:
\begin{equation}
\begin{aligned}
 \gamma^{-2} \sim Ga(2 + qp(p-1)\lambda_s/2, \sum_s{M_s}/(2\sum_s{\lambda_s}) + \frac{1}{2}\sum_s{\sum_{i \neq j}\psi^{ij}_s}).
 \end{aligned}
\end{equation}
\end{itemize}

\section*{C: Derivation of MCMC Sampling} 
In equation $(2)$ of the main text, we condition the precision matrix $\Omega$ on a set of exogenous covariates $X$. Instead of modeling $\gamma^{ij}(x)$, we model the conditional precision function on $\omega^{ij}(x)$ under the regression setting. Since the precision matrix is symmetric, we have $\omega^{ji}(.) = \omega^{ij}(.)$. Hence we can coerce these two functions to have the same form in the sampling scheme. When we regress $\omega^{ij}(.)$ on $X$ in a linear setting:\\
\begin{equation}
\begin{aligned}
\omega^{ij}(x) & = \sum_{s=1}^{q} \beta^{ij}_sX_s.
\end{aligned}
\end{equation}
We have $\beta^{ij}_s = \beta^{ji}_s$ for every $i \neq j$. Then we have a complete likelihood given by equation $(3)$ of the main text. In the context of using a normal-gamma shrinkage prior, the posterior distribution of all parameters in our model can be updated through a Gibbs sampling scheme. We will also follow the normal-gamma paper to talk about updating hyper-parameters for a normal gamma prior through a Metropolis-Hastings step in our model.\\
\paragraph{Update $\beta^{ij}$ for every pair $(i,j), i < j$}
For $\mathbold{\beta^{ij}} = \{\beta^{ij}_s\}^{S}$ of any given pair of vertex $(i, j)$, we derive the full conditional by\\
\begin{equation}
\begin{aligned}
f(\mathbold{\beta^{ij}}|.) & \propto f(\mathbold{Y^i}|\mathbold{Y^{-i}}, \{\beta_s^{i,-(i,j)}\}_{s=1}^q, \{\beta_s^{i,j}\}_{s=1}^q, \omega^{i,i}, \{X_{s,n}\}_{s=1,n=1}^{q,N})\\
& \times f(\mathbold{Y^j}|\mathbold{Y^{-j}}, \{\beta_s^{j,-(i,j)}\}_{s=1}^q, \{\beta_s^{i,j}\}_{s=1}^q, \omega^{j,j}, \{X_{s,n}\}_{s=1,n=1}^{q,N}) \times f(\mathbold{\beta^{i,j}|\psi^{i,j}})\\
& \propto \exp\{-\frac{1}{2}\sum_{n=1}^N[\frac{(Y_n^i + \frac{\sum_{k\neq i}^p\sum_{s=1}^{q} \beta^{ik}_sX_{s,n}Y_n^{k}}{\omega^{ii}} )^2}{(\omega^{ii})^{-1}} + 
\frac{(Y_n^j + \frac{\sum_{k\neq j}^p\sum_{s=1}^{q} \beta^{jk}_sX_{s,n}Y_n^{k}}{\omega^{jj} })^2}{(\omega^{jj})^{-1}}]\}
\times f(\mathbold{\beta^{i,j}|\psi^{i,j}})\\
& \propto \exp\{-\frac{1}{2}\sum_{n=1}^N[ 2Y_n^i\sum_{k\neq i}^p\sum_{s=1}^{q} \beta^{ik}_sX_{s,n}Y_n^{k} + \frac{(\sum_{k\neq i}^p\sum_{s=1}^{q} \beta^{ik}_sX_{s,n}Y_n^{k})^2}{\omega^{ii}}\\
& +2Y_n^j\sum_{k\neq j}^p\sum_{s=1}^{q} \beta^{jk}_sX_{s,n}Y_n^{k} + \frac{(\sum_{k\neq j}^p\sum_{s=1}^{q} \beta^{jk}_sX_{s,n}Y_n^{k})^2}{\omega^{jj}}]\}
\times f(\mathbold{\beta^{i,j}|\psi^{i,j}})\\    
& \propto \exp\{-\frac{1}{2}\sum_{n=1}^N[ 4Y_n^iY_n^j \sum_{s=1}^{q} \beta^{ij}_sX_{s,n} +  \frac{2\sum_{k\neq (i,j),s}\beta^{ik}_sX_{s,n}Y_n^{k}\sum_{s=1}^{q} \beta^{ij}_sX_{s,n}Y_n^{j} + (\sum_{s=1}^{q} \beta^{ij}_sX_{s,n}Y_n^{j})^2 }{\omega^{ii}} \\
& +  \frac{2\sum_{k\neq (i,j),s}\beta^{jk}_sX_{s,n}Y_n^{k}\sum_{s=1}^{q} \beta^{ij}_sX_{s,n}Y_n^{i} + (\sum_{s=1}^{q} \beta^{ij}_sX_{s,n}Y_n^{i})^2 }{\omega^{jj}}
]\}\times f(\mathbold{\beta^{i,j}|\psi^{i,j}})\\         
& \propto f(\mathbold{\beta^{i,j}|\psi^{i,j}})  \times \exp\{-\frac{1}{2}\sum_{n=1}^N[ (\sum_{s=1}^{q} \beta^{ij}_sX_{s,n})^2 (\frac{(Y_n^j)^2}{\omega^{ii}} + \frac{(Y_n^i)^2}{\omega^{jj}})\\
& + 2\sum_{s=1}^{q} \beta^{ij}_sX_{s,n} (2Y_n^iY_n^j + \frac{\sum_{k\neq (i,j),s}\beta^{ik}_sX_{s,n}Y_n^{k}Y_n^{j}}{\omega^{ii}} + \frac{\sum_{k\neq (i,j),s}\beta^{jk}_sX_{s,n}Y_n^{k}Y_n^{i}}{\omega^{jj}} )
 ]\}.\\   
\end{aligned}
\label{eq:10}%
  \end{equation}
In the equation above, we simplify it by denoting:\\
\begin{equation}
\begin{aligned}
S_{1,n} = \frac{(Y_n^j)^2}{\omega^{ii}} + \frac{(Y_n^i)^2}{\omega^{jj}}
\end{aligned}
  \end{equation}
and \\
\begin{equation}
\begin{aligned}
S_{2,n} = 2Y_n^iY_n^j + \sum_{k\neq (i,j),s}\beta^{ik}_sX_{s,n}Y_n^{k}\frac{Y_n^{j}}{\omega^{ii}} + \sum_{k\neq (i,j),s}\beta^{jk}_sX_{s,n}Y_n^{k}\frac{Y_n^{i}}{\omega^{jj}}.
\end{aligned}
  \end{equation}

For simplification of notation we have:\\
\begin{equation}
\begin{aligned}
\sum_{k\neq (i,j),s}\beta^{ik}_sX_{s,n}Y_n^{k} = \mathbold{X_n^T\beta^{i,-j}Y_n^{-(i,j)}}\\
\sum_{k\neq (i,j),s}\beta^{jk}_sX_{s,n}Y_n^{k} = \mathbold{X_n^T\beta^{j,-i}Y_n^{-(j,i)}},
\end{aligned}
  \end{equation}

where $\mathbold{\beta^{i,-j}}$ corresponds to the columns that include $i$ but not $j$ in the superscript in $\mathbold{\beta}$, and $\mathbold{Y_n^{-(i,j)}}$ corresponds to the remaining elements in the vector $\mathbold{Y_n}$ after removing $i$-th and $j$-th element. This way, we rewrite:\\
\begin{equation}
\begin{aligned}
S_{2,n} = 2Y_n^iY_n^j +\mathbold{X_n}^T\mathbold{\beta^{i,-j}Y_n^{-(i,j)}}\frac{\mathbold{Y_n^{j}}}{\omega^{ii}} + \mathbold{X_n}^T\mathbold{\beta^{j,-i}Y_n^{-(j,i)}}\frac{\mathbold{Y_n^{i}}}{\omega^{jj}}.
\end{aligned}
  \end{equation}
We also have:\\
 \begin{equation}
\begin{aligned}
\sum_{s=1}^{q} \beta^{ij}_sX_{s,n} &= (\mathbold{\beta^{ij}})^T\mathbold{X_n}\\
(\sum_{s=1}^{q} \beta^{ij}_sX_{s,n})^2 &= (\mathbold{\beta^{ij}})^T\mathbold{X_nX_n}^T\mathbold{\beta^{ij}}.
\end{aligned}
  \end{equation}
Hence, we have:\\
\begin{equation}
\begin{aligned}
(5) &= f(\mathbold{\beta^{i,j}|\psi^{i,j}})  \times \exp\{-\frac{1}{2}\sum_{n=1}^N[(\mathbold{\beta^{ij}})^T\mathbold{X_nX_n}^T\mathbold{\beta^{ij}}S_{1,n} + 2(\mathbold{\beta^{ij}})^T\mathbold{X_n}S_{2,n} ]\}\\ 
& \propto \exp{-\frac{1}{2}[(\mathbold{\beta^{ij}})^T(\mathbold{\psi^{ij}})^{-1}\mathbold{\beta^{ij}} + \sum_{n=1}^N((\mathbold{\beta^{ij}})^T\mathbold{X_n}S_{1,n}\mathbold{X_n}^T\mathbold{\beta^{ij}} + 2(\mathbold{\beta^{ij}})^T\mathbold{X_n}S_{2,n})]}\\
& = \exp{-\frac{1}{2}[(\mathbold{\beta^{ij}})^T(\mathbold{\psi^{ij}})^{-1}\mathbold{\beta^{ij}} + ((\mathbold{\beta^{ij}})^T\sum_{n=1}^N\mathbold{X_n}S_{1,n}\mathbold{X_n}^T\mathbold{\beta^{ij}} + 2(\mathbold{\beta^{ij}})^T\mathbold{X_n}S_{2,n})]}\\
& = \exp{-\frac{1}{2}[((\mathbold{\beta^{ij}})^T(\sum_{n=1}^N\mathbold{X_n}S_{1,n}\mathbold{X_n}^T+(\mathbold{\psi^{ij}})^{-1})\mathbold{\beta^{ij}} + 2(\mathbold{\beta^{ij}})^T\sum_{n=1}^N\mathbold{X_n}S_{2,n})]}.
\end{aligned}
\end{equation}
According to equation (\ref{eq:10}), we have $\mathbold{\beta^{ij}}|. \sim N(\mathbold{{\widetilde{\mu}}^{ij}}, \mathbold{{\widetilde{\Sigma}}^{ij}})$, where\\
\begin{equation}
\begin{aligned}
\mathbold{{\widetilde{\mu}}^{ij}}& = -(\sum_{n=1}^N\mathbold{X_n}S_{1,n}\mathbold{X_n}^T+(\mathbold{\psi^{ij}})^{-1})^{-1}\sum_{n=1}^N\mathbold{X_n}S_{2,n}\\
\mathbold{{\widetilde{\Sigma}}^{ij}} & = (\sum_{n=1}^N\mathbold{X_n}S_{1,n}\mathbold{X_n}^T+(\mathbold{\psi^{ij}})^{-1})^{-1}.
\end{aligned}
\end{equation}
For further simplification, we formulate $S_1$ and $S_2$ in a matrix form according to \textbf{Appendix A}.
The calculation of element-wise $\mathbold{X_n}^T\mathbold{\beta^{i,-j}}\mathbold{Y_n^{-(i,j)}}$ is complicated in $\mathbold{S_2}$. We calculate this component vector through matrix algebra, where\\
\begin{equation}
\begin{aligned}
\{\mathbold{X_n}^T\mathbold{\beta^{i,-j}}\mathbold{Y_n^{-(i,j)}}\}_{n=1}^N = diag(\mathbold{X}\mathbold{\beta^{i,-j}}(\mathbold{Y^{-(i,j)}})^T).\\
\end{aligned}
\end{equation}

Then we can express $(11)$ as:\\
\begin{equation}
\begin{aligned}
\mathbold{{\widetilde{\mu}}^{ij}}& = -(\mathbold{X}^T\mathbold{S_1}\mathbold{X}+(\mathbold{\psi^{ij}})^{-1})^{-1}\mathbold{X}^T\mathbold{S_2}\\
\mathbold{{\widetilde{\Sigma}}^{ij}}& = (\mathbold{X}^T\mathbold{S_1}\mathbold{X}+(\mathbold{\psi^{ij}})^{-1})^{-1}.
\end{aligned}
\end{equation}

\paragraph{Update $\omega^{ii}, i = 1, 2, \cdots, p$}
By setting $f(\omega^{ii}) \propto 1$, we derive the full conditional by:\\
\begin{equation}
\begin{aligned}
f(\omega^{ii}|.) & \propto f(\mathbold{Y^i}|\mathbold{X},\mathbold{Y^{-i}}, \{\beta_s^{i,-(i,j)}\}_{s=1}^q, \{\beta_s^{i,j}\}_{s=1}^q, \omega^{i,i}, X_{s,n}) \times f(\omega^{i,i})\\
& \propto (\omega^{ii})^{\frac{n}{2}} \exp\{-\frac{1}{2}\sum_{n=1}^N[\frac{(Y_n^i + \frac{\sum_{k\neq i}^p\sum_{s=1}^{q} \beta^{ik}_sX_{s,n}Y_n^{k}}{\omega^{ii}} )^2}{(\omega^{ii})^{-1}}]\}\\
& \propto (\omega^{ii})^{\frac{n}{2}} \exp\{-\frac{1}{2}\sum_{n=1}^N[(Y_n^i)^2\omega^{ii} +  \frac{(\sum_{k\neq i}^p\sum_{s=1}^{q} \beta^{ik}_sX_{s,n}Y_n^{k})^2}{\omega^{ii}}]\}\\
& \propto (\omega^{ii})^{\frac{n}{2}} \exp\{-\frac{1}{2}[\omega^{ii} \sum_{n=1}^N(Y_n^i)^2 +   \sum_{n=1}^N\frac{\mathbold{X_n}^T\mathbold{\beta^{i\cdot}}\mathbold{Y_n^{-i}}(\mathbold{Y_n^{-i}})^T(\mathbold{\beta^{i\cdot}})^T\mathbold{X_n}}{\omega^{ii}}]\}\\
& \propto (\omega^{ii})^{\frac{n}{2}} \exp\{-\frac{1}{2}[\omega^{ii} \sum_{n=1}^N(Y_n^i)^2 +   \frac{diag(\mathbold{X}\mathbold{\beta^{i\cdot}}(\mathbold{Y^{-i}})^T)diag(\mathbold{X}\mathbold{\beta^{i\cdot}}(\mathbold{Y^{-i}})^T)^T}{\omega^{ii}}]\}.\\
\end{aligned}
  \end{equation}
Hence, we can sample $\omega^{ii}|. \sim GIG(\frac{n}{2}+1,  \sum_{n=1}^N(Y_n^i)^2 , diag(\mathbold{X}\mathbold{\beta^{i\cdot}}(\mathbold{Y^{-i}})^T)diag(\mathbold{X}\mathbold{\beta^{i\cdot}}(\mathbold{Y^{-i}})^T)^T)$.\\
\paragraph{Update $\psi^{ij}_s$}
We sample $\psi^{ij}_s$ according to\\
\begin{equation}
\begin{aligned}
f(\psi^{ij}_s|.) & \propto f(\beta^{ij}_s|\psi^{ij}_s) \times f(\psi^{ij}|\lambda_s, \gamma).
\end{aligned}
\end{equation}
That is equally, $\psi^{ij}_s \sim GIG(\lambda_s - \frac{1}{2}, 1/\gamma^2, (\beta^{ij}_s)^2)$.\\

\paragraph{Update $\lambda_s$ and $\gamma$}
Instead of using a cross-validation technique to select $\lambda$ and $\gamma$ in the normal-gamma prior, we choose to sample these parameters by specifying hyper-priors.
We update $\lambda_s$ and $\gamma$ through a Metropolis-Hastings sampling method.\\
If we use $\pi(\lambda_s)$ to denote the prior of $\lambda_s$, we can have the full conditional of $\lambda_s$ as\\
\begin{equation}
\begin{aligned}
f(\lambda_s|\cdot) & \propto \pi(\lambda_s)\frac{1}{(2\gamma^2)^{\frac{p(p-1)}{2}\lambda_s}(\Gamma(\lambda_s))^\frac{p(p-1)}{2}}(\prod_{i \neq j} \psi^{ij}_s)^{\lambda_s},
\end{aligned}
\end{equation}
where we set $\pi(\lambda_s) \sim exp(1)$.\\
We have multiplicative random walk updates on $\lambda_s$ through $\lambda_s^* = exp(\sigma^2_{\lambda_s}z)\lambda_s$, where $z$ satisfies a standard normal. $\sigma^2_{\lambda_s}$ is a tuning parameter for random walk and it is chosen so that the acceptance rate is around $20\%$ to $30\%$. Then the acceptance function is given by:\\
\begin{equation}
\begin{aligned}
min\bigg \{1,  \frac{\lambda^*_s}{\lambda_s}\frac{\pi(\lambda^*_s)}{\pi(\lambda_s)}\frac{(2\gamma^2)^{\frac{p(p-1)}{2}\lambda^*_s}(\Gamma(\lambda^*_s))^\frac{p(p-1)}{2}}{(2\gamma^2)^{\frac{p(p-1)}{2}\lambda_s}(\Gamma(\lambda_s))^\frac{p(p-1)}{2}}(\prod_{i \neq j} \psi^{ij}_s)^{\lambda^*_s - \lambda_s} \bigg\}.
\end{aligned}
\end{equation}
For the scale parameter $\gamma$, we follow the suggested setting in the normal-gamma prior paper, with $\sum_s{\lambda_s}\gamma^{2} \sim Ga(2, \sum{M_s})$. $M_s$ is a hyper-parameter to approximately control the scale of $\lambda_s\gamma^2$ for the $s$-th covariate, so we provide a heuristic solution to obtaining it by calculating the mean square error of elements from zero in a maximum likelihood estimator (MLE) of $\mathbold{\Sigma}$ for each group of samples. 
  $M_s = \frac{\sum_{1 \le i \le j \le p}(\hat \Sigma^{-1}_{s,ij})^2}{p(p-1)/2}$, where $\mathbold{\hat \Sigma_s^{-1}}$ is the inverse of the estimated Gaussian covariance matrix through MLE for samples considering the effect represented by the $s$-th covariate. When $\mathbold{\hat \Sigma_s}$ is singular, we can use the estimated precision matrix $\mathbold{\hat \Omega_s}$ through a regularization method, e.g., graphical lasso \citep{ref7}, instead of $\mathbold{\hat \Sigma_s^{-1}}$. The derivation of $M_s$ is discussed case-by-case in the section of simulation and case study of our main text.\\
Hence we have, $\gamma^{-2} \sim Ga(2 + qp(p-1)\lambda_s/2, \sum_s{M_s}/(2\sum_s{\lambda_s}) + \frac{1}{2}\sum_s{\sum_{i \neq j}\psi^{ij}_s})$.
 
\section*{D: More simulation settings and results}

\textit{NFMGGM} learns a nonparametric mixture of Gaussian graphical models that depend on a univariate covariate. In practice, we fix the mixture number to be one and consider $\pi$ as the univariate covariate to run the algorithm for only estimating a single varying graphical model. We select tuning parameters as recommended in \cite{ref3} by searching over a grid of possible values for tuning parameter $\lambda_1$ and $\lambda_2$, and we choose the combination that minimizes the approximate $AIC(\lambda_1, \lambda_2)$ for the fused and group graphical lassos. We adopt a similar search algorithm for tuning parameter $\lambda_1$ and $\lambda_2$ in \textit{LASICH}, but through minimizing the approximate $BIC(\lambda_1, \lambda_2)$ as suggested by the authors. \textit{NFMGGM} is implemented with one regularization parameter $\lambda$ for the graphical lasso and the other tuning parameter $h$ for the bandwidth in the kernel regression. We follow the two-step procedure from their paper to first select $\lambda$ with the minimum $BIC$ score and then select $h$ with a cross validation to minimize the negative log-likelihood \citep{lee2018nonparametric}. The paper of \textit{BIMGGM} suggests a similar posterior inference using Bayesian FDR considerations as our method. Thus,  using the marginal PPI output by \textit{BIMGGM}, we adopt a model selection procedure that is consistent with our consideration.

When running fused lasso, group lasso, and BIMGGM, we use the default setting accordingly from the provided R functions and source code. For running NFMGGM, considering $\pi \in [0,1]$, we let $\mathcal{U} =\{0, 0.1, 0.2, \cdots, 1\}$ be the set of grid points of the covariate $\pi$. Given that we have many normal samples with $\pi=0$ in our simulation, to better initialize the functional parameters with each grid point, in initialization we assign all the simulated normal samples to the grid point $u=0.0$. Then we divide the simulated tumor samples equally, after sorting them based on their values of $\pi$, so that these samples can be evenly assigned to the remaining grid points $\{0.1, \cdots, 1\}$. In the likelihood calculation, it was suggested to us that we implement linear interpolation to approximate parameter estimates of a tumor sample from the parameter estimates of its nearest grid points. For running LASICH, we get the graph Laplacian matrix by using the provided R functions in the \textit{LASICH} package. The estimated graph Laplacian matrix is added with a small perturbation on the diagonal elements with $0.001$ to facilitate the computation.

For running our Bayesian edge regression method in a group case, we adopt a different parameterization of \textit{CPF}:  with the grouped data in the simulation, we first use two binary covariates, $X_1$ and $X_2$, to denote the group membership for each sample, i.e., $\mathbold{Y_n}$ is from group $k$; $k\in \{1,2\} \Leftrightarrow X_{k,n} = 1, X_{-k,n} = 0$ for observation $n$, where $k$ denotes the index of tumor or the normal group. Then we add an interaction term to borrow strength between these two groups. The conditional dependence function $\omega(\mathbold{X})$ for the group case is finally formulated as:\\
 \begin{equation}
\omega^{ij}(\mathbold{X}) = \beta_1^{ij}X_1 + \beta_2^{ij}X_2 +  \beta_{12}^{ij}X_1X_2. 
\label{sup:D1}
\end{equation}
$\beta_1^{ij}$ and $\beta_2^{ij}$ measure the unshared strength in each group, and $\beta_{12}^{ij}$ models the shared strength between the two groups. Here is how the values of covariates are assigned for the data in each group.\\
\begin{table}[htbp!]
\centering
\caption*{}
\begin{tabular}{llll}
Group  & $X_1$ & $X_2$ & $X_1X_2$  \\
Normal    & 1 & 0 & 1   \\
Tumor     & 0 & 1 & 1  \\
\end{tabular}
\end{table}

\begin{figure}[H] 
\centering
\includegraphics[width=1\textwidth]{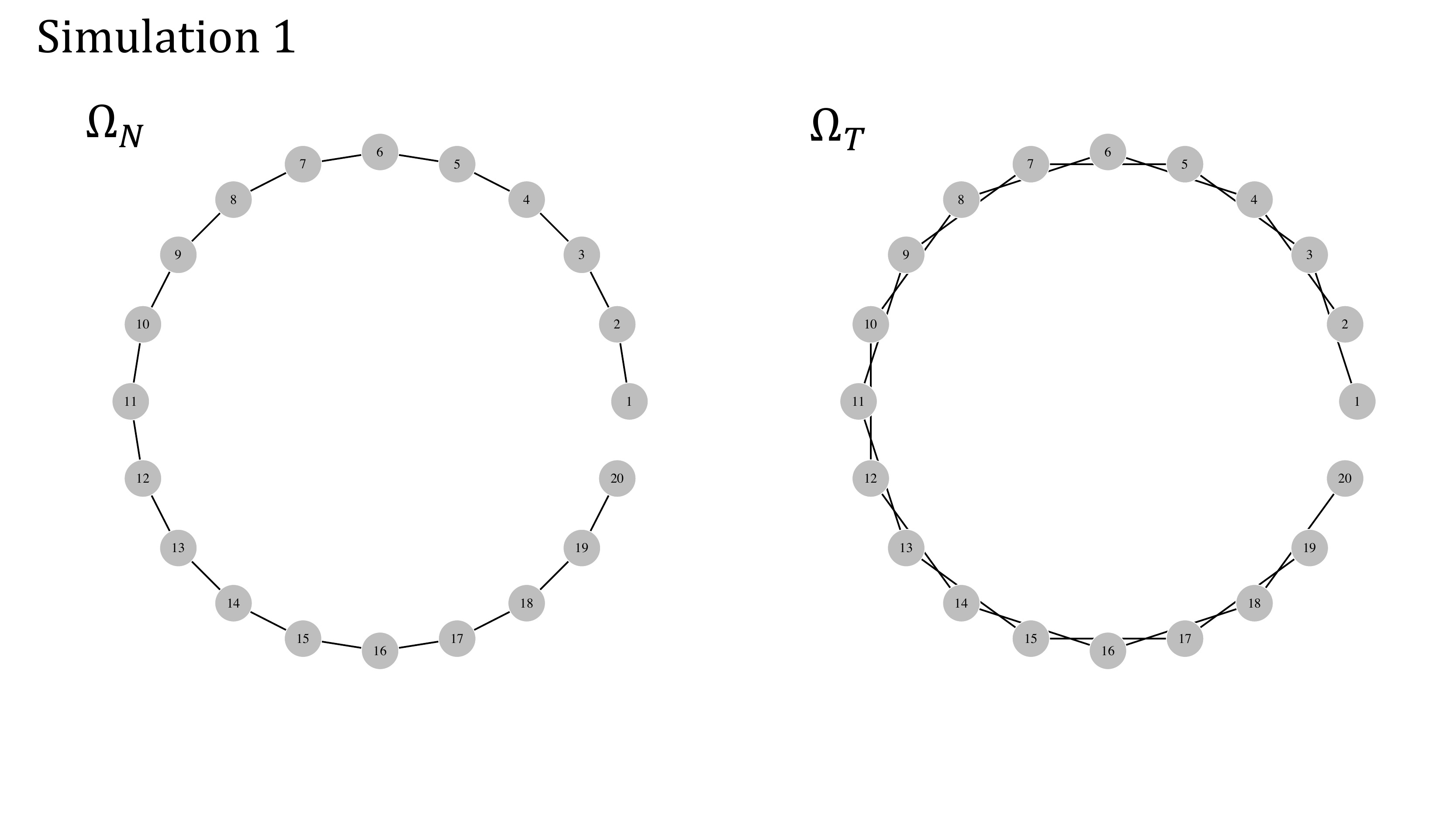}\\
\includegraphics[width=1\textwidth]{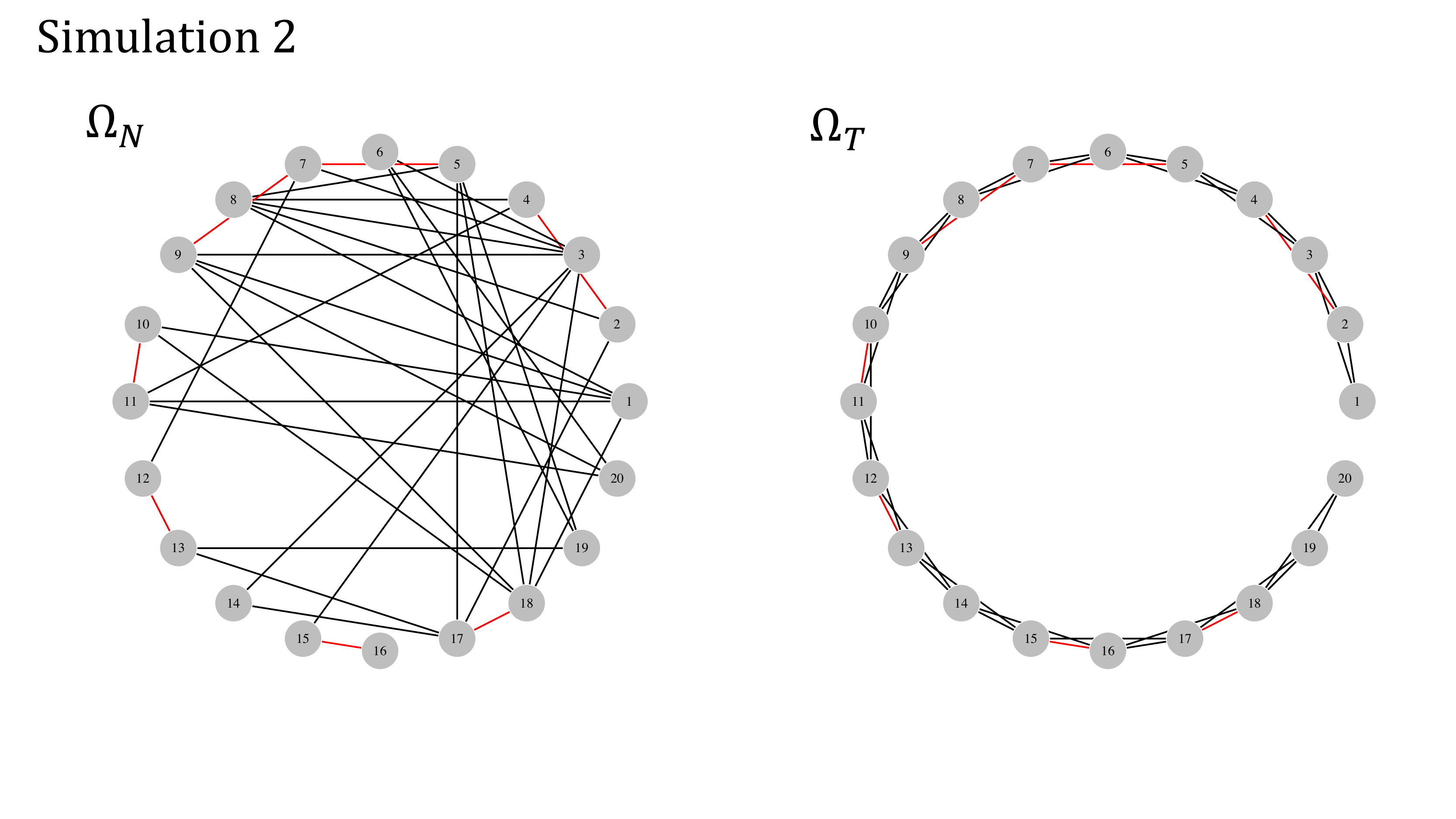}
\caption{\small Graph structures for $\Omega_N$ and $\Omega_T$ in \textit{Simulation 1} and \textit{Simulation 2}. The shared edges are colored with red.}
\label{figsimstruct}
\end{figure}

 \begin{table}[H]
\centering
\caption{\small Results of edge selection for \textit{Simulation 1} in terms of True Positive Rate (TPR), False Positive Rate (FPR), and bAUC. The reported TPR and FPR are calculated from the result using the selection rule of tuning parameters for each method. bAUC is the bivariate AUC that is computed through varying both of the two tuning parameters (regularization parameters) for each method. The numbers are averaged across 100 simulated sets, and the standard deviations are given within the parentheses.}
\label{TBres2}
\begin{threeparttable}
\begin{tabular}{lllll}
  \hline
Method & & Normal ($\mathbold{\Omega_N}$)  & Tumor ($\mathbold{\Omega_T}$) & Overall\\ 
  \hline
Fused graphical lasso &&&&\\
&TPR & 0.992 (0.019) & 0.837 (0.111) & 0.915 (0.058) \\ 
&  FPR & 0.424 (0.091) & 0.383 (0.106) & 0.403 (0.096) \\ 
&  bAUC & 0.948 (0.010) & 0.706 (0.055) & 0.827 (0.027) \\ 
Group graphical lasso &&&&\\
&TPR & 0.992 (0.019) & 0.844 (0.108) & 0.918 (0.056) \\ 
 & FPR & 0.433 (0.086) & 0.394 (0.105) & 0.414 (0.093) \\ 
 & bAUC & 0.941 (0.012) & 0.793 (0.050) & 0.867 (0.025) \\ 
LASICH &&&&\\
&TPR & 0.889 (0.099) & 0.726 (0.127) & 0.808 (0.092) \\ 
 & FPR & 0.039 (0.029) & 0.216 (0.053) & 0.127 (0.038) \\ 
 & bAUC & 0.955 (0.013) & 0.808 (0.049) & 0.882 (0.023) \\ 
NFMGGM &&&&\\
&TPR & 1.000 (0.000) & 0.989 (0.027) & 0.995 (0.014) \\ 
&  FPR & 0.611 (0.102) & 0.629 (0.095) & 0.620 (0.095) \\ 
&  bAUC & 0.950 (0.007) & 0.810 (0.031) & 0.880 (0.015) \\ 
BIMGGM &&&&\\
&TPR & 0.779 (0.111) & 0.463 (0.120) & 0.621 (0.085) \\ 
&  FPR & 0.007 (0.006) & 0.056 (0.016) & 0.032 (0.009) \\ 
&  AUC & 0.981 (0.013) & 0.847 (0.052) & 0.914 (0.027) \\ 
Bayesian edge regression (group case) &&&&\\
&  TPR & 0.968 (0.037) & 0.650 (0.148) & 0.809 (0.081) \\ 
&  FPR & 0.112 (0.020) & 0.132 (0.024) & 0.122 (0.019) \\ 
 & bAUC & 0.942 (0.011) & 0.812 (0.059) & 0.877 (0.027) \\ 
Bayesian edge regression &&&&\\
&  TPR & 0.982 (0.028) & 0.838 (0.095) & 0.910 (0.049) \\ 
&  FPR & 0.094 (0.027) & 0.083 (0.027) & 0.089 (0.019) \\ 
&  bAUC& 0.947 (0.011) & 0.916 (0.030) & 0.932 (0.016) \\ 
\hline
\end{tabular}
 \begin{tablenotes}
 \footnotesize
 \item[*] A univariate AUC is reported for BIMGGM, while bivariate AUCs are reported for all the other methods.
\end{tablenotes}
\end{threeparttable}
\end{table}
 \begin{table}[H]
\centering
\caption{\small Results of edge selection for \textit{Simulation 1} in terms of True Positive Rate (TPR), False Positive Rate (FPR), and univariate AUC. The reported TPR and FPR are calculated from the result that leads to an FPR that is closest to 0.1 for each method. The univariate AUC is reported by taking the maximum value from the AUC scores that are computed through varying one tuning parameter (or regularization parameter) while fixing the other with different values, and we report them with AUC1 and AUC2. The numbers are averaged across 100 simulated sets and the standard deviations are given within the parentheses.}
\label{TBres22}
\begin{threeparttable}
\footnotesize
\begin{tabular}{lllll}
  \hline
Method & & Normal ($\mathbold{\Omega_N}$)  & Tumor ($\mathbold{\Omega_T}$) & Overall\\ 
  \hline
Fused graphical lasso &  &  &  &  \\ 
   & TPR & 0.965 (0.042) & 0.490 (0.120) & 0.727 (0.064) \\ 
   & FPR & 0.099 (0.002) & 0.099 (0.001) & 0.099 (0.001) \\ 
   & AUC1 & 0.985 (0.010) & 0.827 (0.051) & 0.906 (0.025) \\ 
   & AUC2 & 0.966 (0.016) & 0.748 (0.048) & 0.857 (0.025) \\ 
  Group graphical lasso &  &  &  &  \\ 
   & TPR & 0.959 (0.045) & 0.533 (0.120) & 0.746 (0.064) \\ 
   & FPR & 0.099 (0.001) & 0.099 (0.000) & 0.099 (0.001) \\ 
   & AUC1 & 0.983 (0.012) & 0.834 (0.053) & 0.908 (0.026) \\ 
   & AUC2 & 0.980 (0.012) & 0.805 (0.047) & 0.893 (0.024) \\ 
  LASICH &  &  &  &  \\ 
   & TPR & 0.961 (0.044) & 0.564 (0.110) & 0.763 (0.060) \\ 
   & FPR & 0.099 (0.000) & 0.099 (0.000) & 0.099 (0.000) \\ 
   & AUC1 & 0.989 (0.009) & 0.838 (0.050) & 0.914 (0.025) \\ 
   & AUC2 & 0.973 (0.016) & 0.796 (0.048) & 0.884 (0.025) \\ 
  NFMGGM &  &  &  &  \\ 
   & TPR & 0.945 (0.064) & 0.233 (0.126) & 0.589 (0.070) \\ 
   & FPR & 0.099 (0.000) & 0.099 (0.000) & 0.099 (0.000) \\ 
   & AUC1 & 0.993 (0.006) & 0.958 (0.024) & 0.976 (0.012) \\ 
   & AUC2 & 0.980 (0.014) & 0.864 (0.041) & 0.922 (0.022) \\ 
BIMGGM &  &  &  &  \\ 
   & TPR & 0.957 (0.046) & 0.590 (0.121) & 0.773 (0.065) \\ 
   & FPR & 0.099 (0.000) & 0.099 (0.000) & 0.099 (0.000) \\ 
   & AUC & 0.981 (0.013) & 0.847 (0.052) & 0.914 (0.027) \\ 
  Bayesian edge regression (group case) &  &  &  &  \\ 
   & TPR & 0.934 (0.063) & 0.567 (0.138) & 0.750 (0.076) \\ 
   & FPR & 0.099 (0.000) & 0.099 (0.000) & 0.099 (0.000) \\ 
   & AUC1 & 0.979 (0.015) & 0.849 (0.047) & 0.914 (0.024) \\ 
   & AUC2 & 0.979 (0.015) & 0.852 (0.046) & 0.915 (0.024) \\   
  Bayesian edge regression &  &  &  &  \\ 
   & TPR & 0.982 (0.027) & 0.854 (0.087) & 0.918 (0.045) \\ 
   & FPR & 0.099 (0.000) & 0.099 (0.000) & 0.099 (0.000) \\ 
   & AUC1 & 0.994 (0.008) & 0.957 (0.028) & 0.976 (0.015) \\ 
   & AUC2 & 0.994 (0.008) & 0.957 (0.028) & 0.975 (0.015) \\ 
   \hline
\end{tabular}
 \begin{tablenotes}
 \scriptsize
\item[*] AUC1: $\max_{\lambda_2} AUC_{\lambda_1}$ for Fused/Group graphical lasso and LASICH; $\max_{h} AUC_{\lambda}$ for NFMGGM; $\max_{\kappa} AUC_{\alpha}$ for Bayesian edge regression.
\item[*] AUC2: $\max_{\lambda_1} AUC_{\lambda_2}$ for Fused/Group graphical lasso and LASICH; $\max_{\lambda} AUC_{h}$ for NFMGGM; $\max_{\alpha} AUC_{\kappa}$ for Bayesian edge regression.
\end{tablenotes}
\end{threeparttable}
\end{table}

\begin{table}[H]
\centering
\caption{\small Results of edge selection for \textit{Simulation 2} in terms of True Positive Rate (TPR), False Positive Rate (FPR), and bAUC. The reported TPR and FPR are calculated from the result using the selection rule of tuning parameters for each method. bAUC is the bivariate AUC that is computed through varying both two tuning parameters (regularization parameters) for each method. The numbers are averaged across 100 simulated sets and the standard deviations are given within the parentheses.}\label{TBres3}
\begin{threeparttable}
\begin{tabular}{lllll}
  \hline
Method & & Normal ($\mathbold{\Omega_N}$) & Tumor ($\mathbold{\Omega_T}$)  & Overall \\ 
  \hline
  Fused graphical lasso &&&&\\
&TPR & 0.904 (0.057) & 0.982 (0.026) & 0.943 (0.033) \\ 
 & FPR & 0.594 (0.104) & 0.566 (0.090) & 0.580 (0.094) \\ 
 & bAUC & 0.748 (0.040) & 0.758 (0.029) & 0.753 (0.020) \\ 
  Group graphical lasso &&&&\\
 & TPR & 0.892 (0.063) & 0.981 (0.025) & 0.937 (0.037) \\ 
&  FPR & 0.577 (0.116) & 0.547 (0.103) & 0.562 (0.107) \\ 
 & bAUC & 0.770 (0.040) & 0.813 (0.020) & 0.791 (0.020) \\ 
 LASICH &&&&\\
&TPR & 0.431 (0.129) & 0.898 (0.058) & 0.665 (0.083) \\ 
&  FPR & 0.045 (0.030) & 0.264 (0.047) & 0.154 (0.035) \\ 
&  bAUC & 0.751 (0.039) & 0.852 (0.019) & 0.801 (0.020) \\ 
 NFMGGM &&&&\\
&TPR & 0.956 (0.036) & 0.999 (0.006) & 0.977 (0.019) \\ 
&  FPR & 0.786 (0.124) & 0.835 (0.083) & 0.811 (0.100) \\ 
&  bAUC & 0.738 (0.035) & 0.799 (0.014) & 0.769 (0.019) \\ 
BIMGGM &&&&\\
&TPR & 0.164 (0.06) & 0.645 (0.096) & 0.404 (0.059) \\ 
 & FPR & 0.002 (0.004) & 0.081 (0.018) & 0.042 (0.009) \\ 
&  AUC & 0.809 (0.044) & 0.907 (0.032) & 0.858 (0.026) \\ 
Bayesian edge regression (group case) &&&&\\
&  TPR & 0.411 (0.067) & 0.865 (0.044) & 0.638 (0.040) \\ 
&  FPR & 0.051 (0.024) & 0.192 (0.035) & 0.122 (0.025) \\ 
 & bAUC & 0.773 (0.030) & 0.870 (0.023) & 0.822 (0.021) \\ 
  Bayesian edge regression &&&&\\
  &TPR & 0.489 (0.090) & 0.959 (0.032) & 0.724 (0.048) \\ 
  &FPR & 0.047 (0.022) & 0.269 (0.044) & 0.158 (0.022) \\ 
  &bAUC & 0.803 (0.035) & 0.913 (0.019) & 0.858 (0.019) \\ 
   \hline
\end{tabular}
 \begin{tablenotes}
 \footnotesize
 \item[*] A univariate AUC is reported for BIMGGM, while bivariate AUCs are reported for all the other methods.
\end{tablenotes}
\end{threeparttable}
\end{table}
\begin{table}[H]
\centering
\caption{\small Results of edge selection for \textit{Simulation 2} in terms of True Positive Rate (TPR), False Positive Rate (FPR), and univariate AUC. The reported TPR and FPR are calculated from the result that leads to an FPR that is closest to 0.1 for each method. The univariate AUC is reported by taking the maximum value from the AUC scores that are computed through varying one tuning parameter (or regularization parameter) while fixing the other with different values, and we report them with AUC1 and AUC2. The numbers are averaged across 100 simulated sets, and the standard deviations are given within the parentheses.}
\label{TBres32}
\begin{threeparttable}
\footnotesize
\begin{tabular}{lllll}
  \hline
Method & & Normal ($\mathbold{\Omega_N}$) & Tumor ($\mathbold{\Omega_T}$)  & Overall \\ 
  \hline
Fused graphical lasso &  &  &  &  \\ 
   & TPR & 0.532 (0.118) & 0.405 (0.080) & 0.469 (0.070) \\ 
   & FPR & 0.099 (0.002) & 0.098 (0.002) & 0.099 (0.001) \\ 
   & AUC1 & 0.838 (0.039) & 0.846 (0.019) & 0.842 (0.020) \\ 
   & AUC2 & 0.761 (0.033) & 0.778 (0.021) & 0.770 (0.018) \\ 
  Group graphical lasso &  &  &  &  \\ 
   & TPR & 0.517 (0.099) & 0.399 (0.081) & 0.458 (0.063) \\ 
   & FPR & 0.098 (0.001) & 0.098 (0.001) & 0.098 (0.001) \\ 
   & AUC1 & 0.824 (0.042) & 0.846 (0.019) & 0.835 (0.022) \\ 
   & AUC2 & 0.799 (0.039) & 0.822 (0.021) & 0.810 (0.021) \\ 
  LASICH &  &  &  &  \\ 
   & TPR & 0.582 (0.092) & 0.486 (0.081) & 0.534 (0.064) \\ 
   & FPR & 0.098 (0.000) & 0.098 (0.000) & 0.098 (0.000) \\ 
   & AUC1 & 0.845 (0.037) & 0.872 (0.018) & 0.859 (0.020) \\ 
   & AUC2 & 0.783 (0.034) & 0.836 (0.022) & 0.809 (0.019) \\ 
  NFMGGM &  &  &  &  \\ 
   & TPR & 0.575 (0.105) & 0.192 (0.060) & 0.383 (0.059) \\ 
   & FPR & 0.098 (0.000) & 0.098 (0.000) & 0.098 (0.000) \\ 
   & AUC1 & 0.863 (0.037) & 0.857 (0.014) & 0.860 (0.020) \\ 
   & AUC2 & 0.771 (0.034) & 0.806 (0.019) & 0.789 (0.020) \\ 
   BIMGGM &  &  &  &  \\ 
   & TPR & 0.564 (0.092) & 0.685 (0.129) & 0.625 (0.079) \\ 
   & FPR & 0.098 (0.000) & 0.098 (0.000) & 0.098 (0.000) \\ 
   & AUC & 0.809 (0.044) & 0.907 (0.032) & 0.858 (0.026) \\ 
  Bayesian edge regression (group case) &  &  &  &  \\ 
   & TPR & 0.534 (0.085) & 0.658 (0.113) & 0.596 (0.067) \\ 
   & FPR & 0.098 (0.000) & 0.098 (0.000) & 0.098 (0.000) \\ 
   & AUC1 & 0.806 (0.043) & 0.897 (0.031) & 0.852 (0.025) \\ 
   & AUC2 & 0.807 (0.043) & 0.899 (0.031) & 0.853 (0.025) \\ 
  Bayesian edge regression &  &  &  &  \\ 
   & TPR & 0.600 (0.086) & 0.839 (0.087) & 0.720 (0.060) \\ 
   & FPR & 0.098 (0.000) & 0.098 (0.000) & 0.098 (0.000) \\ 
   & AUC1 & 0.843 (0.036) & 0.948 (0.022) & 0.896 (0.020) \\ 
   & AUC2 & 0.845 (0.035) & 0.948 (0.022) & 0.896 (0.020) \\ 
   \hline\end{tabular}
    \begin{tablenotes}
 \scriptsize
\item[*] AUC1: $\max_{\lambda_2} AUC_{\lambda_1}$ for Fused/Group graphical lasso and LASICH; $\max_{h} AUC_{\lambda}$ for NFMGGM; $\max_{\kappa} AUC_{\alpha}$ for Bayesian edge regression.
\item[*] AUC2: $\max_{\lambda_1} AUC_{\lambda_2}$ for Fused/Group graphical lasso and LASICH; $\max_{\lambda} AUC_{h}$ for NFMGGM; $\max_{\alpha} AUC_{\kappa}$ for Bayesian edge regression.
\end{tablenotes}
\end{threeparttable}
\end{table}
\begin{table}[H]
\centering
\caption{\small In our simulation, we examine the estimated subject-level precision matrices over a densely-partitioned grid of $\pi$ through equal partitioning [0,1] over 101 grid points (e.g., $0, 0.01, \cdots, 1$) and performing the tests of positive definiteness (using the R package \textit{matrixcalc}), so for each simulation setting, we test $10100$ estimated matrices. We find that a vast majority are positive definite, especially in Simulation 1, more than 99\% of generated subject-level matrices are tested positive definite. We report the results respectively in terms of before posterior thresholding and after posterior thresholding.}
\label{TBrespd}
\begin{tabular}{llll}
  \hline
 &Simulation 1 & Simulation 2  & Overall \\ 
   \hline
Before posterior thresholding & 99.73\%& 92.15\%& 95.94\%\\
After posterior thresholding ($\alpha = 0.1, \kappa = 0.1$) & 99.84\% & 89.01\% & 94.43\%\\
  \hline
\end{tabular}
\end{table}
\newpage

In our supplementary, we further include a group of simulations containing $100$ variables with a comparison to the group/fussed Lasso method, LASICH, and NFMGGM in the same simulation setup. We include two more simulations, respectively corresponding to our simulations 1 and 2, each over 10 datasets with $p = 100$ instead of $p = 20$ and report the model performance in Supplementary Figure \ref{figroc2}, Supplementary Tables \ref{TBres2_100}, \ref{TBres22_100}, \ref{TBres3_100} and \ref{TBres32_100}. Our method is still reported with the highest bAUC (0.950 versus 0.848, 0.902, 0.920 and 0.887) and the highest TPR when FPR$\approx 0.1$ (0.966 versus 0.784, 0.848, 0.830, 0.744) for the tumor graphs in Simulation 1. Similarly, in Simulation 2 our method still outperforms most of the compared approaches for the estimation of tumor graphs (bAUC: 0.944 versus 0.881, 0.919, 0.961 and 0.886; TPR when FPR$\approx 0.1$: 0.947 versus 0.842, 0.820, 0.921, and 0.518).  Note that a large increase of $p$ generates overly sparse graphs (with around 4\% of possible edges included for $p = 100$), which saturates the model performance across all the methods, especially in Simulation 2.

\begin{figure}[H] 
\centering
\includegraphics[width=0.3\textwidth]{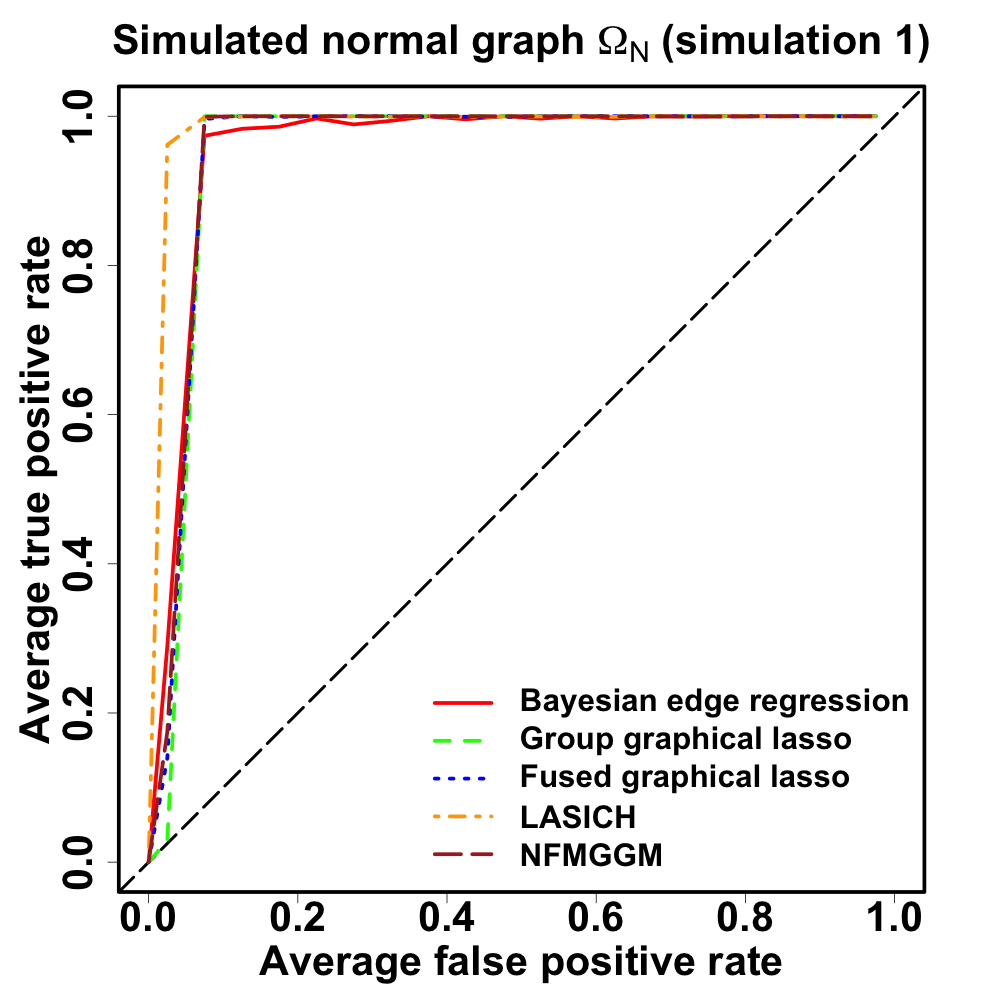}
\includegraphics[width=0.3\textwidth]{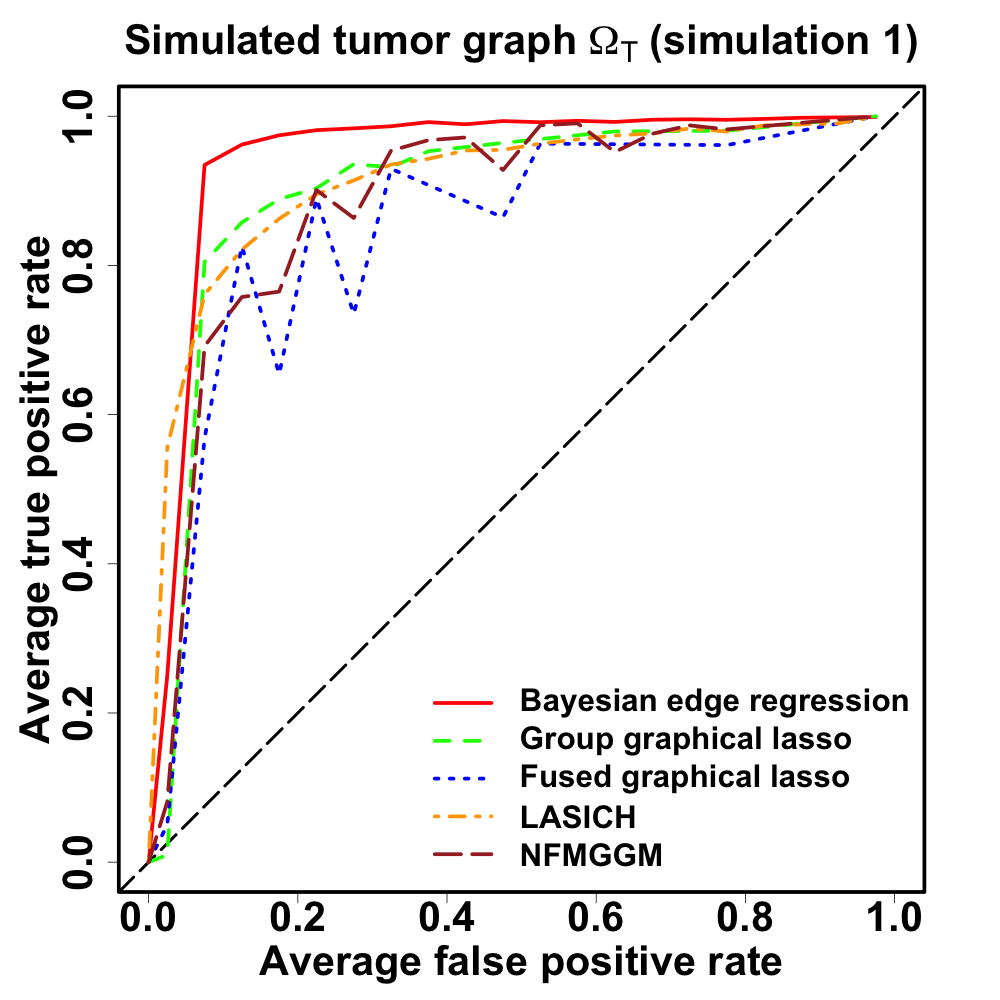}\\
\includegraphics[width=0.3\textwidth]{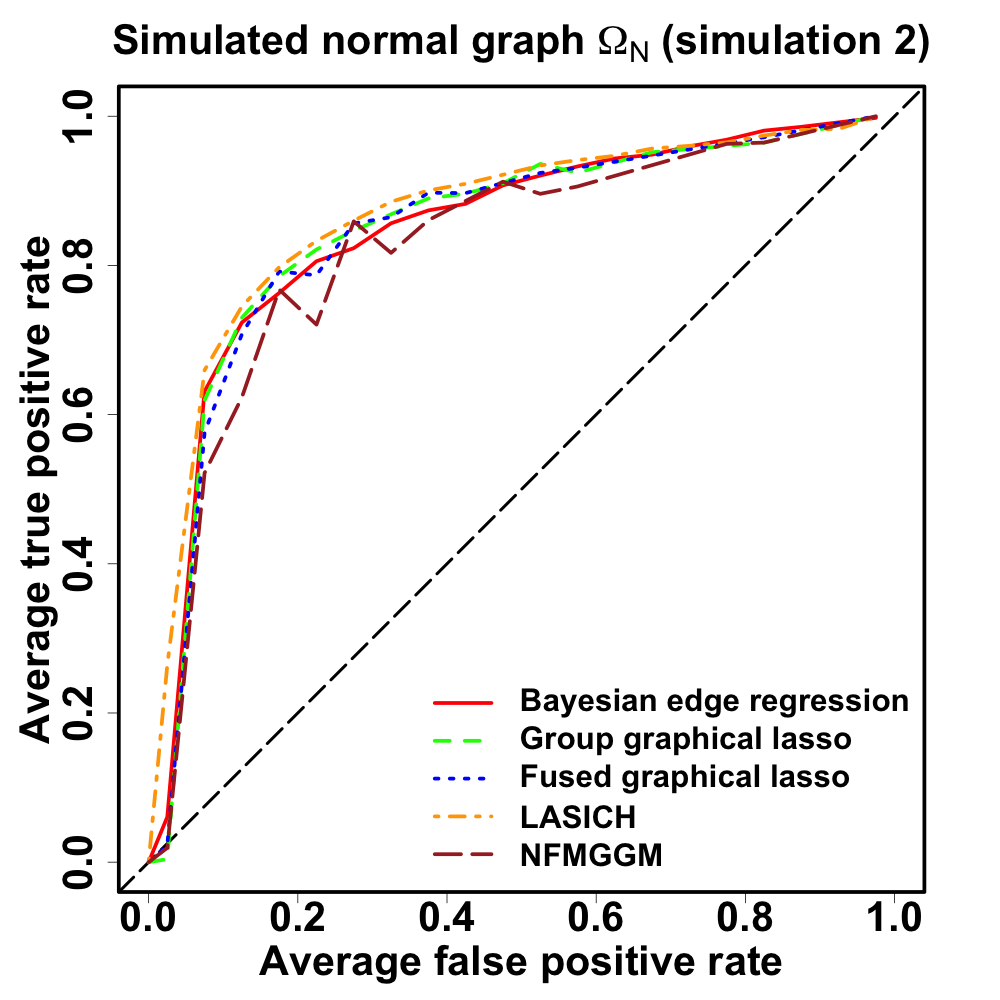}
\includegraphics[width=0.3\textwidth]{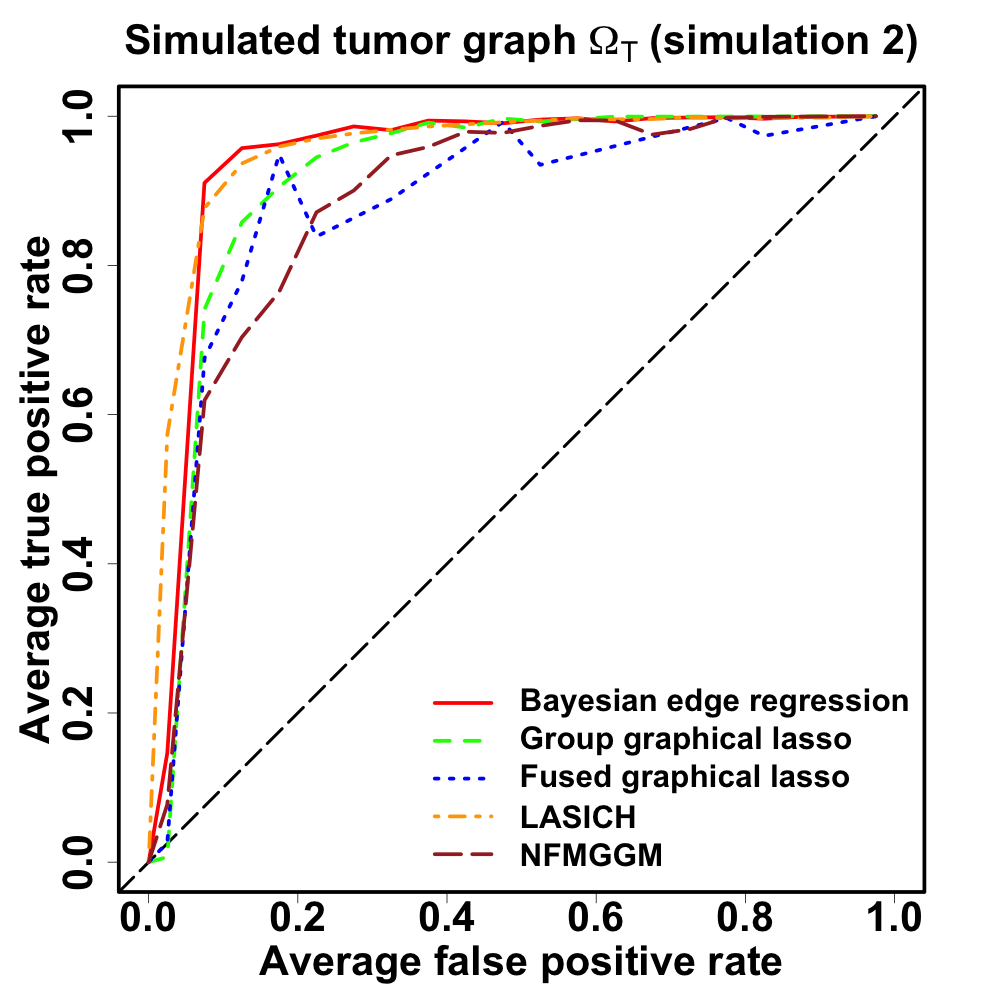}
\caption{\small ROC curves for the structure learning of simulated normal (w.r.t $\mathbold{\Omega_N}$) and tumor graphs (w.r.t $\mathbold{\Omega_T}$) when $p=100$ with the simulation setup in \textit{Simulation 1} and  \textit{Simulation 2}.}
\label{figroc2}
\end{figure}
\begin{table}[H]
\centering
\caption{\small Results of edge selection when $p=100$ with the simulation setup for \textit{Simulation 1} in terms of True Positive Rate (TPR), False Positive Rate (FPR), and bAUC. The reported TPR and FPR are calculated from the result using the selection rule of tuning parameters for each method. bAUC is the bivariate AUC that is computed through varying both of the two tuning parameters (regularization parameters) for each method. The numbers are averaged across ten simulated sets, and the standard deviations are given within the parentheses.}
\label{TBres2_100}
\begin{threeparttable}
\begin{tabular}{lllll}
  \hline
Method & & Normal ($\mathbold{\Omega_N}$)  & Tumor ($\mathbold{\Omega_T}$) & Overall\\ 
  \hline
Fused graphical lasso &&&&\\
& TPR & 1.000 (0.000) & 0.821 (0.028) & 0.911 (0.014) \\ 
& FPR & 0.144 (0.027) & 0.121 (0.026) & 0.133 (0.026) \\ 
& bAUC & 0.957 (0.001) & 0.848 (0.019) & 0.902 (0.009) \\ 
Group graphical lasso &&&&\\
& TPR & 1.000 (0.000) & 0.822 (0.029) & 0.911 (0.014) \\ 
& FPR & 0.142 (0.026) & 0.120 (0.025) & 0.131 (0.026) \\ 
& bAUC & 0.951 (0.000) & 0.902 (0.017) & 0.927 (0.008) \\ 
LASICH &&&&\\
& TPR & 1.000 (0.000) & 0.745 (0.034) & 0.872 (0.017) \\ 
& FPR & 0.011 (0.003) & 0.048 (0.005) & 0.030 (0.003) \\ 
& bAUC & 0.998 (0.001) & 0.920 (0.014) & 0.959 (0.007) \\ 
NFMGGM &&&&\\
& TPR & 1.000 (0.000) & 0.994 (0.010) & 0.997 (0.005) \\ 
& FPR & 0.211 (0.025) & 0.262 (0.023) & 0.236 (0.023) \\ 
& bAUC & 0.959 (0.000) & 0.887 (0.007) & 0.923 (0.003) \\ 
Bayesian edge regression &&&&\\
& TPR & 0.998 (0.004) & 0.884 (0.024) & 0.941 (0.012) \\ 
& FPR & 0.020 (0.002) & 0.020 (0.003) & 0.020 (0.002) \\ 
& bAUC & 0.960 (0.001) & 0.950 (0.009) & 0.955 (0.005) \\ 
\hline
\end{tabular}
\end{threeparttable}
\end{table}
 \begin{table}[H]
\centering
\caption{\small Results of edge selection when $p=100$ with the simulation setup for \textit{Simulation 1} in terms of True Positive Rate (TPR), False Positive Rate (FPR), and univariate AUC. The reported TPR and FPR are calculated from the result that leads to an FPR that is closest to 0.1 for each method. The univariate AUC is reported by taking the maximum value from the AUC scores that are computed through varying one tuning parameter (or regularization parameter) while fixing the other with different values, and we report them with AUC1 and AUC2. The numbers are averaged across ten simulated sets and the standard deviations are given within the parentheses.}
\label{TBres22_100}
\begin{threeparttable}
\begin{tabular}{lllll}
  \hline
Method & & Normal ($\mathbold{\Omega_N}$)  & Tumor ($\mathbold{\Omega_T}$) & Overall\\ 
  \hline
Fused graphical lasso &  &  &  &  \\ 
   & TPR & 1.000 (0.000) & 0.784 (0.050) & 0.892 (0.025) \\ 
   & FPR & 0.100 (0.004) & 0.099 (0.003) & 0.099 (0.002) \\ 
   & AUC1 & 1.000 (0.000) & 0.924 (0.017) & 0.962 (0.009) \\ 
   & AUC2 & 0.996 (0.001) & 0.857 (0.016) & 0.926 (0.008) \\ 
  Group graphical lasso &  &  &  &  \\ 
   & TPR & 1.000 (0.000) & 0.848 (0.037) & 0.924 (0.018) \\ 
   & FPR & 0.099 (0.003) & 0.098 (0.004) & 0.098 (0.002) \\ 
   & AUC1 & 1.000 (0.000) & 0.934 (0.017) & 0.967 (0.009) \\ 
   & AUC2 & 0.999 (0.001) & 0.934 (0.016) & 0.966 (0.008) \\ 
  LASICH &  &  &  &  \\ 
   & TPR & 1.000 (0.000) & 0.830 (0.020) & 0.915 (0.010) \\ 
   & FPR & 0.100 (0.001) & 0.100 (0.000) & 0.100 (0.001) \\ 
   & AUC1 & 1.000 (0.000) & 0.932 (0.018) & 0.966 (0.009) \\ 
   & AUC2 & 0.997 (0.001) & 0.890 (0.013) & 0.944 (0.006) \\ 
  NFMGGM &  &  &  &  \\ 
   & TPR & 1.000 (0.000) & 0.744 (0.312) & 0.872 (0.156) \\ 
   & FPR & 0.100 (0.000) & 0.100 (0.000) & 0.100 (0.000) \\ 
   & AUC1 & 0.999 (0.000) & 0.989 (0.006) & 0.994 (0.003) \\ 
   & AUC2 & 0.997 (0.001) & 0.916 (0.014) & 0.957 (0.007) \\ 
  Bayesian edge regression &  &  &  &  \\ 
   & TPR & 0.965 (0.112) & 0.966 (0.015) & 0.965 (0.052) \\ 
   & FPR & 0.100 (0.001) & 0.100 (0.001) & 0.100 (0.001) \\ 
   & AUC1 & 1.000 (0.000) & 0.986 (0.008) & 0.993 (0.004) \\ 
   & AUC2 & 1.000 (0.000) & 0.987 (0.006) & 0.993 (0.003) \\ 
   \hline
\end{tabular}
 \begin{tablenotes}
 \scriptsize
\item[*] AUC1: $\max_{\lambda_2} AUC_{\lambda_1}$ for Fused/Group graphical lasso and LASICH; $\max_{h} AUC_{\lambda}$ for NFMGGM; $\max_{\kappa} AUC_{\alpha}$ for Bayesian edge regression.
\item[*] AUC2: $\max_{\lambda_1} AUC_{\lambda_2}$ for Fused/Group graphical lasso and LASICH; $\max_{\lambda} AUC_{h}$ for NFMGGM; $\max_{\alpha} AUC_{\kappa}$ for Bayesian edge regression.
\end{tablenotes}
\end{threeparttable}
\end{table}
\begin{table}[H]
\centering
\caption{\small Results of edge selection when $p=100$ with the simulation setup for \textit{Simulation 2} in terms of True Positive Rate (TPR), False Positive Rate (FPR), and bAUC. The reported TPR and FPR are calculated from the result using the selection rule of tuning parameters for each method. bAUC is the bivariate AUC that is computed through varying both two tuning parameters (regularization parameters) for each method. The numbers are averaged across ten simulated sets and the standard deviations are given within the parentheses.}\label{TBres3_100}
\begin{threeparttable}
\begin{tabular}{lllll}
  \hline
Method & & Normal ($\mathbold{\Omega_N}$) & Tumor ($\mathbold{\Omega_T}$)  & Overall \\ 
  \hline
  Fused graphical lasso &&&&\\
& TPR & 0.810 (0.030) & 0.931 (0.023) & 0.871 (0.022) \\ 
 & FPR & 0.202 (0.005) & 0.150 (0.004) & 0.176 (0.003) \\ 
 & bAUC & 0.844 (0.015) & 0.881 (0.012) & 0.862 (0.012) \\ 
  Group graphical lasso &&&&\\
 & TPR & 0.804 (0.028) & 0.928 (0.018) & 0.866 (0.019) \\ 
 & FPR & 0.201 (0.006) & 0.147 (0.004) & 0.174 (0.003) \\ 
 & bAUC & 0.848 (0.015) & 0.919 (0.004) & 0.884 (0.009) \\ 
 LASICH &&&&\\
& TPR & 0.421 (0.045) & 0.866 (0.021) & 0.644 (0.031) \\ 
 & FPR & 0.012 (0.003) & 0.048 (0.004) & 0.030 (0.003) \\ 
& bAUC & 0.871 (0.016) & 0.961 (0.005) & 0.916 (0.009) \\ 
 NFMGGM &&&&\\
 & TPR & 0.792 (0.027) & 0.996 (0.004) & 0.894 (0.015) \\ 
 & FPR & 0.165 (0.007) & 0.362 (0.015) & 0.263 (0.007) \\ 
 & bAUC & 0.823 (0.014) & 0.886 (0.004) & 0.855 (0.008) \\ 
  Bayesian edge regression &&&&\\
 & TPR & 0.449 (0.027) & 0.895 (0.028) & 0.672 (0.017) \\ 
& FPR & 0.017 (0.002) & 0.048 (0.003) & 0.032 (0.002) \\ 
 & bAUC & 0.847 (0.019) & 0.944 (0.005) & 0.895 (0.010) \\ 
   \hline
\end{tabular}
\end{threeparttable}
\end{table}
\begin{table}[H]
\centering
\caption{\small Results of edge selection when $p=100$ with the simulation setup for \textit{Simulation 2} in terms of True Positive Rate (TPR), False Positive Rate (FPR), and univariate AUC. The reported TPR and FPR are calculated from the result that leads to an FPR that is closest to 0.1 for each method. The univariate AUC is reported by taking the maximum value from the AUC scores that are computed through varying one tuning parameter (or regularization parameter) while fixing the other with different values, and we report them with AUC1 and AUC2. The numbers are averaged across ten simulated sets, and the standard deviations are given within the parentheses.}
\label{TBres32_100}
\begin{threeparttable}
\begin{tabular}{lllll}
  \hline
Method & & Normal ($\mathbold{\Omega_N}$) & Tumor ($\mathbold{\Omega_T}$)  & Overall \\ 
  \hline
Fused graphical lasso &  &  &  &  \\ 
   & TPR & 0.697 (0.062) & 0.842 (0.056) & 0.769 (0.056) \\ 
   & FPR & 0.100 (0.002) & 0.099 (0.004) & 0.100 (0.003) \\ 
   & AUC1 & 0.886 (0.018) & 0.949 (0.004) & 0.918 (0.010) \\ 
   & AUC2 & 0.824 (0.016) & 0.887 (0.010) & 0.856 (0.010) \\ 
  Group graphical lasso &  &  &  &  \\ 
   & TPR & 0.706 (0.030) & 0.820 (0.036) & 0.763 (0.028) \\ 
   & FPR & 0.099 (0.002) & 0.099 (0.001) & 0.099 (0.002) \\ 
   & AUC1 & 0.880 (0.017) & 0.950 (0.004) & 0.915 (0.009) \\ 
   & AUC2 & 0.865 (0.016) & 0.941 (0.005) & 0.903 (0.009) \\ 
  LASICH &  &  &  &  \\ 
   & TPR & 0.722 (0.026) & 0.921 (0.027) & 0.822 (0.018) \\ 
   & FPR & 0.100 (0.001) & 0.100 (0.000) & 0.100 (0.000) \\ 
   & AUC1 & 0.893 (0.017) & 0.968 (0.003) & 0.931 (0.009) \\ 
   & AUC2 & 0.829 (0.019) & 0.938 (0.008) & 0.883 (0.011) \\ 
  NFMGGM &  &  &  &  \\ 
   & TPR & 0.564 (0.073) & 0.518 (0.244) & 0.541 (0.142) \\ 
   & FPR & 0.100 (0.000) & 0.100 (0.000) & 0.100 (0.000) \\ 
   & AUC1 & 0.886 (0.014) & 0.948 (0.004) & 0.917 (0.007) \\ 
   & AUC2 & 0.796 (0.014) & 0.879 (0.013) & 0.838 (0.008) \\ 
  Bayesian edge regression &  &  &  &  \\ 
   & TPR & 0.690 (0.046) & 0.947 (0.020) & 0.818 (0.028) \\ 
   & FPR & 0.100 (0.001) & 0.100 (0.000) & 0.100 (0.000) \\ 
   & AUC1 & 0.873 (0.020) & 0.979 (0.005) & 0.926 (0.010) \\ 
   & AUC2 & 0.873 (0.020) & 0.979 (0.005) & 0.926 (0.010) \\ 
   \hline
   \end{tabular}
    \begin{tablenotes}
 \scriptsize
\item[*] AUC1: $\max_{\lambda_2} AUC_{\lambda_1}$ for Fused/Group graphical lasso and LASICH; $\max_{h} AUC_{\lambda}$ for NFMGGM; $\max_{\kappa} AUC_{\alpha}$ for Bayesian edge regression.
\item[*] AUC2: $\max_{\lambda_1} AUC_{\lambda_2}$ for Fused/Group graphical lasso and LASICH; $\max_{\lambda} AUC_{h}$ for NFMGGM; $\max_{\alpha} AUC_{\kappa}$ for Bayesian edge regression.
\end{tablenotes}
\end{threeparttable}
\end{table}


\newpage 
\section*{E: HCC Data Result Analysis} \label{hccsup}
In our simulations, we chose $\kappa=0.1$, which has been suggested as a threshold for conditional dependency in previous works \citep{peterson2013inferring,bashir2019post}. In a real data analysis, the choice of $\kappa$ can be quite subjective, because it determines the graph sparsity and the associations we will further investigate. The final choice of $\kappa$ should be contingent on the empirical background and the expectation of graph sparsity. In our case study, we chose $\kappa=0.15$ because, compared with other thresholds, it obtains graphs with a moderate sparsity, and presented the results for $\kappa=0.1$ and $\kappa=0.2$ in Supplementary Figures 4, 5, 8 and 9.
 
 In addition to our discussion about interesting connections in Section 4 of the main text, we include more differentially connected edges between the tumor and normal graphs with their linearly varying effects in Supplementary Figure \ref{figedge2}. We present edges with: positive connections and negative connections that exist in the tumor graph, but not the normal graph; positive connection in the normal but not tumor graph; and positive connections appearing on both the tumor and normal graphs, but with significant difference of edge strength. In addition to \textit{6Ckine}/\textit{MIP-3, $\beta$} we discuss in the main text, \textit{IL-22}/\textit{TWEAK} and \textit{IP-10}/\textit{MIG} are another two pairs of transcription factors that are not connected for samples with low \textit{HepatoScore}, but connected with the increase of \textit{HepatoScore}.\textit{IP-10} and \textit{MIG} suppress tumorigenesis in human ovarian cancer through recruiting tumor-infiltrating lymphocytes \citep{bref6}. As we mention \textit{CA-15-3} in the main text,  \textit{CA-15-3}/\textit{MMP-10} is another pair that shows negative correlation for $\pi = 0$ and positive correlation for $\pi = 1$, where \textit{MMP-10} participates in tissue repair and promotes a pro-tumorigenic microenvironment \citep{bref3}. \textit{IGFBP-2} has been considered as a tumor marker for HCC patients \citep{sbref1}. In our study, \textit{HCC-4}/\textit{IGFBP-2} begins to have positive connection after $\pi > 0.3$. We find \textit{DKK-1} is positively correlated with \textit{HER-2} and with \textit{IGBP1} with high \textit{HepatoScore}, which respectively correspond to a similar result in breast cancer and human endometrium \citep{sbref3,sbref4}. \textit{MMP-1} is similarly observed to positively correlate with \textit{VEGF} ($\pi > 0.8$), which has been shown to play a critical role in the cancer and immune environment \citep{sbref5}. \textit{IGFBP5} and \textit{IGFBP6}, which have been reported to be two critical tumor suppressors, are positively correlated from $\pi = 0.4$ \citep{sbref7,sbref8}. \textit{MDC} (also known as \textit{CCL22}) and \textit{TARC} (known as \textit{CCL17}) have been reported to be closely related, and both have significantly higher levels in tumor than normal tissue for gastric cancer, which agrees with our observation \citep{sbref9,sbref10}. The strong connection of \textit{ITAC}/\textit{MIG} also reflects previously observed associations between \textit{ITAC} and human monokine \citep{sbref11}. \textit{IL-8} has been shown to be a stimulator for \textit{MPO} in human neutrophil, and in our results these two genes show strong positive correlation from $\pi > 0.2$ \citep{sbref12}. The connection \textit{IP-10}/\textit{MPIF-1} begins to appear with negative correlation from the medium \textit{HepatoScore}. \textit{IGFBP-3}/\textit{MMP-3} has the same feature, where \textit{MMP-3} has been reported to cleave \textit{IGFBP-3} \citep{sbref13}. \textit{IL-8}/\textit{MIP-1, $\beta$} and \textit{IL-22}/\textit{TARC} are two connections that appear on both tumor and normal graphs, but have been drastically weakened from $\pi = 0$ to $\pi = 1$, which corresponds to some previous work on inflammatory effects present \textit{IL-22}'s regulation on \textit{TARC} and constitutive production of \textit{IL-8} and \textit{MIP-1, $\beta$} are simultaneously reduced in neonatal neutrophils \citep{sbref14,sbref15}. Additionally, we provide the information about all the connected edges corresponding to Main Figure 3, Supplementary Figures \ref{fighccm1} and \ref{fighccm2} in the Supplementary File.

 For the purpose of comparison, we also ran the algorithm of \textit{NFMGGM}, which similarly constructs a non-static graph but dependent on just univariate covariate, through modeling \textit{HepatoScore} $\pi$ as the univariate covariate. We presented the estimated graphs for the same choices of \textit{HepatoScore} (0, 0.5, 1) in Supplementary Figure \ref{fighccm3} based on the model selection with the aforementioned two-step procedure, and the number of connections within each pathway and between each pair of pathways estimated from \textit{NFMGGM} as a function of \textit{HepatoScore} $\pi$ in Supplementary Figure \ref{figheat3}. These results show that the \textit{NFMGGM} yields similar varying behavior of edge connectedness across different $\pi$, but the generated graph has a very high density, where the proportion of connected edges exceeds $50\%$ in the high \textit{HepatoScore} groups, as reflected with the reported high FPR in the simulations.
  
 \begin{figure}[H] 
\centering
\includegraphics[width=0.75\textwidth]{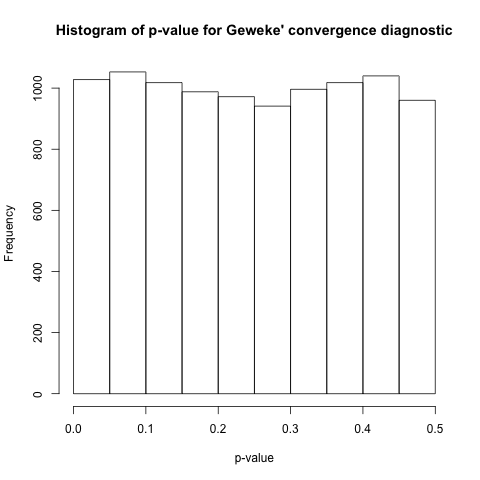}
\caption{\small Histogram of $p-$values under Geweke convergence diagnostic for all the parameters we sample from the MCMC chain.}
\label{fighist}
\end{figure}

 \begin{figure}[H]
 \begin{subfigure}[H]{0.475\textwidth}
\includegraphics[width=1\textwidth]{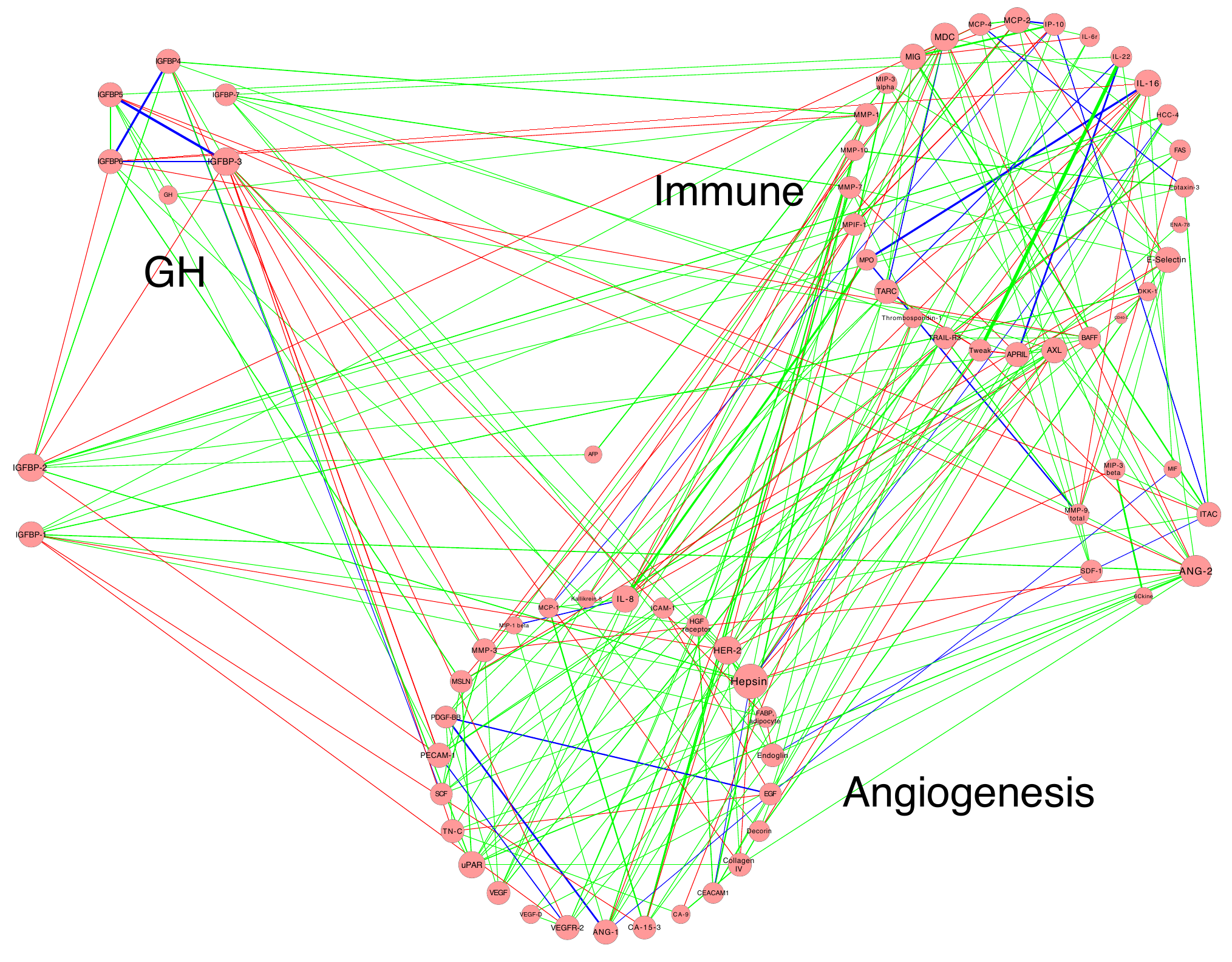}
\caption{\small High HepatoScore}
\end{subfigure}
 \begin{subfigure}[H]{0.475\textwidth}
\includegraphics[width=1\textwidth]{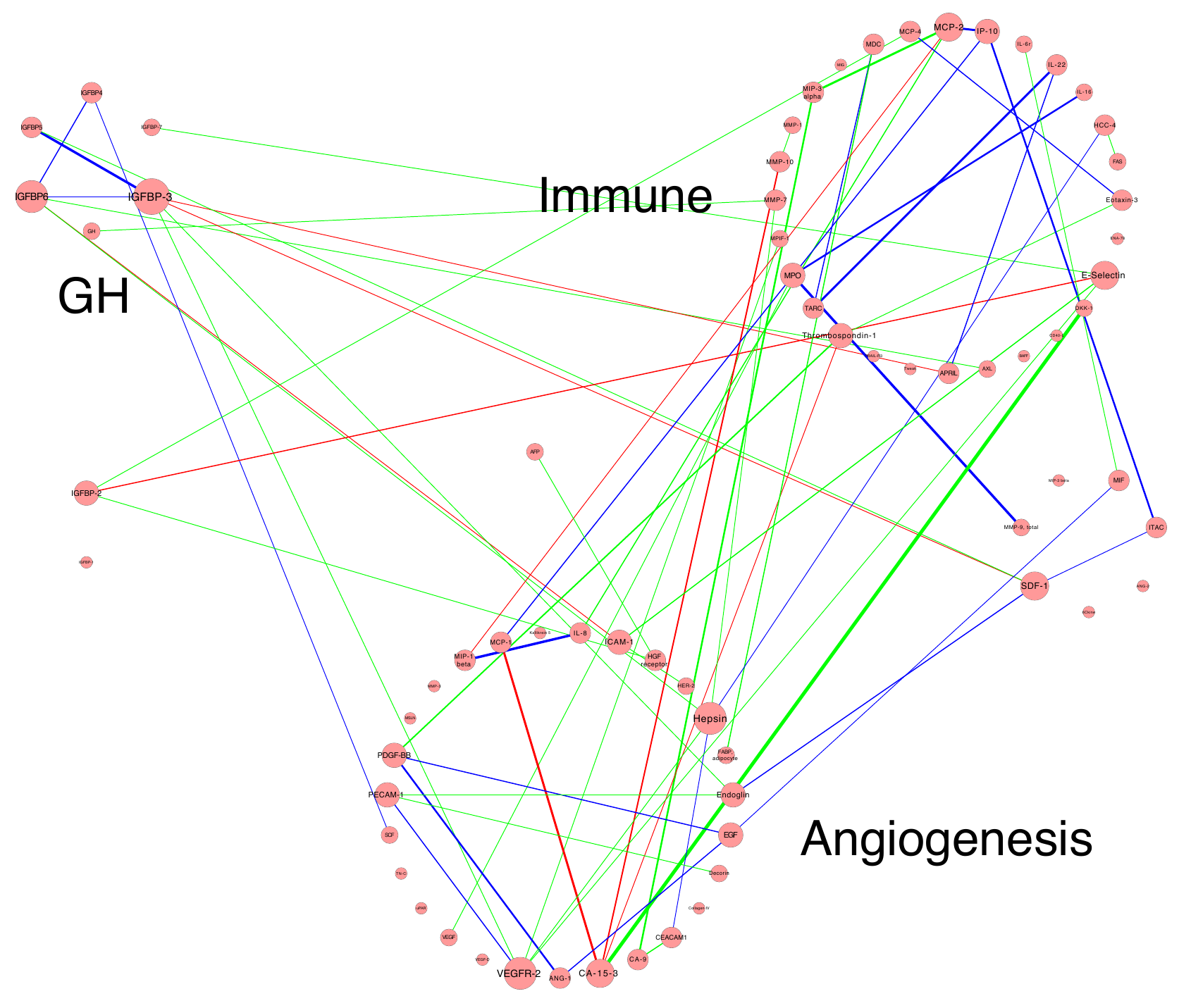}
\caption{\small Low HepatoScore}
\end{subfigure}%
\\
 \begin{subfigure}[H]{0.475\textwidth}
\includegraphics[width=1\textwidth]{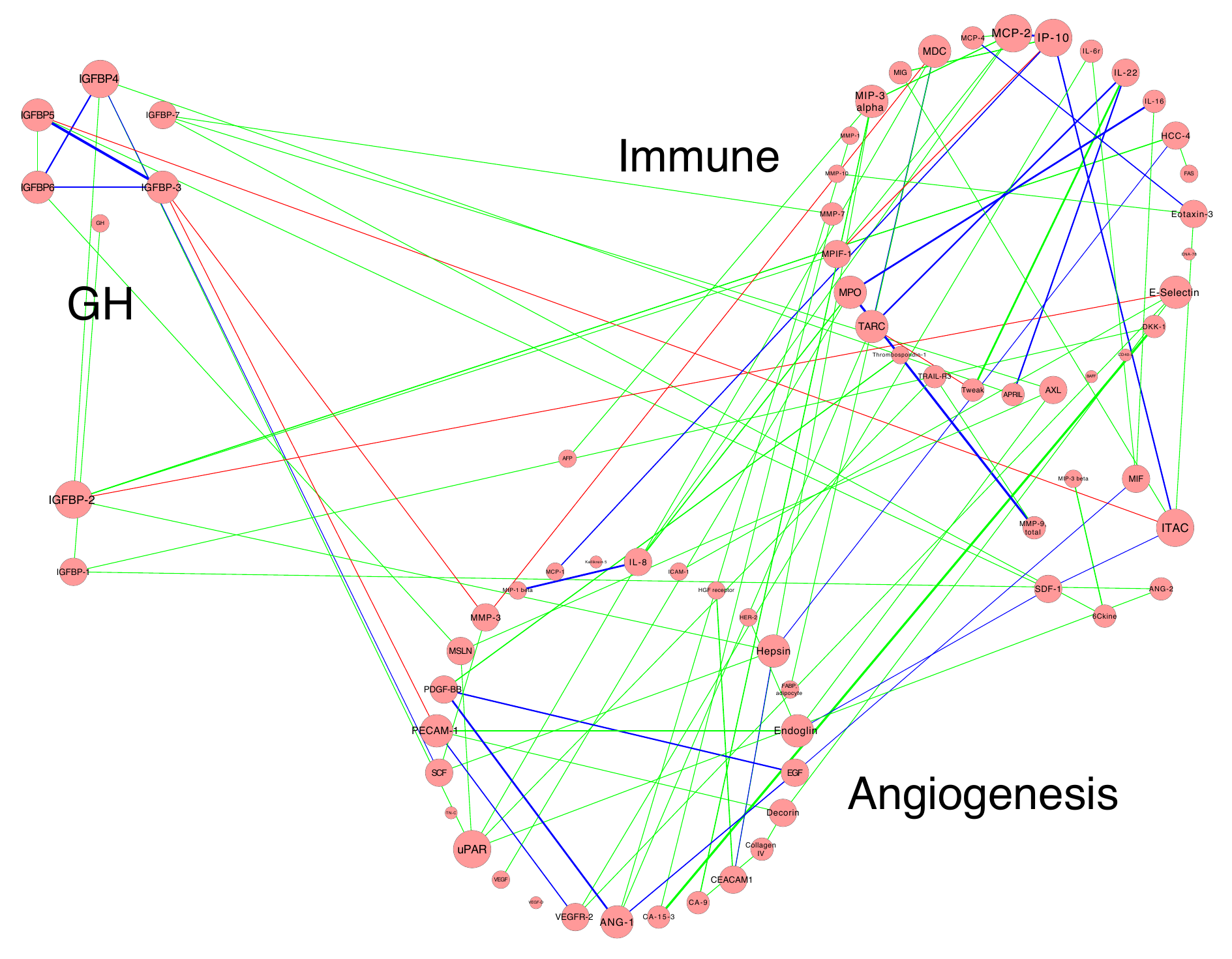}
\caption{\small Medium HepatoScore}
\end{subfigure}%
 \begin{subfigure}[H]{0.475\textwidth}
\includegraphics[width=1\textwidth]{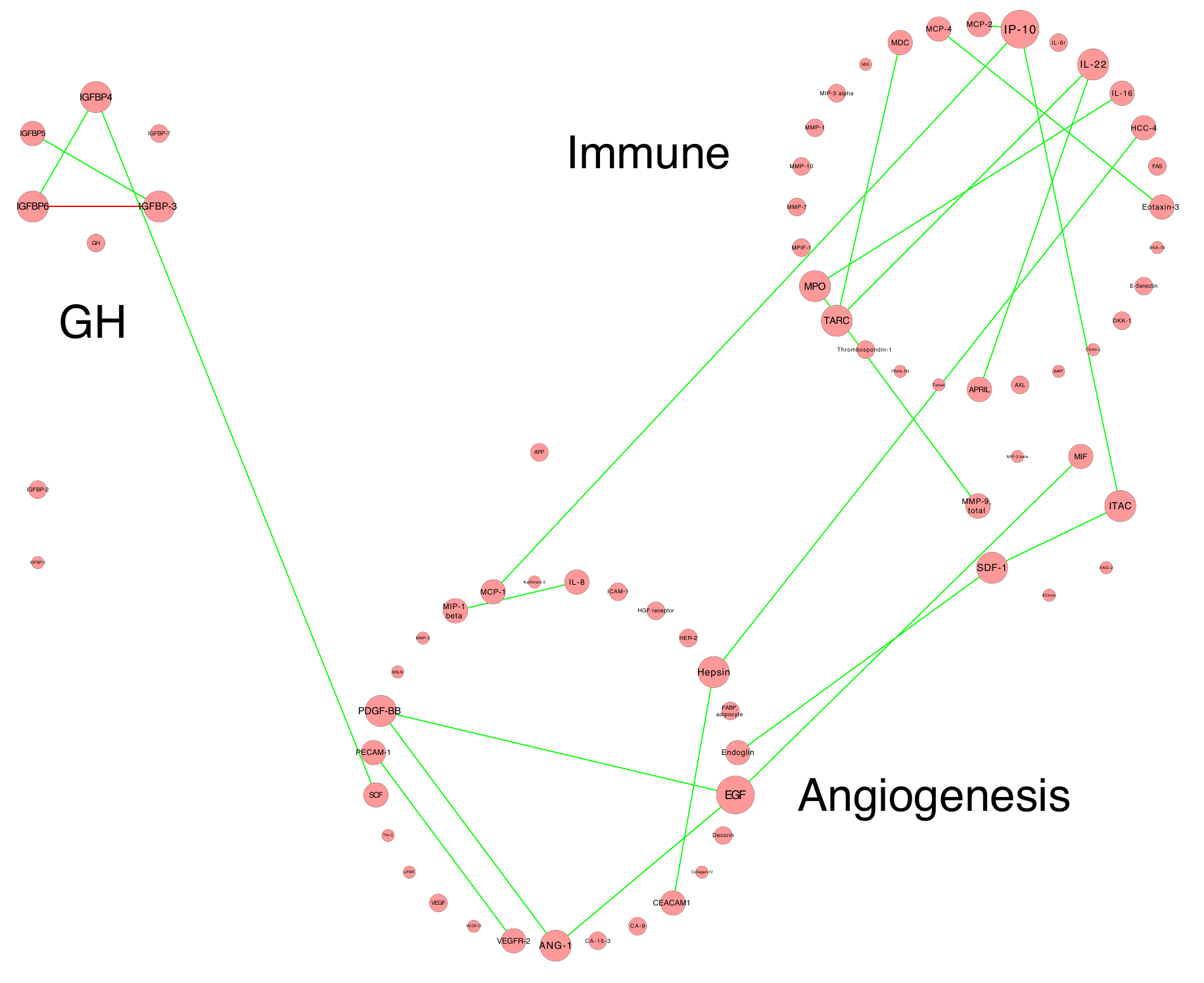}
\caption{\small Shared edges}
\end{subfigure}%
\caption{\small Estimated graphs ($\alpha=0.10, \kappa=0.1$) from the Bayesian edge regression for the GH, immune, and angiogenesis pathways with (a)  $\pi = 0$; (b) $\pi = 0.5$; (c) $\pi = 1$; (d) Common edges. Colors indicate positive (green), negative (red), and common (blue) edges. The thickness of edge is proportional to $\hat \rho^{ij}$ for the edge $(i, j)$, and the size of node is proportional to its degree.}
\label{fighccm1}
\end{figure}
 \begin{figure}[H]
 \begin{subfigure}[H]{0.475\textwidth}
\includegraphics[width=1\textwidth]{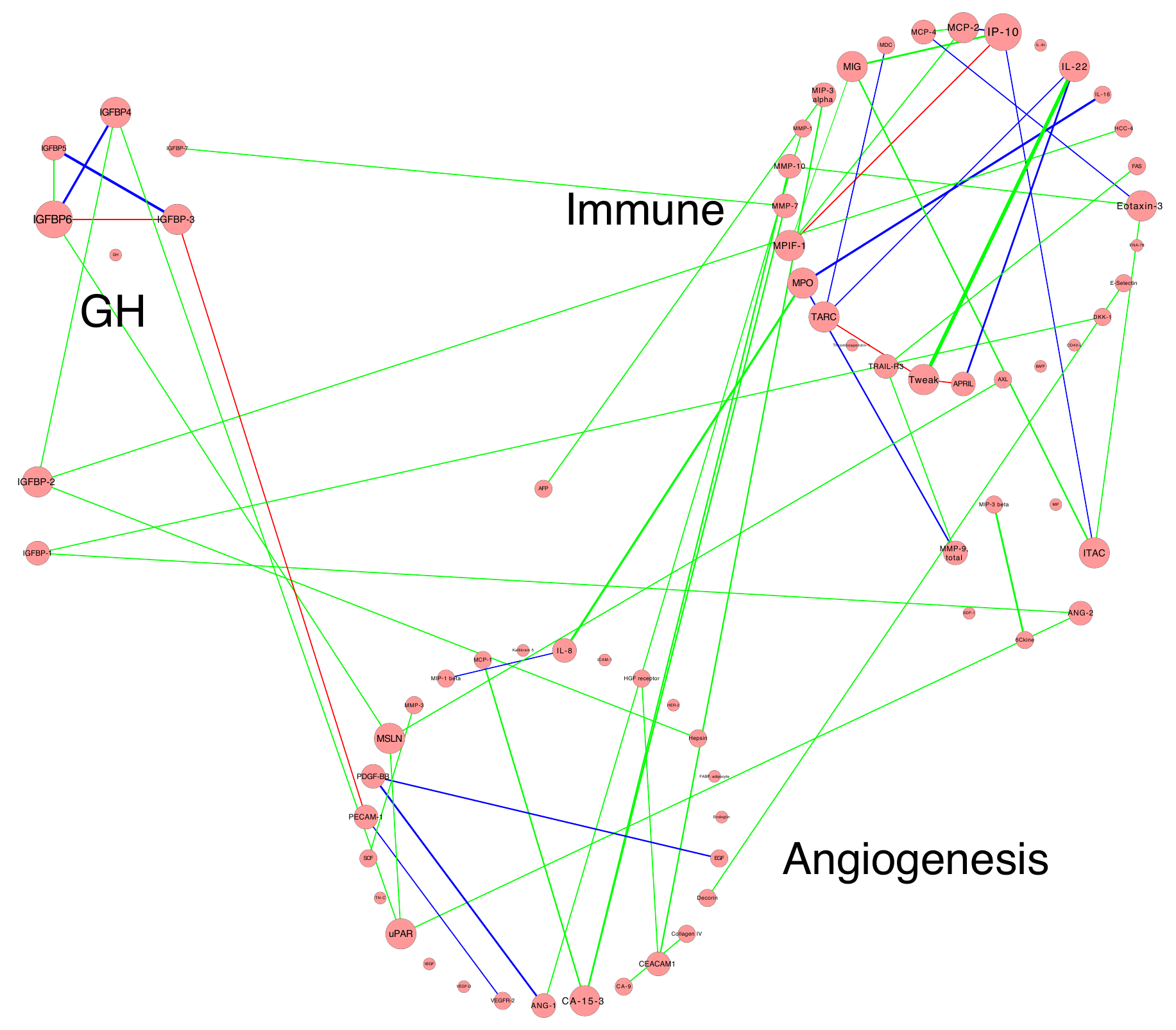}
\caption{\small High HepatoScore}
\end{subfigure}
 \begin{subfigure}[H]{0.475\textwidth}
\includegraphics[width=1\textwidth]{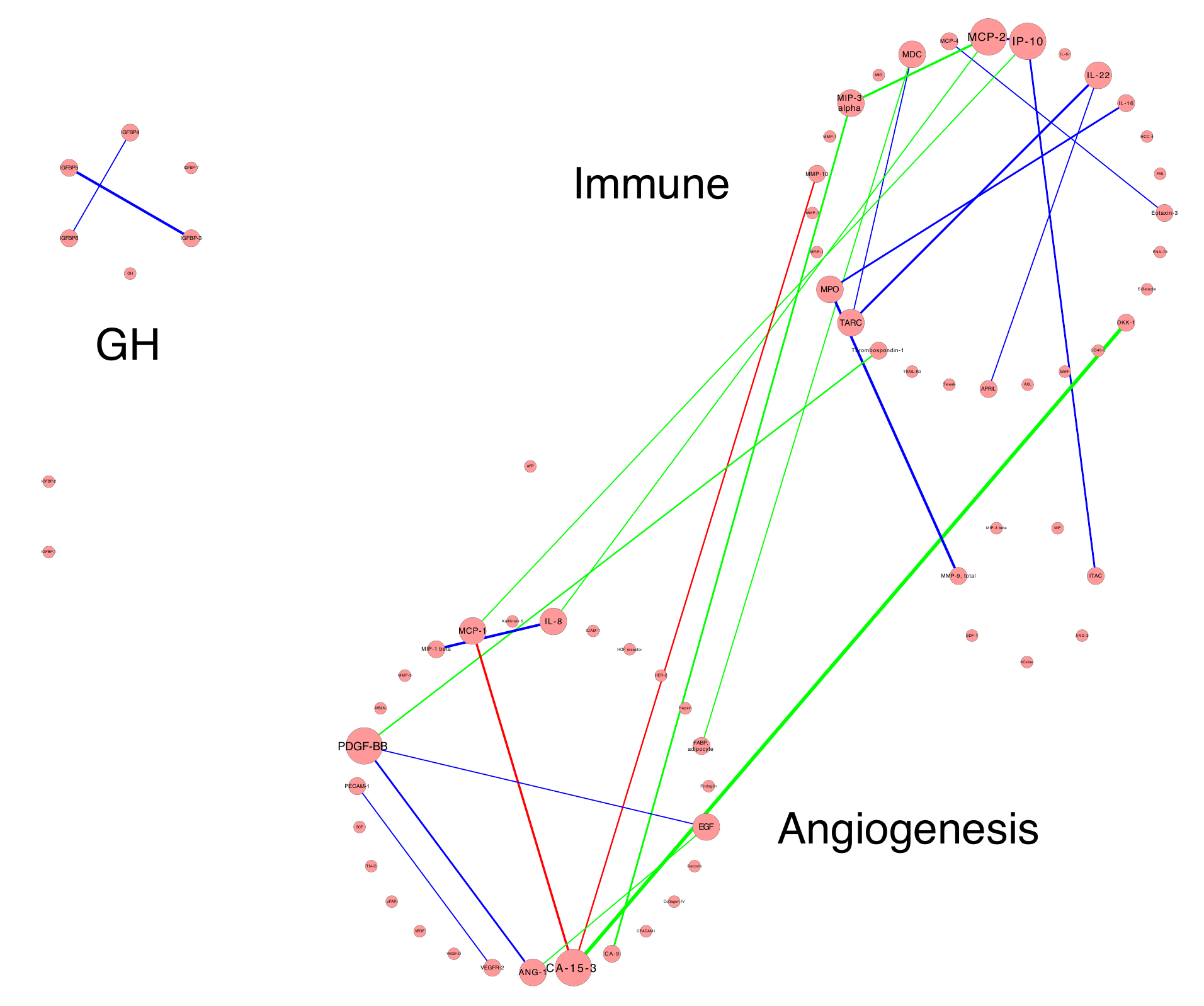}
\caption{\small Low HepatoScore}
\end{subfigure}%
\\
 \begin{subfigure}[H]{0.475\textwidth}
\includegraphics[width=1\textwidth]{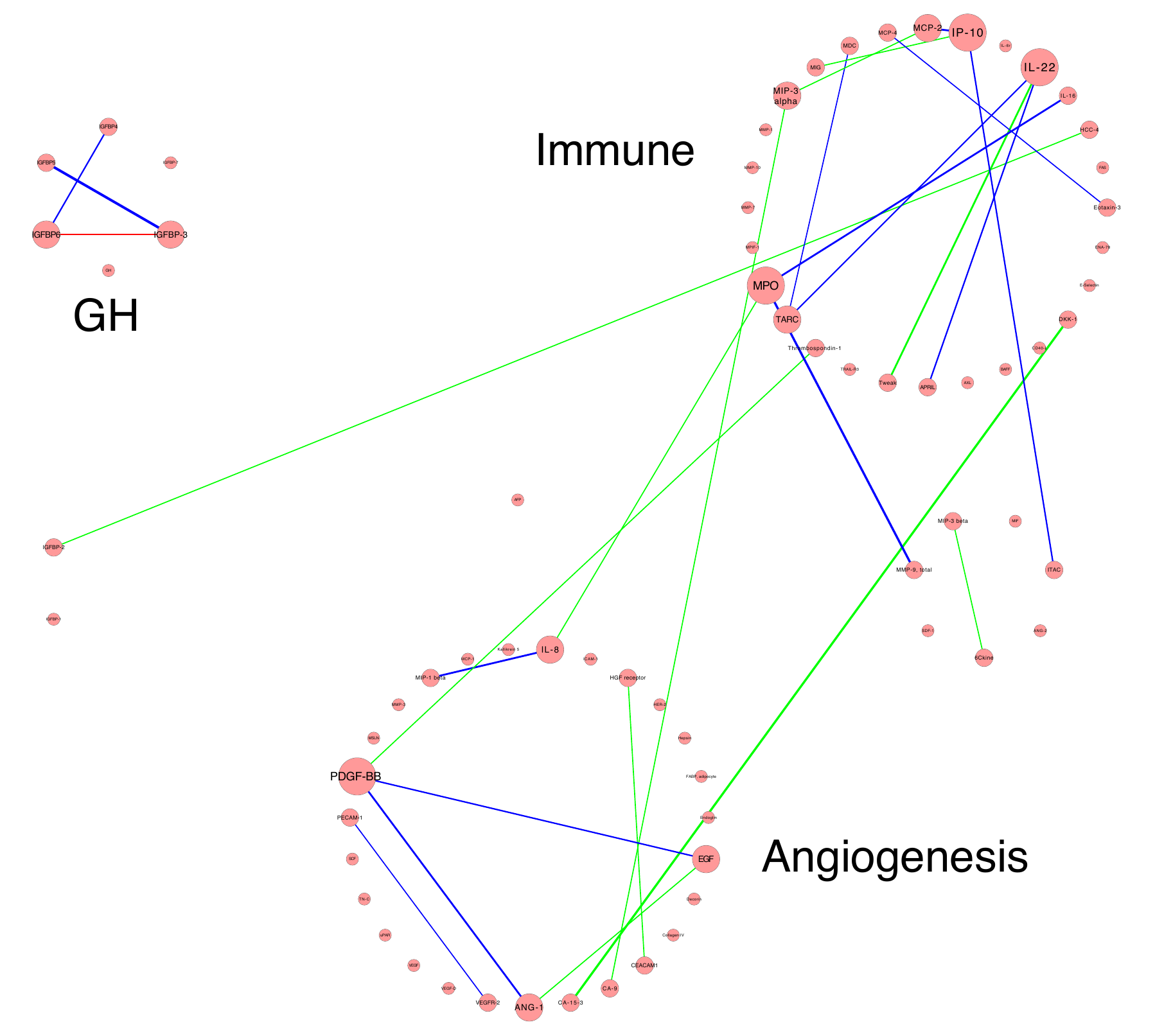}
\caption{\small Medium HepatoScore}
\end{subfigure}%
 \begin{subfigure}[H]{0.475\textwidth}
\includegraphics[width=1\textwidth]{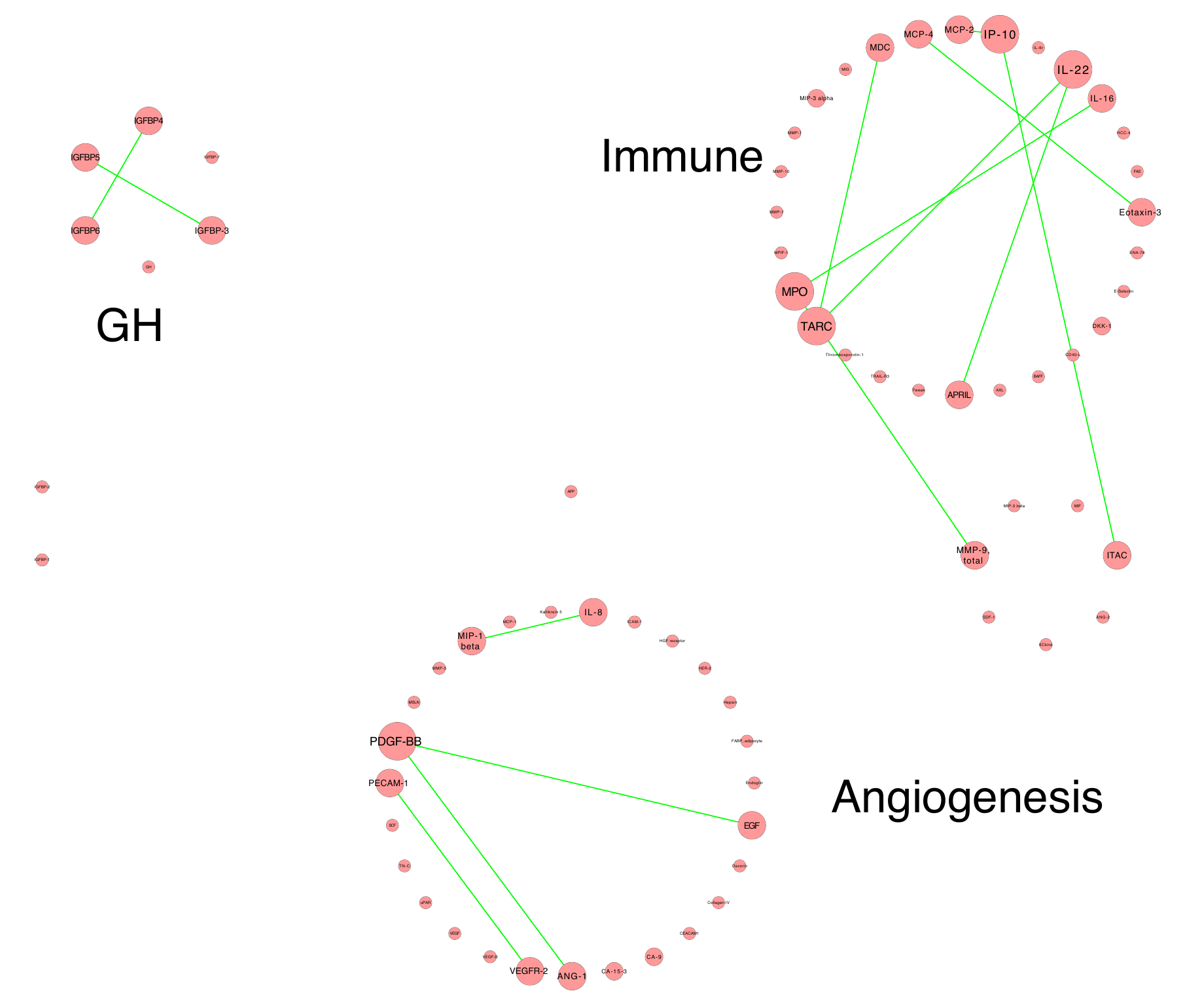}
\caption{\small Shared edges}
\end{subfigure}%
\caption{\small Estimated graphs ($\alpha=0.10, \kappa=0.2$) from the Bayesian edge regression for the GH, immune, and angiogenesis pathways with (a)  $\pi = 0$; (b) $\pi = 0.5$; (c) $\pi = 1$; (d) Common edges. Colors indicate positive (green), negative (red), and common (blue) edges. The thickness of edge is proportional to $\hat \rho^{ij}$ for the edge $(i, j)$, and the size of node is proportional to its degree.}
\label{fighccm2}
\end{figure}
 \begin{figure}[H]
 \begin{subfigure}[H]{0.475\textwidth}
\includegraphics[width=1\textwidth]{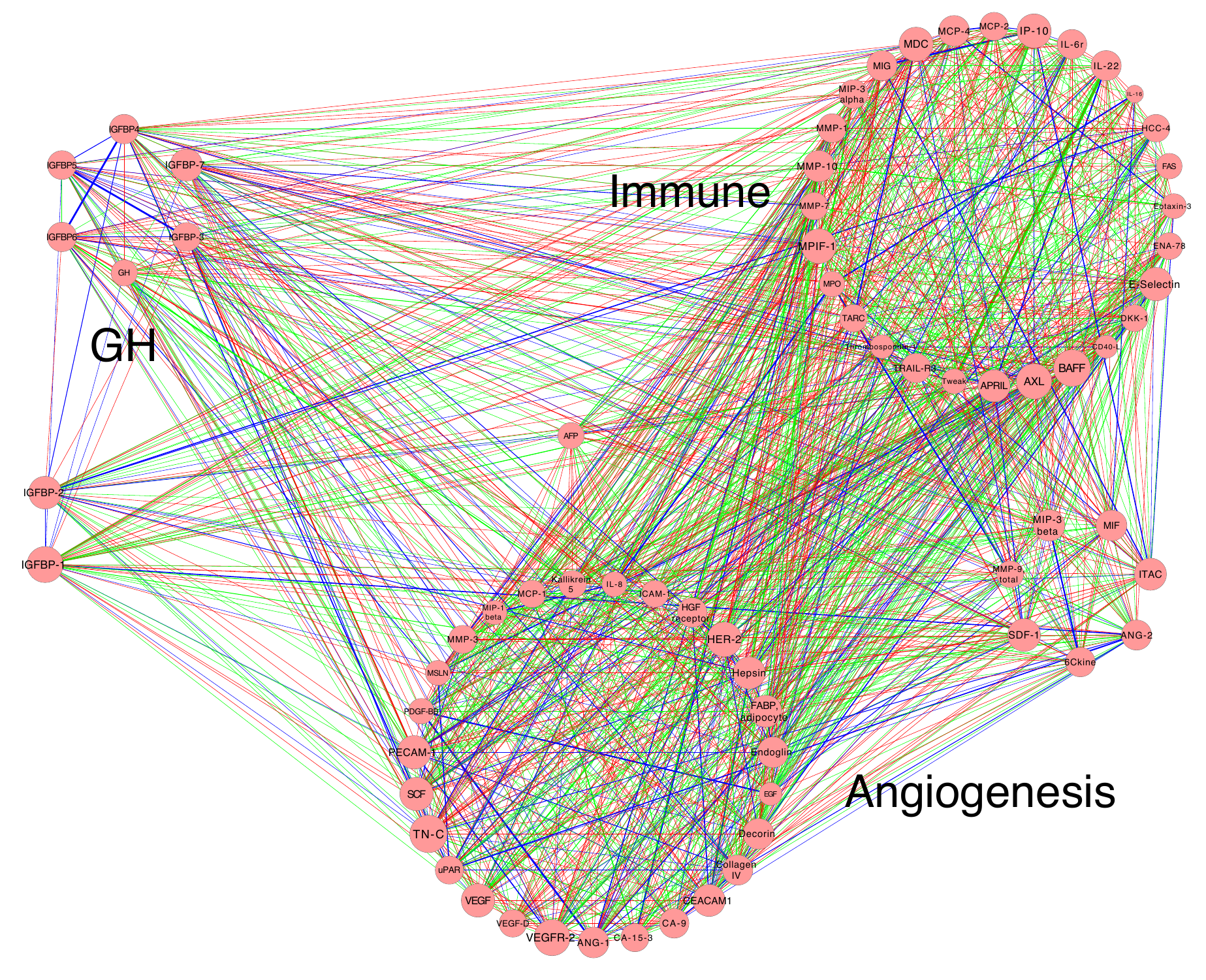}
\caption{\small High HepatoScore}
\end{subfigure}
 \begin{subfigure}[H]{0.475\textwidth}
\includegraphics[width=1\textwidth]{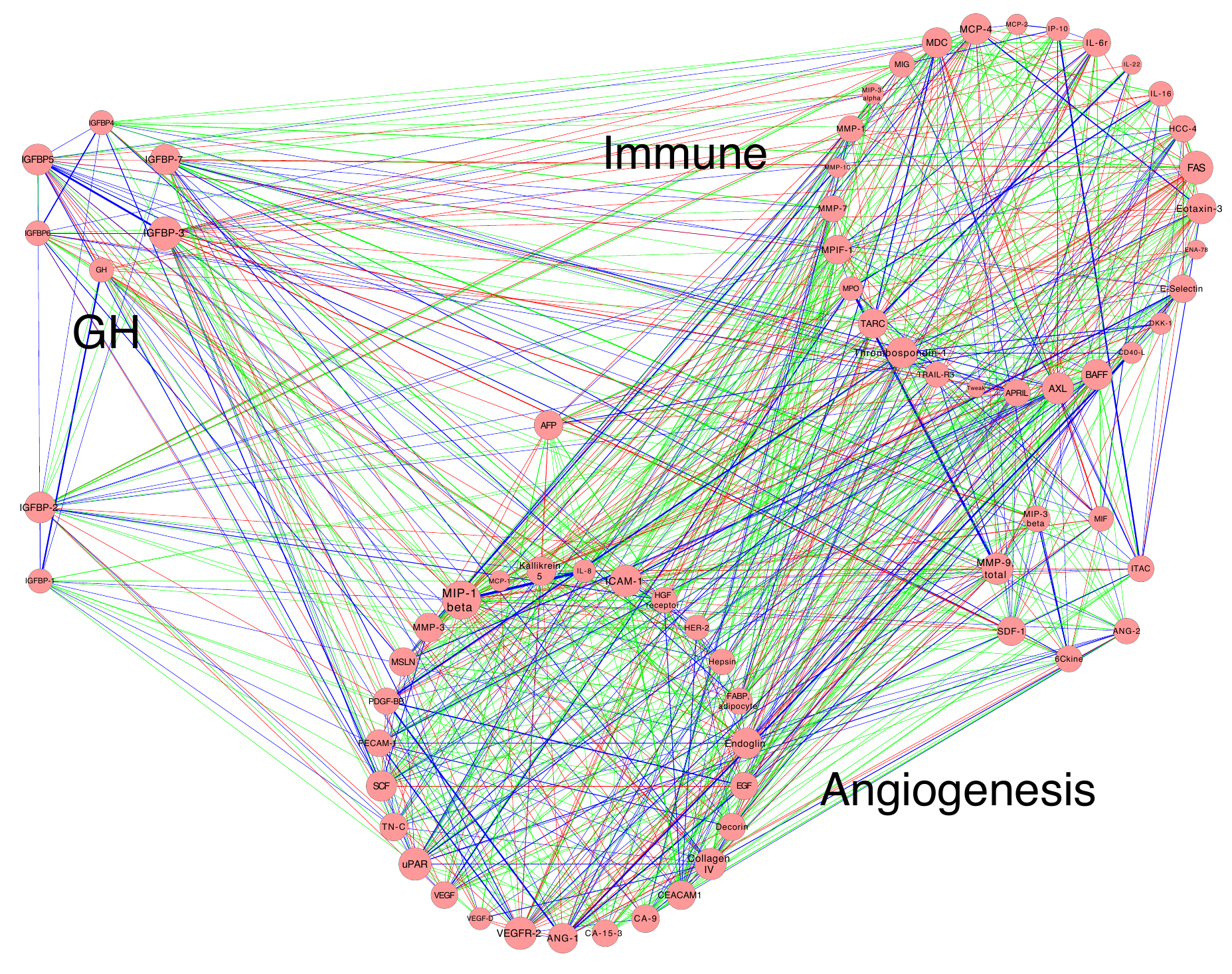}
\caption{\small Low HepatoScore}
\end{subfigure}%
\\
 \begin{subfigure}[H]{0.475\textwidth}
\includegraphics[width=1\textwidth]{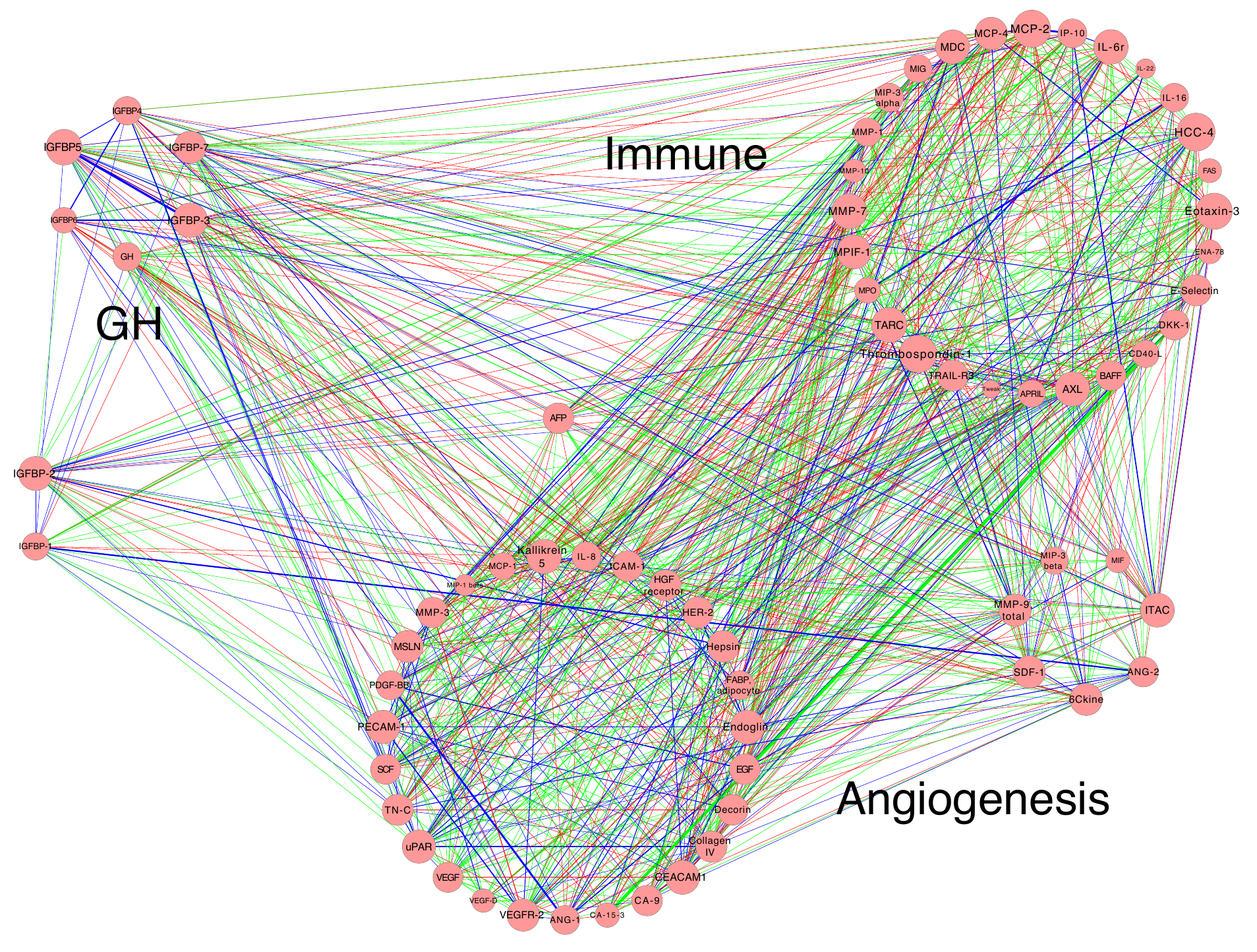}
\caption{\small Medium HepatoScore}
\end{subfigure}%
 \begin{subfigure}[H]{0.475\textwidth}
\includegraphics[width=1\textwidth]{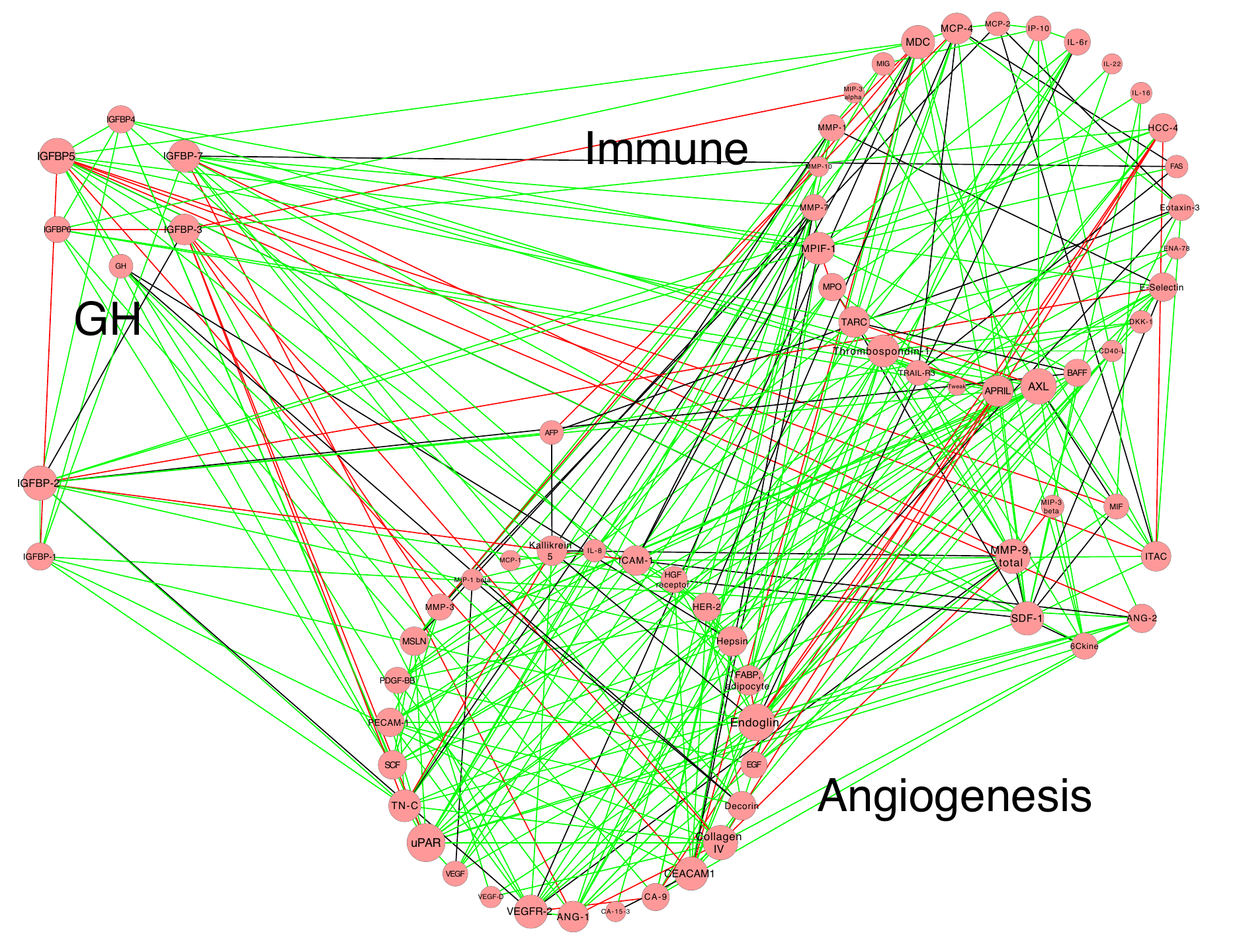}
\caption{\small Shared edges}
\end{subfigure}%
\caption{\small 
Estimated graphs from \textit{NFMGGM} for the GH, immune, and angiogenesis pathways with (a)  $\pi = 0$; (b) $\pi = 0.5$; (c) $\pi = 1$; (d) Common edges. Colors indicate positive (green) and negative (red). Common edges across graphs (a), (b) and (c) are colored with blue in each graph. Those with consistent signs across graphs are colored following the same rule (green=positive, red=negative) in (d), while those with different signs are colored with black. The thickness of edge is proportional to $\hat \rho^{ij}$ for the edge $(i, j)$, and the size of node is proportional to its degree.}
\label{fighccm3}
\end{figure}
%
 \begin{landscape}
 \begin{table}[H]
\centering
\begin{tabular}{|r|r|r|r|r|r|r|r|r|r|}
\hline
\multicolumn{1}{|c|}{}&
\multicolumn{3}{c|}{Low}&
\multicolumn{3}{c|}{Medium}&
\multicolumn{3}{c|}{High}\\
  \hline
  \multicolumn{1}{|c}{}&
\multicolumn{1}{c}{GH}&
\multicolumn{1}{c}{Angio}&
\multicolumn{1}{c}{Immune}&
\multicolumn{1}{c}{GH}&
\multicolumn{1}{c}{Angio}&
\multicolumn{1}{c}{Immune}&
\multicolumn{1}{c}{GH}&
\multicolumn{1}{c}{Angio}&
\multicolumn{1}{c|}{Immune}\\
  \hline
 \multicolumn{10}{|c|}{$\kappa = 0.1$}\\
 \hline
  GH & 3 & 8 & 8 & 6 & 12 & 11 & 9 & 30 & 22 \\ 
   &  (0.11) &  (0.03) &  (0.03) &  (0.21) &  (0.04) &  (0.04) &  (0.32) &  (0.1) &  (0.08) \\ 
   \hline
Angio & 8 & 16 & 24 & 12 & 22 & 41 & 30 & 65 & 106 \\ 
    &  (0.03) &  (0.03) &  (0.02) &  (0.04) &  (0.03) &  (0.03) &  (0.1) &  (0.1) &  (0.08) \\ 
   \hline
Immune & 8 & 24 & 14 & 11 & 41 & 24 & 22 & 106 & 58 \\ 
     &  (0.03) &  (0.02) &  (0.02) &  (0.04) &  (0.03) &  (0.04) &  (0.08) &  (0.08) &  (0.1) \\ 
   \hline
   \multicolumn{10}{|c|}{$\kappa = 0.15$}\\
   \hline
  & GH & Angiogenesis & Immune & GH & Angiogenesis & Immune & GH & Angiogenesis & Immune \\ 
 GH & 2 & 2 & 3 & 4 & 2 & 2 & 7 & 14 & 11 \\ 
   &  (0.07) &  (0.01) &  (0.01) &  (0.14) &  (0.01) &  (0.01) &  (0.25) &  (0.05) &  (0.04) \\ 
   \hline
Angio & 2 & 8 & 15 & 2 & 10 & 17 & 14 & 26 & 40 \\ 
    &  (0.01) &  (0.01) &  (0.01) &  (0.01) &  (0.02) &  (0.01) &  (0.05) &  (0.04) &  (0.03) \\ 
   \hline
Immune & 3 & 15 & 10 & 2 & 17 & 16 & 11 & 40 & 30 \\ 
     &  (0.01) &  (0.01) &  (0.02) &  (0.01) &  (0.01) &  (0.03) &  (0.04) &  (0.03) &  (0.05) \\ 
   \hline
  \multicolumn{10}{|c|}{$\kappa = 0.2$}\\
\hline
GH & 2 & 0 & 0 & 3 & 0 & 1 & 5 & 6 & 4 \\ 
   &  (0.07) &  (0) &  (0) &  (0.11) &  (0) &  (0) &  (0.18) &  (0.02) &  (0.01) \\ 
   \hline
Angio & 0 & 6 & 9 & 0 & 7 & 9 & 6 & 13 & 18 \\ 
    &  (0) &  (0.01) &  (0.01) &  (0) &  (0.01) &  (0.01) &  (0.02) &  (0.02) &  (0.01) \\ 
   \hline
Immune & 0 & 9 & 9 & 1 & 9 & 12 & 4 & 18 & 22 \\ 
     &  (0) &  (0.01) &  (0.02) &  (0) &  (0.01) &  (0.02) &  (0.01) &  (0.01) &  (0.04) \\ 
   \hline
\end{tabular}
\caption{\small Table that summarizes edge connectedness within/across different pathways for the constructed graph of the low \textit{HepatoScore} ($\pi = 0$), medium \textit{HepatoScore} ($\pi = 0.5$), and high \textit{HepatoScore} ($\pi = 1$).}
\label{edgecn}
\end{table}
 \end{landscape}

\newpage
{\tiny
\renewcommand{\arraystretch}{0.5}
\begin{longtable}{|r|r|r|r|r|r|r|r|}
  \hline
  \multicolumn{2}{|c|}{High HepatoScore}&
  \multicolumn{2}{c|}{Medium HepatoScore}&
  \multicolumn{2}{c|}{Low HepatoScore}&
  \multicolumn{2}{c|}{Common}\\
  \hline
Gene & Degree & Gene & Degree & Gene & Degree & Gene & Degree \\ 
  \hline
IGFBP-3 & 8 & IP-10 & 4 & CA-15-3 & 4 & PDGF-BB & 2 \\ 
  Hepsin & 7 & PDGF-BB & 3 & PDGF-BB & 3 & IP-10 & 2 \\ 
  IGFBP4 & 7 & ITAC & 3 & E-Selectin & 3 & MPO & 2 \\ 
  IGFBP-2 & 6 & IGFBP6 & 3 & SDF-1 & 3 & IL-22 & 2 \\ 
  IGFBP6 & 6 & MCP-2 & 3 & IP-10 & 3 & TARC & 2 \\ 
  ANG-1 & 5 & MPO & 3 & MCP-2 & 3 & ANG-1 & 1 \\ 
  MDC & 5 & IL-22 & 3 & MPO & 3 & Eotaxin-3 & 1 \\ 
  APRIL & 5 & ANG-1 & 2 & ANG-1 & 2 & MCP-4 & 1 \\ 
  MIP-3 alpha & 4 & EGF & 2 & EGF & 2 & EGF & 1 \\ 
  PDGF-BB & 4 & CA-15-3 & 2 & MCP-1 & 2 & IGFBP-3 & 1 \\ 
  TARC & 4 & MIP-3 alpha & 2 & Thrombospondin-1 & 2 & IGFBP5 & 1 \\ 
  IGFBP-1 & 4 & IGFBP-2 & 2 & CA-9 & 2 & IGFBP4 & 1 \\ 
  uPAR & 4 & Eotaxin-3 & 2 & MIP-3 alpha & 2 & IGFBP6 & 1 \\ 
  CA-15-3 & 4 & MCP-4 & 2 & VEGFR-2 & 2 & ITAC & 1 \\ 
  Collagen IV & 4 & MDC & 2 & MDC & 2 & MCP-2 & 1 \\ 
  HER-2 & 4 & IGFBP-3 & 2 & IGFBP-3 & 2 & IL-8 & 1 \\ 
  MPO & 4 & IGFBP5 & 2 & IGFBP5 & 2 & MIP-1 beta & 1 \\ 
  TRAIL-R3 & 4 & IGFBP4 & 2 & IL-8 & 2 & IL-16 & 1 \\ 
  SCF & 4 & SCF & 2 & IL-22 & 2 & APRIL & 1 \\ 
  MPIF-1 & 4 & MIG & 2 & TARC & 2 & MDC & 1 \\ 
  IP-10 & 4 & IL-8 & 2 & AFP & 1 & MMP-9, total & 1 \\ 
  MCP-2 & 4 & TARC & 2 & HGF receptor & 1 & PECAM-1 & 1 \\ 
  IL-8 & 4 & MMP-9, total & 2 & DKK-1 & 1 & VEGFR-2 & 1 \\ 
  VEGF & 4 & 6Ckine & 1 & MMP-10 & 1 &  &  \\ 
  Tweak & 4 & MIP-3 beta & 1 & CEACAM1 & 1 &  &  \\ 
  MMP-1 & 3 & AXL & 1 & HCC-4 & 1 &  &  \\ 
  AXL & 3 & IGFBP-7 & 1 & FAS & 1 &  &  \\ 
  IGFBP-7 & 3 & DKK-1 & 1 & IGFBP-2 & 1 &  &  \\ 
  MSLN & 3 & MMP-7 & 1 & ICAM-1 & 1 &  &  \\ 
  MMP-7 & 3 & CA-9 & 1 & Endoglin & 1 &  &  \\ 
  Decorin & 3 & CEACAM1 & 1 & Eotaxin-3 & 1 &  &  \\ 
  E-Selectin & 3 & HGF receptor & 1 & MCP-4 & 1 &  &  \\ 
  MCP-4 & 3 & HCC-4 & 1 & FABP, adipocyte & 1 &  &  \\ 
  Eotaxin-3 & 3 & Decorin & 1 & IGFBP4 & 1 &  &  \\ 
  ITAC & 3 & E-Selectin & 1 & IGFBP6 & 1 &  &  \\ 
  FAS & 3 & Endoglin & 1 & ITAC & 1 &  &  \\ 
  MMP-9, total & 3 & SDF-1 & 1 & MIP-1 beta & 1 &  &  \\ 
  MMP-3 & 3 & FABP, adipocyte & 1 & IL-16 & 1 &  &  \\ 
  MIG & 3 & Hepsin & 1 & APRIL & 1 &  &  \\ 
  IL-22 & 3 & MCP-1 & 1 & MMP-9, total & 1 &  &  \\ 
  Thrombospondin-1 & 2 & MIP-1 beta & 1 & VEGF & 1 &  &  \\ 
  ANG-2 & 2 & IL-16 & 1 & PECAM-1 & 1 &  &  \\ 
  BAFF & 2 & Tweak & 1 &  &  &  &  \\ 
  HCC-4 & 2 & APRIL & 1 &  &  &  &  \\ 
  MMP-10 & 2 & MMP-3 & 1 &  &  &  &  \\ 
  CA-9 & 2 & TRAIL-R3 & 1 &  &  &  &  \\ 
  CEACAM1 & 2 & PECAM-1 & 1 &  &  &  &  \\ 
  HGF receptor & 2 & VEGFR-2 & 1 &  &  &  &  \\ 
  DKK-1 & 2 & Thrombospondin-1 & 1 &  &  &  &  \\ 
  Endoglin & 2 &  &  &  &  &  &  \\ 
  FABP, adipocyte & 2 &  &  &  &  &  &  \\ 
  EGF & 2 &  &  &  &  &  &  \\ 
  SDF-1 & 2 &  &  &  &  &  &  \\ 
  IGFBP5 & 2 &  &  &  &  &  &  \\ 
  PECAM-1 & 2 &  &  &  &  &  &  \\ 
  IL-6r & 2 &  &  &  &  &  &  \\ 
  VEGFR-2 & 2 &  &  &  &  &  &  \\ 
  IL-16 & 2 &  &  &  &  &  &  \\ 
  6Ckine & 1 &  &  &  &  &  &  \\ 
  MIP-3 beta & 1 &  &  &  &  &  &  \\ 
  AFP & 1 &  &  &  &  &  &  \\ 
  MCP-1 & 1 &  &  &  &  &  &  \\ 
  TN-C & 1 &  &  &  &  &  &  \\ 
  Kallikrein 5 & 1 &  &  &  &  &  &  \\ 
  ICAM-1 & 1 &  &  &  &  &  &  \\ 
  MIP-1 beta & 1 &  &  &  &  &  &  \\ 
  MIF & 1 &  &  &  &  &  &  \\ 
  VEGF-D & 1 &  &  &  &  &  &  \\ 
   \hline
      \caption{\small Table that summarizes the degree of all the genes for the constructed graphs with low \textit{HepatoScore} ($\pi = 0$), medium \textit{HepatoScore} ($\pi = 0.5$), high \textit{HepatoScore} ($\pi = 1$), and common graph given $\kappa = 0.15$. Genes are ranked decreasingly according to their degree. } 
      \label{fighub}
\end{longtable}
}
\begin{figure}[H] 
\centering
\includegraphics[width=0.185\textwidth]{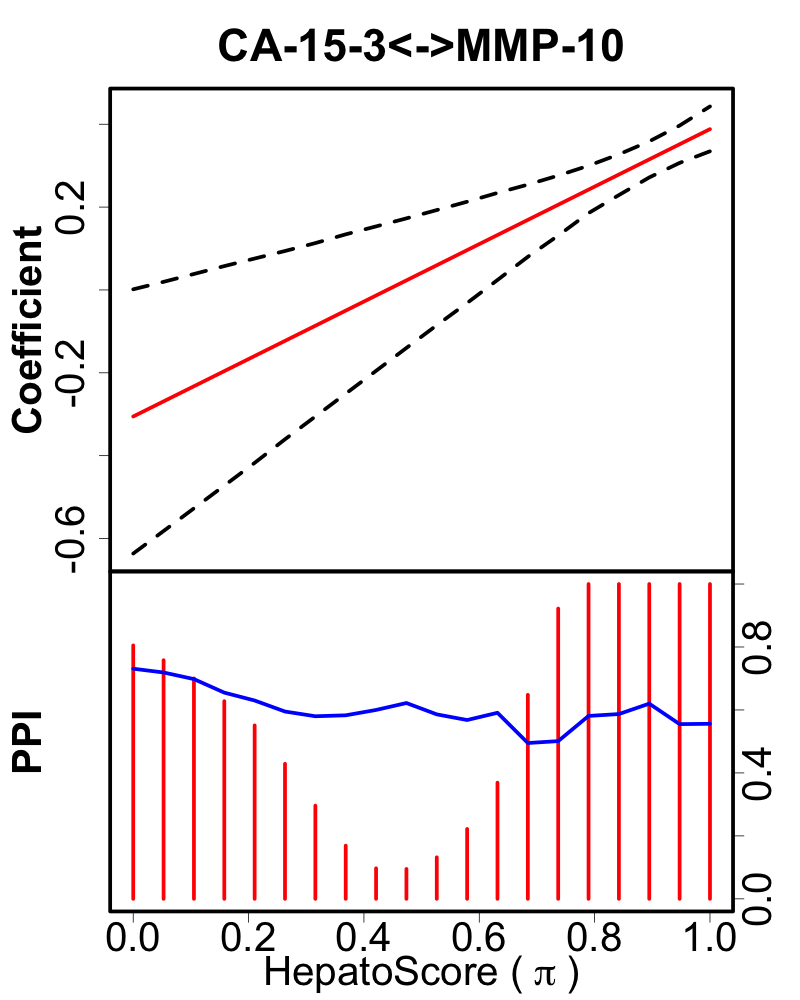}
\includegraphics[width=0.185\textwidth]{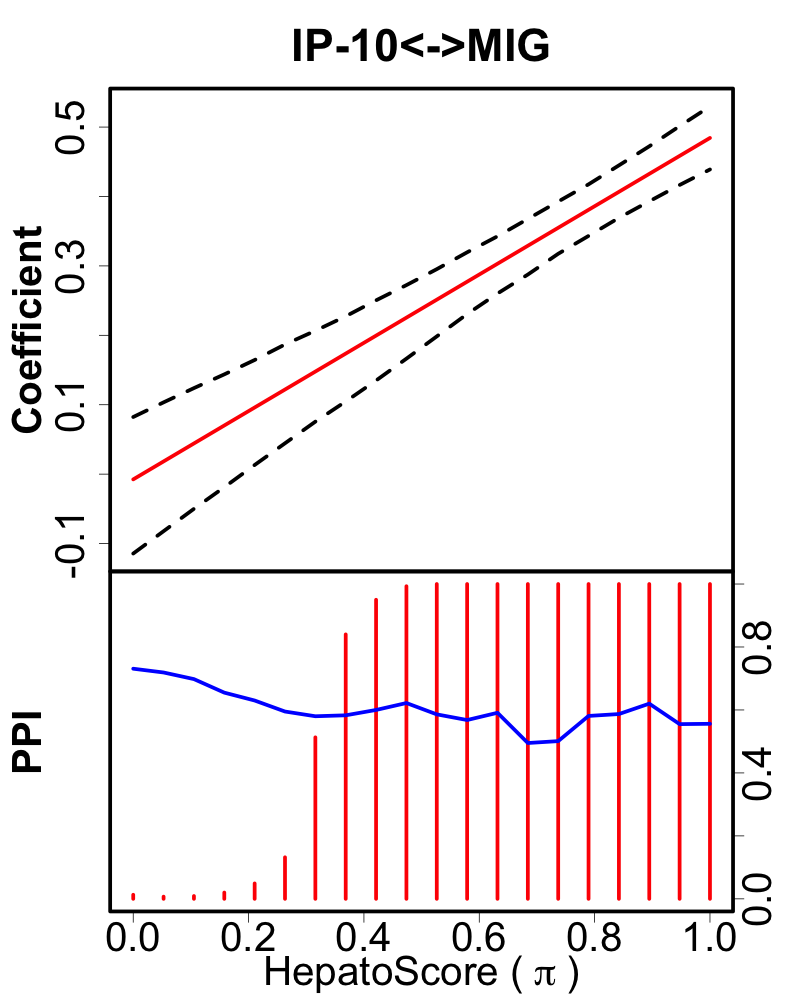}
\includegraphics[width=0.185\textwidth]{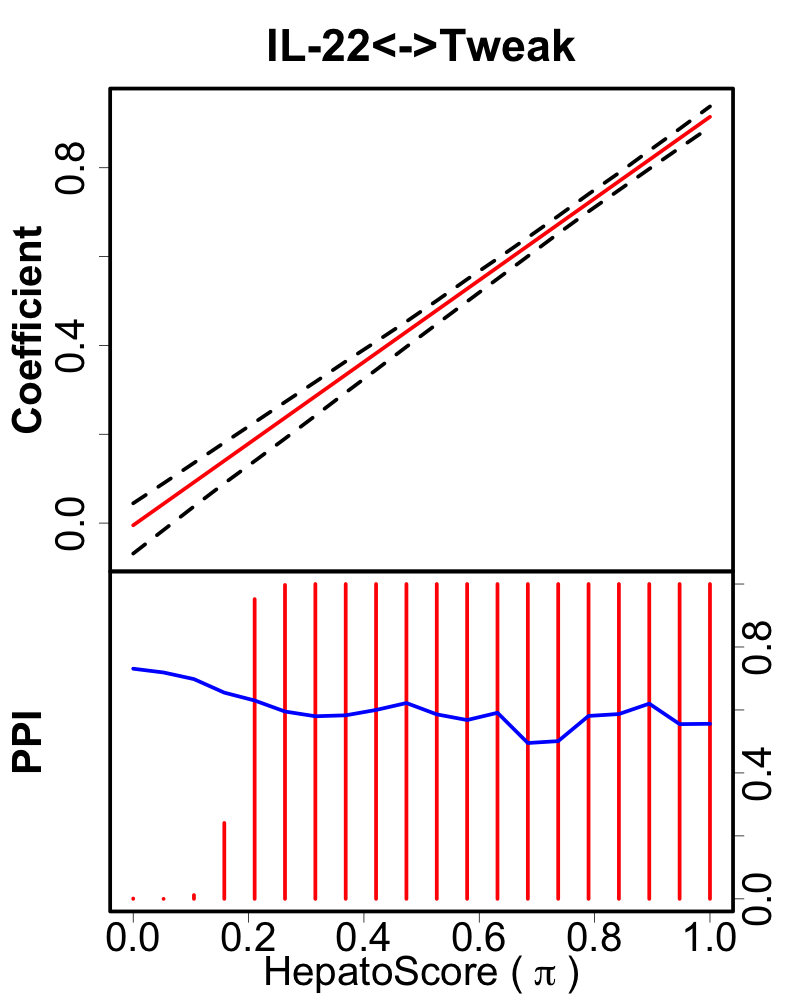}
\includegraphics[width=0.185\textwidth]{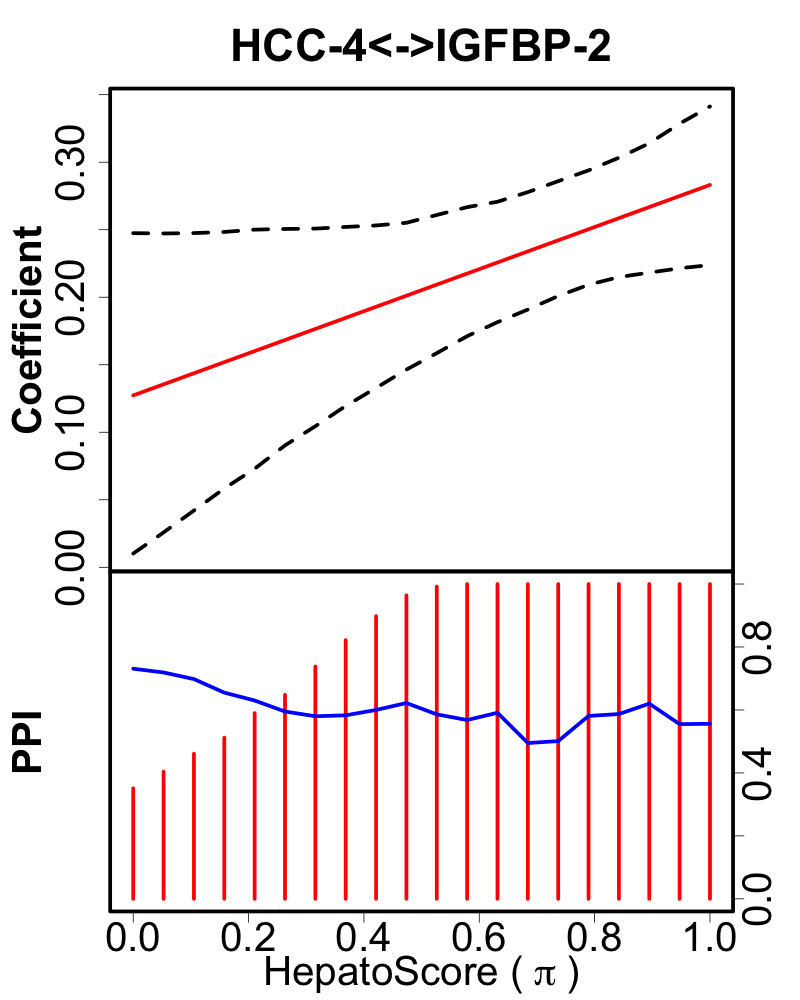}\\
\includegraphics[width=0.185\textwidth]{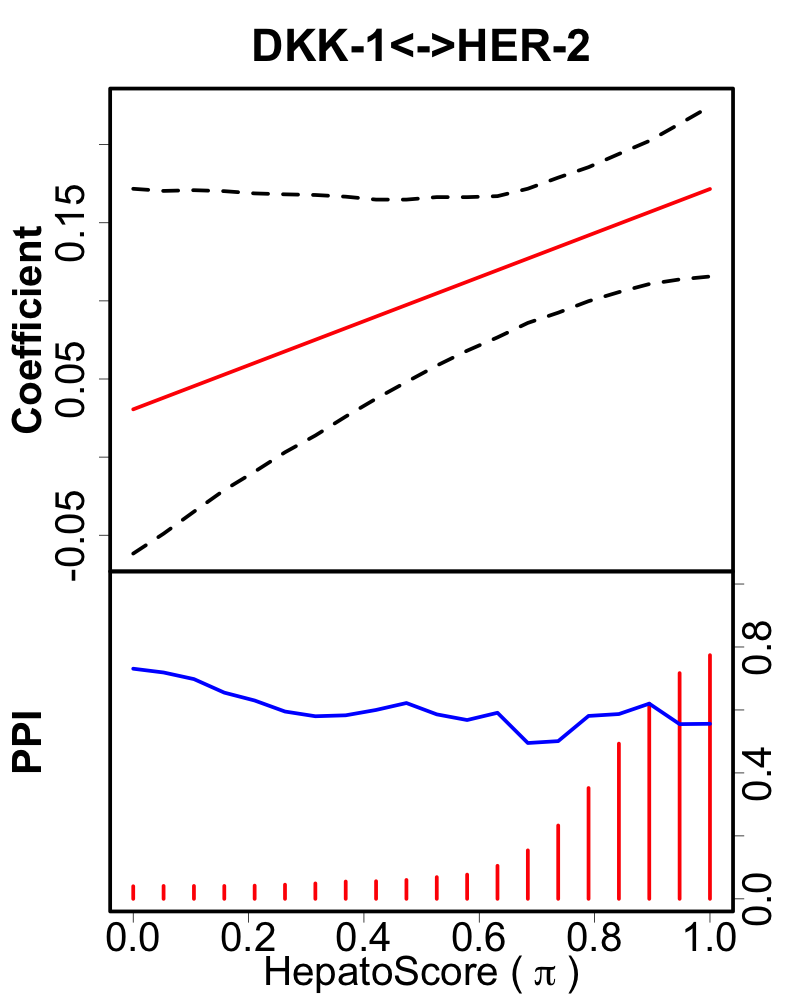}
\includegraphics[width=0.185\textwidth]{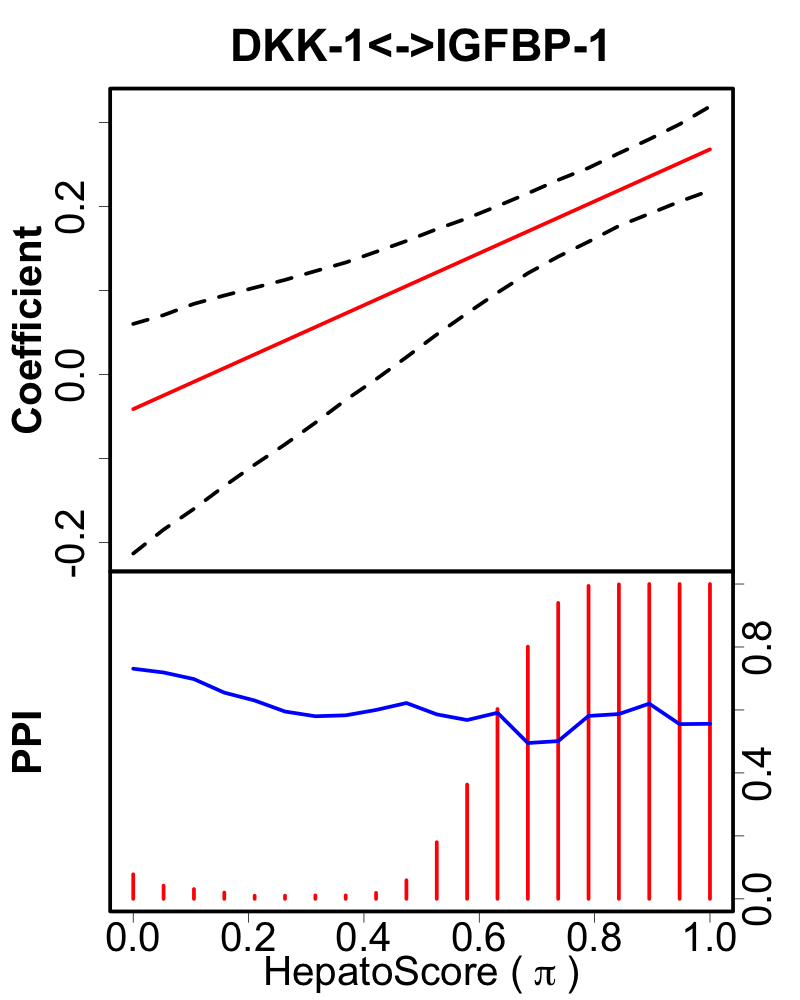}
\includegraphics[width=0.185\textwidth]{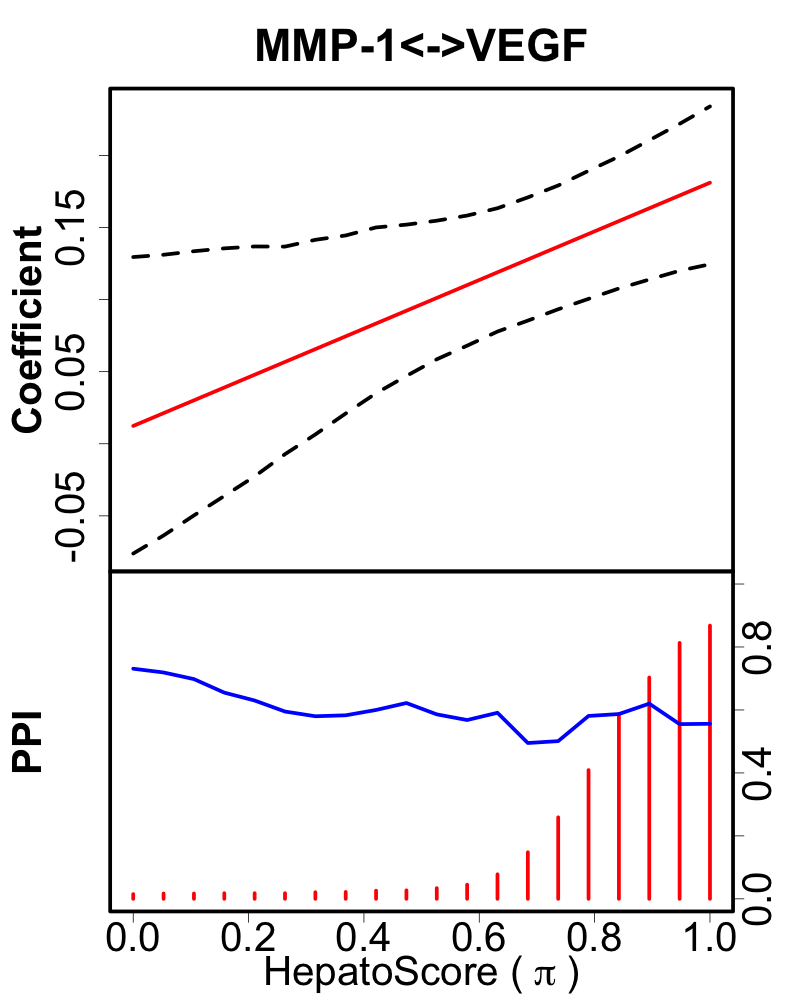}
\includegraphics[width=0.185\textwidth]{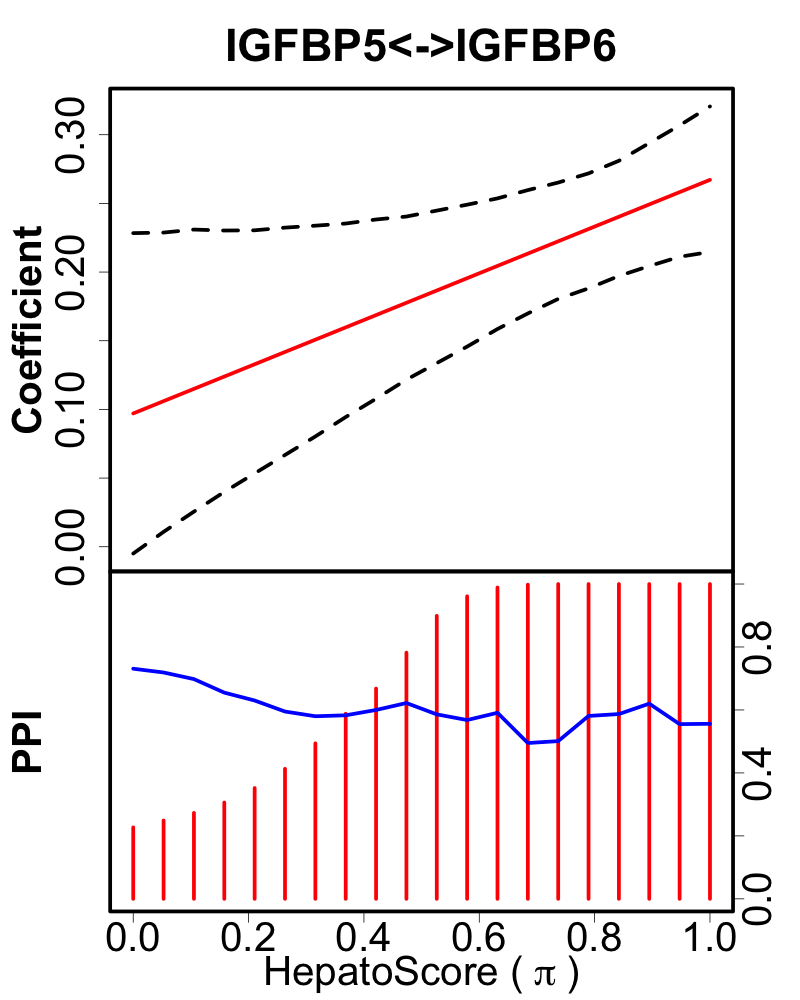}\\
\includegraphics[width=0.185\textwidth]{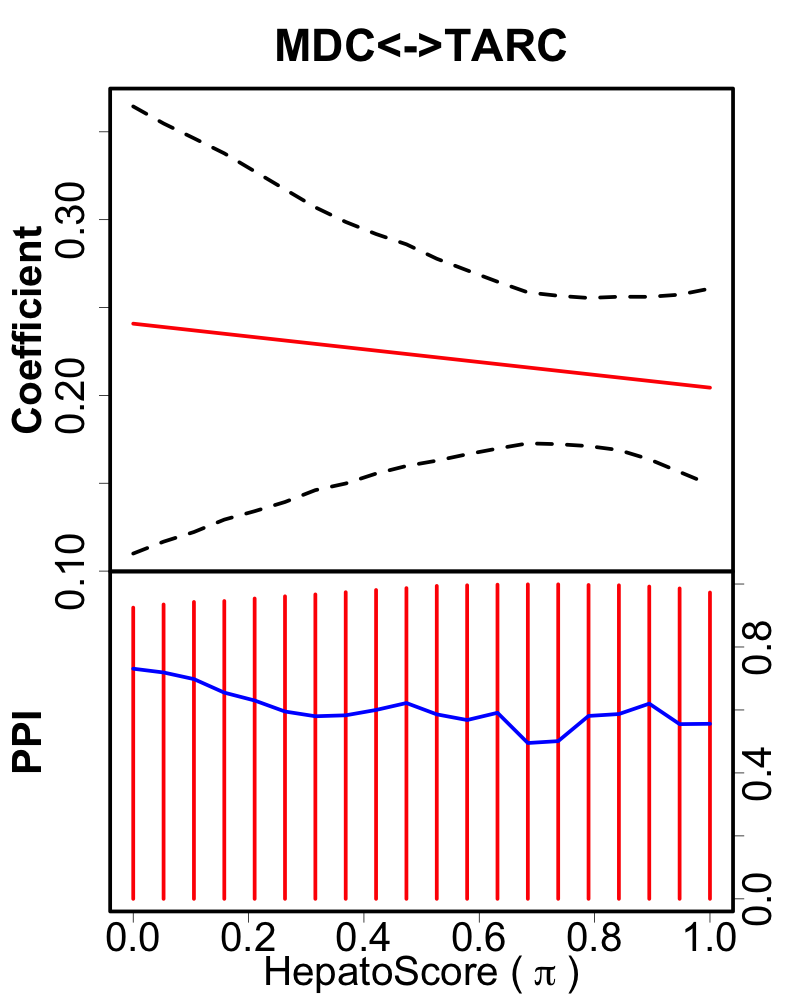}
\includegraphics[width=0.185\textwidth]{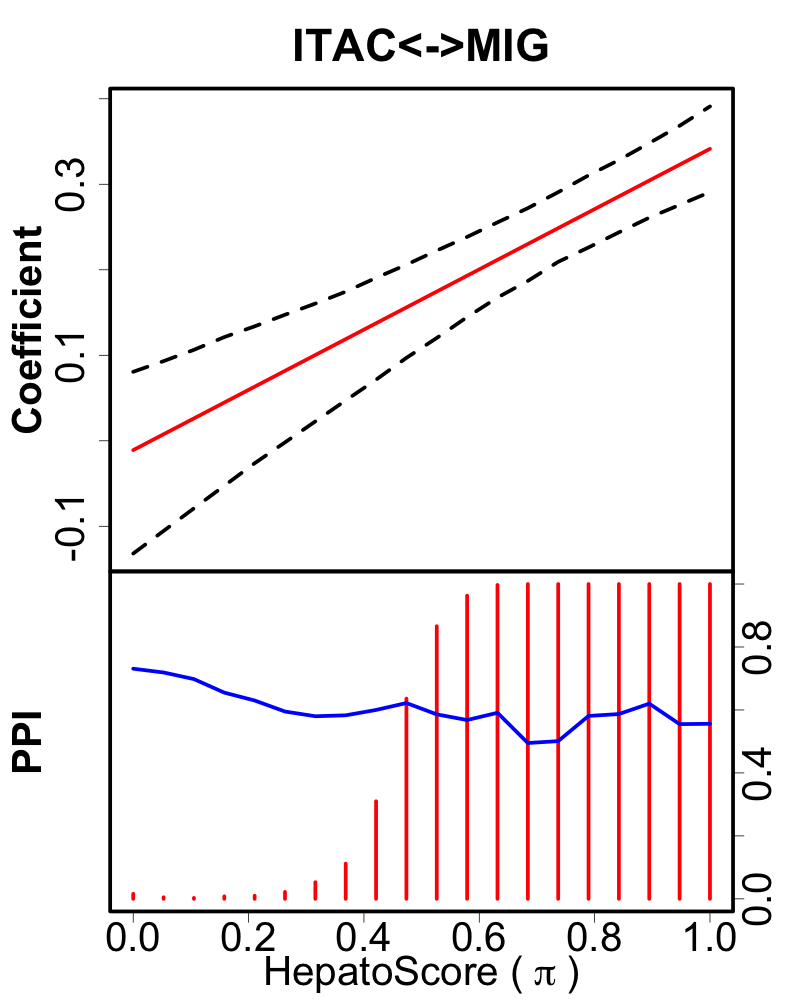}
\includegraphics[width=0.185\textwidth]{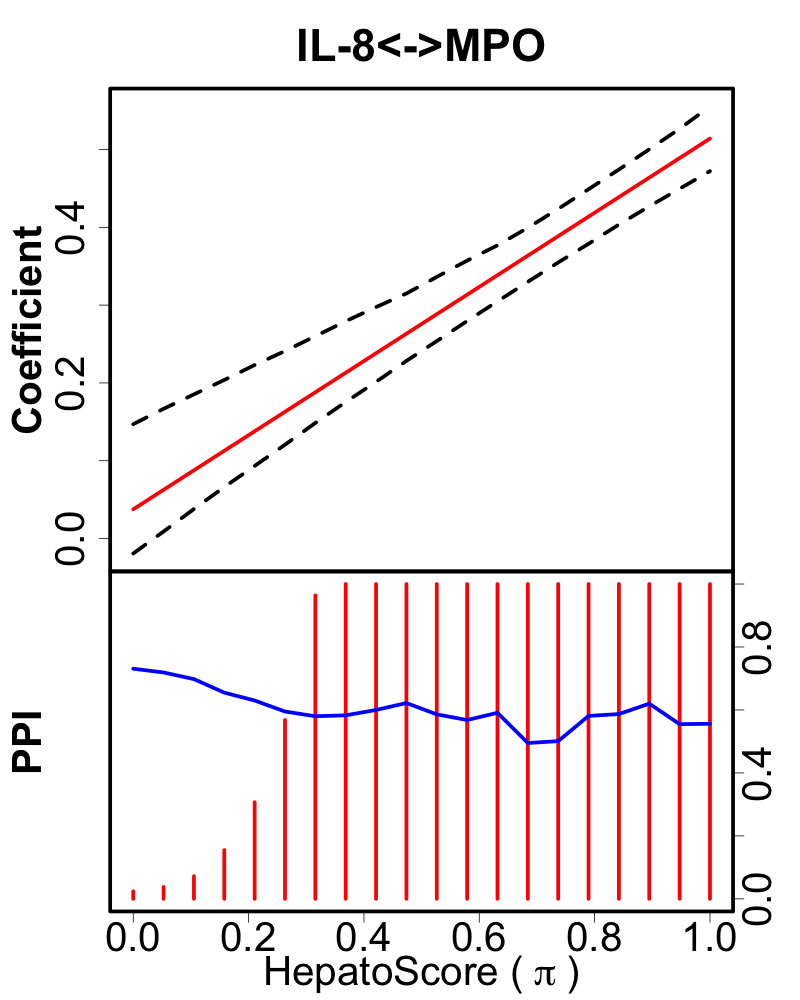}
\includegraphics[width=0.185\textwidth]{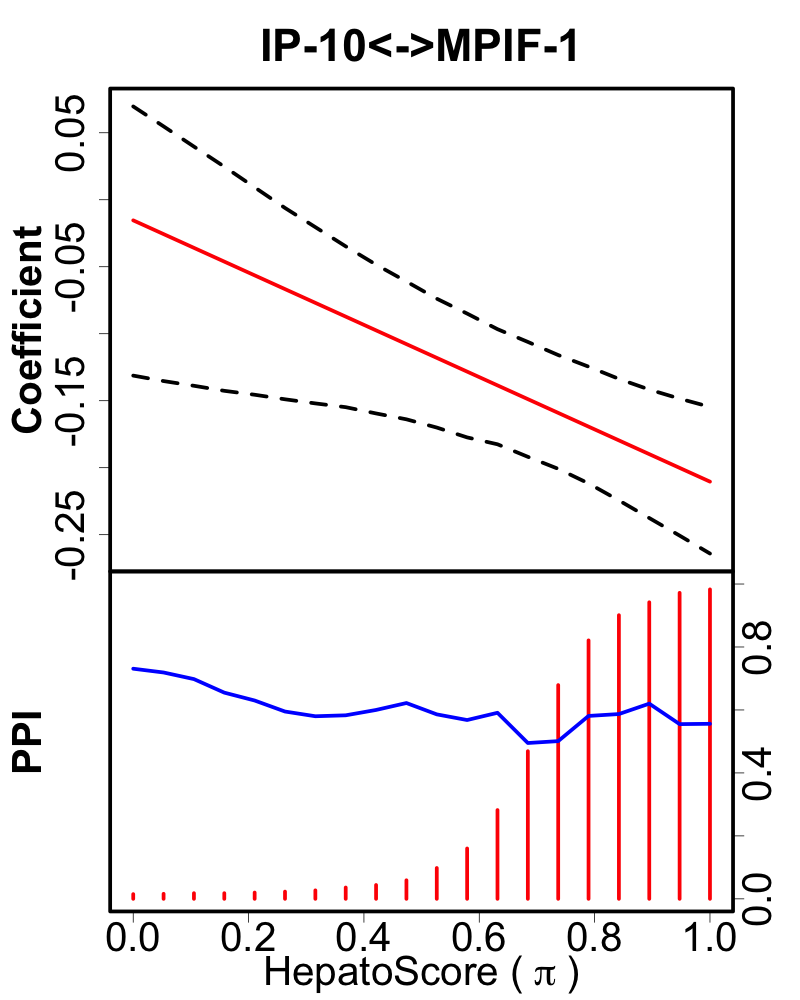}\\
\includegraphics[width=0.185\textwidth]{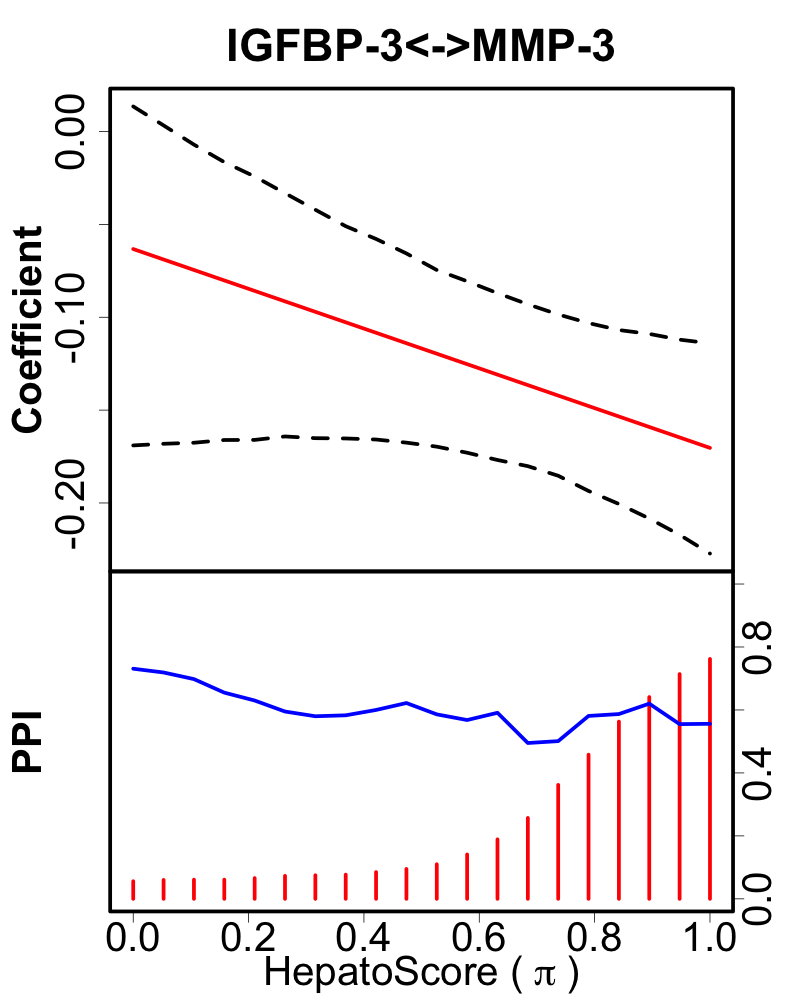}
\includegraphics[width=0.185\textwidth]{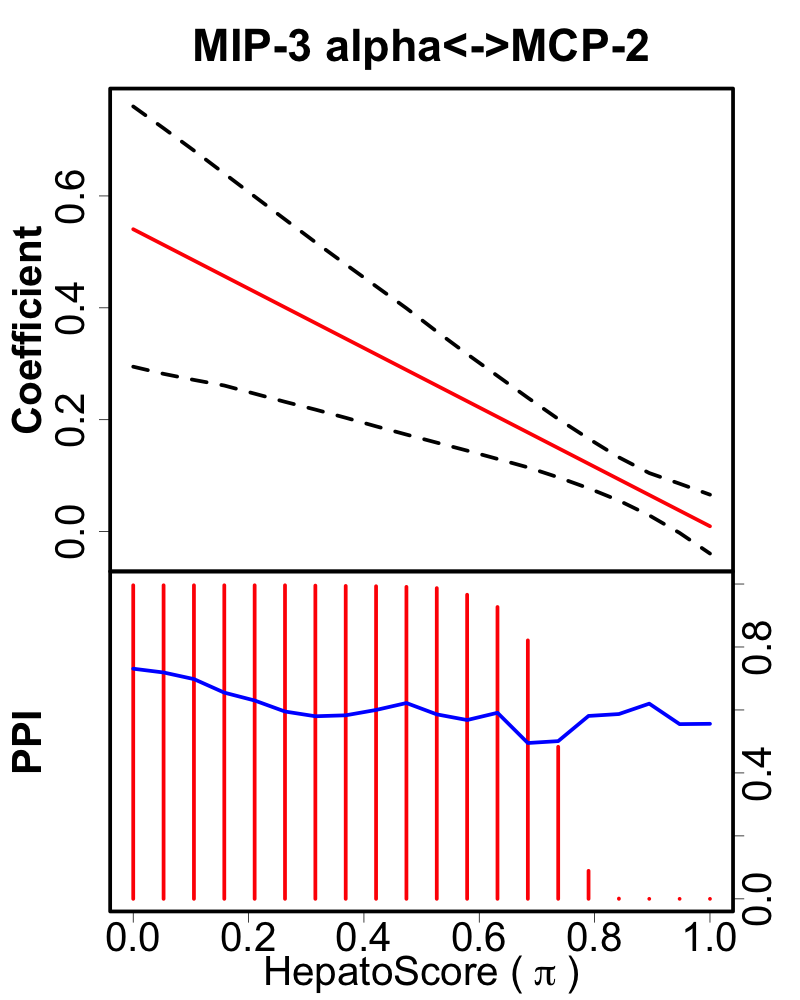}
\includegraphics[width=0.185\textwidth]{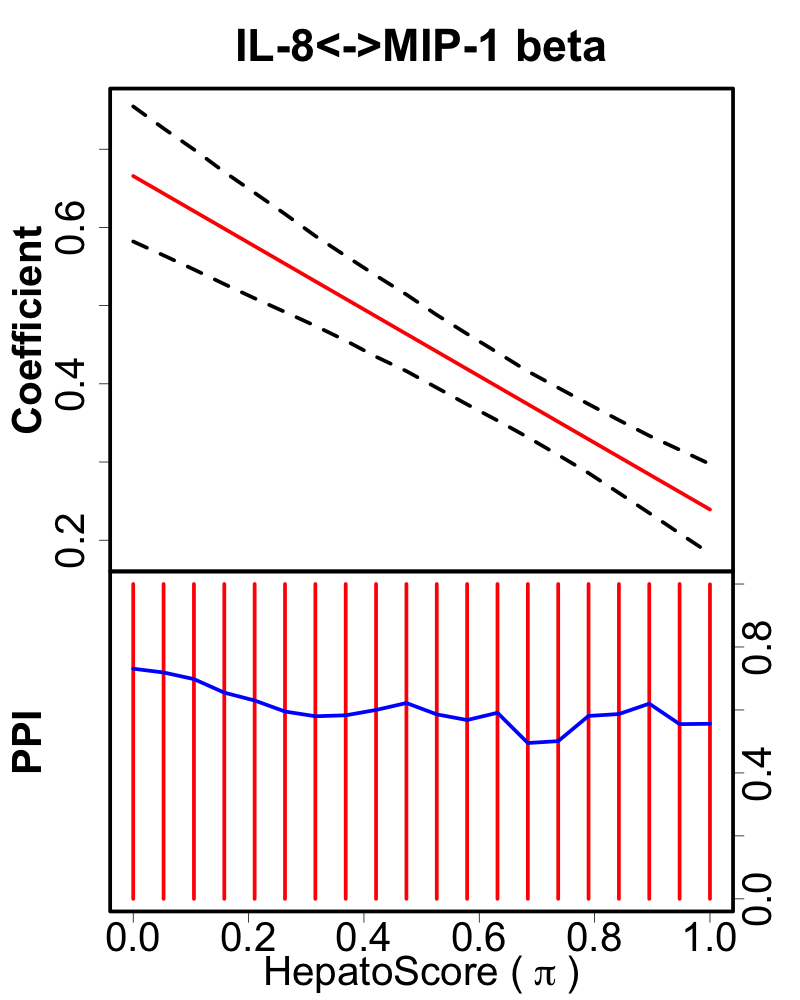}
\includegraphics[width=0.185\textwidth]{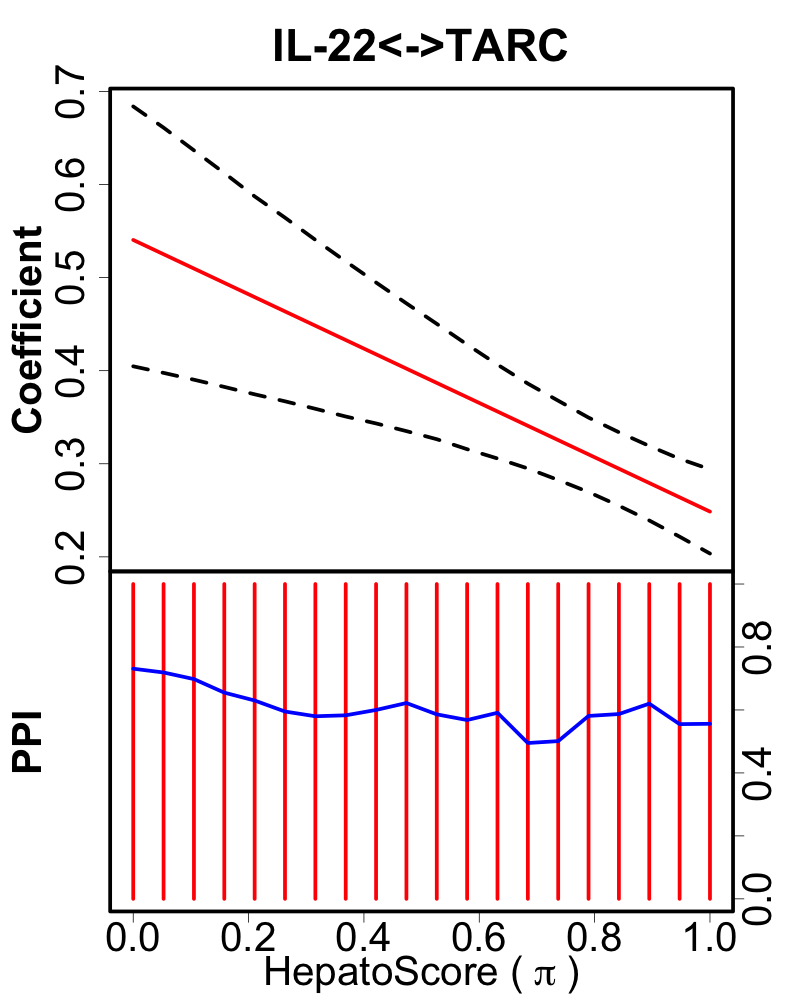}
\caption{\small Top panels show predicted edge strengths (posterior point estimate of $\rho^{ij}$) with 95\% credible interval, with bottom panels plotting posterior probabilities of inclusion (PPI) with blue lines indicating FDR=0.10 thresholds. For a value of $\pi$, the head of the corresponding red line above the blue curve indicates a nonzero edge strength}
\label{figedge2}
\end{figure}
\begin{figure}[H] 
\centering
\includegraphics[width=1\textwidth]{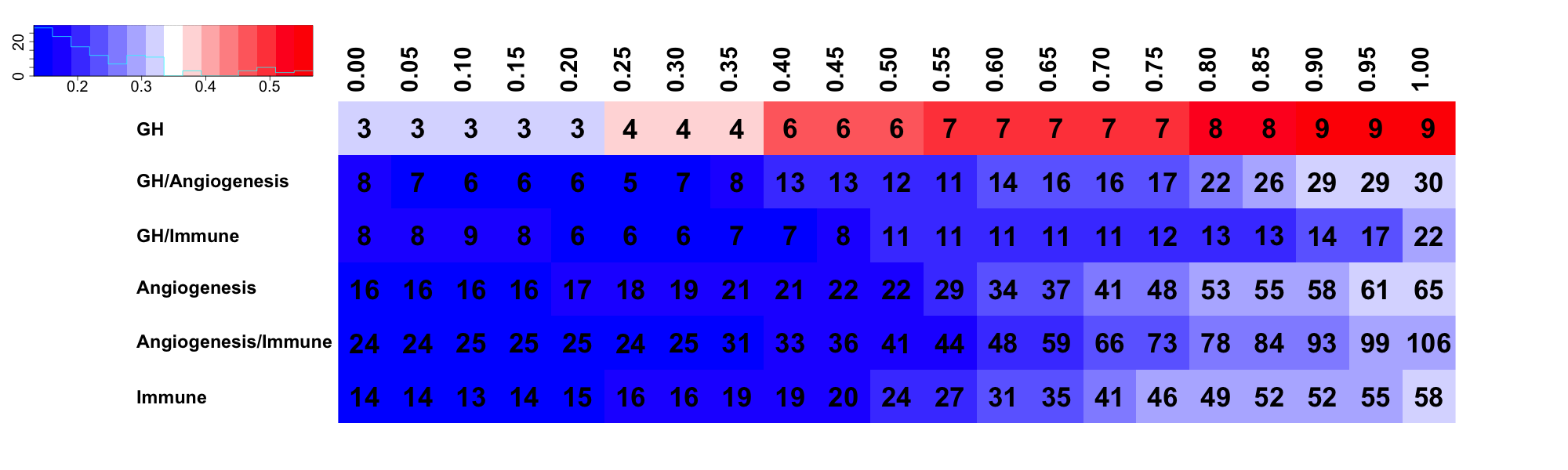}
\caption{\small Bayesian edge regression ($\kappa = 0.1$). The number of connected edges within each pathway/across different pathways varies with different \textit{HepatoScore}s (from $\pi = 0$ to $1$). The color gradient for each table cell changes with the square root of the proportion of connected edges on the whole possibly connected edge set. }
\label{figheat1}
\end{figure}
\begin{figure}[H] 
\centering
\includegraphics[width=1\textwidth]{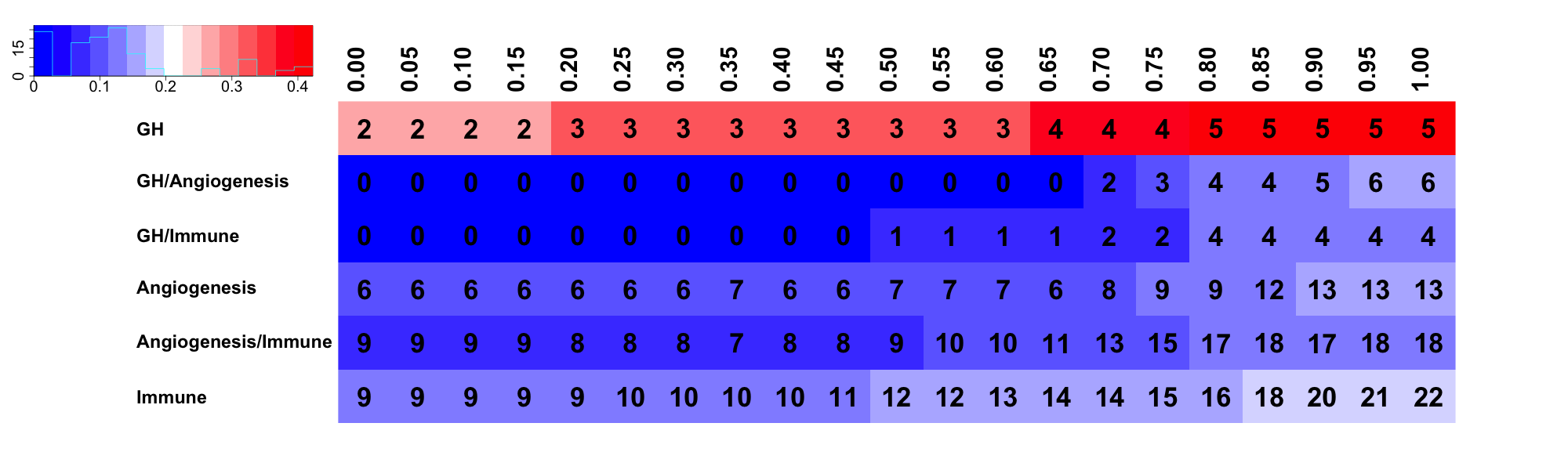}
\caption{\small Bayesian edge regression ($\kappa = 0.2$). The number of connected edges within each pathway/across different pathways varies with different \textit{HepatoScore}s (from $\pi = 0$ to $1$). The color gradient for each table cell changes with the square root of the proportion of connected edges on the whole possibly connected edge set.}
\label{figheat2}
\end{figure}
 \begin{figure}[H] 
\centering
\includegraphics[width=1\textwidth]{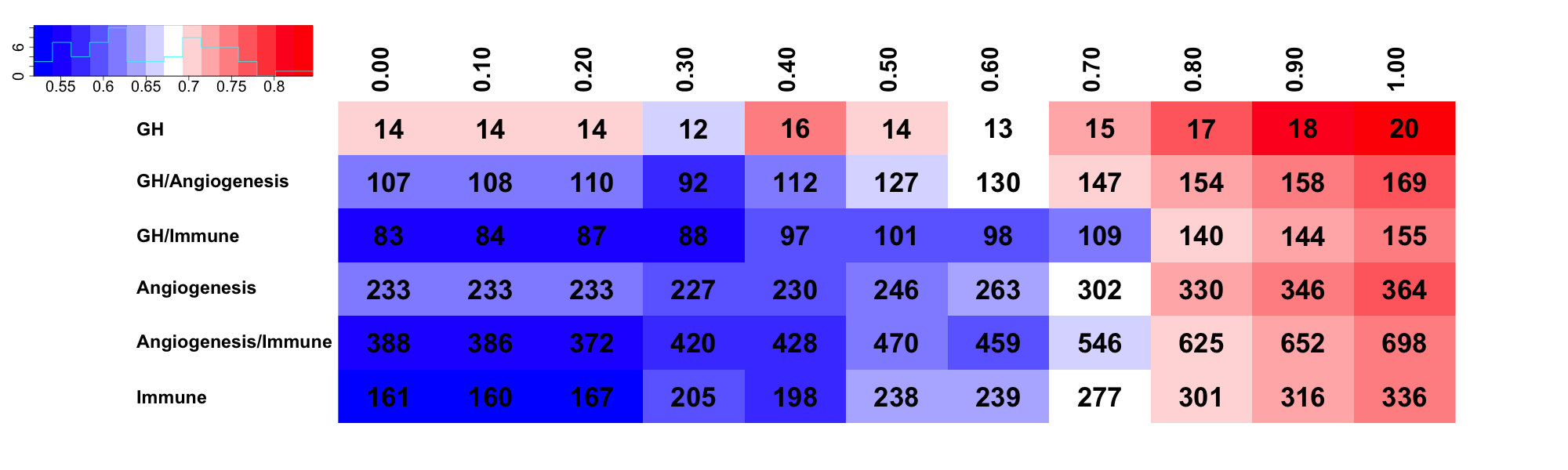}
\caption{\small NFMGGM: The number of connected edges within each pathway/across different pathways varies with different \textit{HepatoScore}s (from $\pi = 0$ to $1$). The color gradient for each table cell changes with the square root of the proportion of connected edges on the whole possibly connected edge set.}
\label{figheat3}
\end{figure}

\end{document}